\documentclass[fleqn,usenatbib]{mnras}

\usepackage{newtxtext,newtxmath}

\usepackage[T1]{fontenc}

\DeclareRobustCommand{\VAN}[3]{#2}
\let\VANthebibliography\thebibliography
\def\thebibliography{\DeclareRobustCommand{\VAN}[3]{##3}\VANthebibliography}

\usepackage{graphicx}	
\usepackage{amsmath}	
\usepackage{hyperref}
\hypersetup{colorlinks=True,citecolor=blue}
\usepackage{xcolor}

\def\ms{\hbox{\,m\,s$^{-1}$}}         
\def\cms{\hbox{\,cm\,s$^{-1}$}}       
\def\m2s2{\hbox{\,m$^{2}$\,s$^{-2}$}} 
\def\kms{\hbox{\,km\,s$^{-1}$}}       

\def\S/N{$S/N_{cont}$}

\def\ang{\text{\AA}}

\def\Halpha{H\hspace{0.1ex}$\alpha$ }

\def\Ca IIK{Ca\,II\,\,K }
\def\Ca IIH{Ca\,II\,\,H }
\def\Ca IIHK{Ca\,II\,\,H {\&} K }
\def\K3{K$_3$ } 
\def\K2{K$_2$ }
\def\K2v{K$_{2V}$ }
\def\K2r{K$_{2R}$ }
\def\Mghk{Mg\,II\,\,h {\&} k }

\def\Sidx{$S_{\text{index}}$ }

\title[Stellar surface information from Ca H\&K lines. I.]{Stellar surface information from the Ca II H\&K lines \newline I. Intensity profiles of the solar activity components}

\author[M. Cretignier]{
M. Cretignier,$^{1}$\thanks{E-mail: michael.cretignier@physics.ox.ac.uk}
A.G.M. Pietrow,$^{2}$
and S. Aigrain$^{1}$
\\
$^{1}$Department of Physics, University of Oxford, OX13RH Oxford, UK \\
$^{2}$Leibniz-Institut für Astrophysik Potsdam (AIP), An der Sternwarte 16, 14482 Potsdam, Germany\\
}

\date{Accepted XXX. Received YYY; in original form ZZZ}

\pubyear{2023}

\begin{document}
\label{firstpage}
\pagerange{\pageref{firstpage}--\pageref{lastpage}}

\maketitle

\begin{abstract}
   The detection of Earth-like planets with the radial-velocity method is currently limited by the presence of stellar activity signatures. On rotational timescales, spots and plages (or faculae) are known to introduce different RV signals, but their corrections require better activity proxies. The best-known chromospheric activity proxies in the visible are the Ca II H \& K  lines, but the physical quantities measured by their profiles need to be clarified. We first investigate resolved images of the Sun in order to better understand the spectrum of plages, spots, and the network using the Meudon spectroheliogram. We show that distinct line profiles are produced by plages, spots, and by the network component and we also derived the center-to-limb variations of the three profiles. Some care is required to disentangle their contributions due to their similarities. By combining disk-integrated spectra from the ISS high-resolution spectrograph with SDO direct images of the Sun, we managed to extract a high-resolution emission spectrum of the different components, which tend to confirm the spectra extracted from the Meudon spectroheliogram datacubes. Similar results were obtained with the HARPS-N Sun-as-a-star spectra. We concluded using a three-component model that the temporal variation of the popular \Sidx contains, on average for the $24^{\rm th}$ solar cycle: $70\pm12\%$ of plage, $26\pm12\%$ of network and $4\pm4 \%$ of spots. This preliminary investigation suggests that a detailed study of the Ca II H \& K profiles may provide rich information about the filling factor and distribution of different types of active regions. 
\end{abstract}




\section{Introduction}

Stellar activity is today the main limitation for the upcoming ultra-stable spectroscopic era, which targets a radial velocity (RV) precision below 50 \cms{}. Among the stellar activity components, faculae/plages and spots are known to dominate in the RV budget of Sun-like stars \citep{Meunier(2022)}. 

Sunspots are the most readily visible, and perhaps best known, manifestations of inhomogeneities in the solar magnetic field concentration. A recent detailed review on sunspots were given by \citet{Solanki2003} and \citet{Borrero(2011)}. They tend to emerge from a region with enhanced magnetic activity within the span of several days and have been observed to persist up to two months \citep{Schrijver(2000)}. They are characterized by a dark core known as the umbra, surrounded by a brighter, radially extending structure called the penumbra. A spot that lacks the penumbra is called a pore \citep{Verma2014} and has a shorter lifetime \citep{Keppens00}. 

The intensity in the visible of umbra and penumbra as compared to the quiet Sun continuum tends to be 40\% and 80\% respectively \citep{Lites(1993),Beck(1993)}, with a magnetic field orientation vertical in the umbra and mostly horizontal in the penumbra \citep{Lites(1993),Skumanich(1994)}. The photospheric magnetic field in the umbra varies gradually between 1800 and 3700~G, and then lowers to a much lower range of 700 to 1000~G for the penumbra \citep{Brants(1982),Beck(1993),Solanki2003,Borrero(2011)}, far exceeding the average quiet Sun magnetic field of 46~G \citep{Arjona2021}.


The definition of "plage" has always been, and to some degree still is, a loose observational term. In solar physics it tends to be used to describe both the photospheric and chromospheric parts of a region with a quiet-Sun-like photosphere in the continuum that is covered by a hot and bright chromospheric canopy that connects regions of opposite polarity. In white light, plage manifests itself on the disk as abnormal granulation and becomes bright close to the limb, where it is also known as 'faculae'. Plage can also be seen as bright inter-granular structures on the disk when using a narrow band filter. These structures have historically also been called faculae, which likely contributed to the muddiness of the definition \citep{Stix2002,Buehler2015,Carlsson2019,Pietrow2020}.

Plages accompany sunspots and pores, sometimes forming as a precursor to them, and also tend to outlive the spots, with lifetimes up to two or three times longer \citep{Foukal2003}. They are magnetic phenomena with photospheric fields around 1500~G \citep{Buehler2015}, stronger than the quiet Sun, but weaker than those of sunspots to the point where convection is hardly affected. Together with enhanced networks, plage accounts for about half of the total magnetic flux on the Sun, and are responsible for the majority of the variations in the UV flux, as well as the total solar irradiance \citep{Morosin2020,Pietrow2022,Joao2022, Chowdhury2022}. Faculae are often described as the photospheric counterpart of plage \citep[e.g.][]{Dumusque(2014)}, however they can also form in the quiet Sun as part of the enhanced network, or at the boundary of super granules \citep{Muller1983,Buehler2019}, and thus the word should be used with caution (e.g. all plages correspond to faculae, but not all faculae correspond to plage). In this work, we define plages as large (>8 millionth of solar hemisphere MSH), bright, continuous regions in chromospheric diagnostics, and faculae as the more concentrated bright areas underneath, excluding the darker quiet Sun-like pattern in between.

Network can be found in the quiet Sun, and typically presents itself as elongated chains of bright points nested within the intergranular lanes. Network tends to be much more spread out than plage and not connected to active regions \citep{Buehler2019}. While plage and network share many characteristics, such as being identified as bright points in the photosphere and expanding into a bright canopy in the chromosphere, several differences have been found between the two that can be used to distinguish them. The primary difference is that network tends to consist of narrower flux tubes than those found in the plage, resulting on average in temperatures that are about 200 K higher, a magnetic field that is about 150~G stronger, a 5\% increase in grain contrast, and a more vertically oriented field \citep{Solanki1986, Solanki1992, Buehler2019}. Despite the higher temperature, magnetic field strength and inclination of the individual flux tubes, the relatively lower fill-factor of said tubes results in a less bright chromospheric canopy compared to plage \citep[e.g.][]{Worden1998,Worden1999}. In this work we define network as bright regions in the chromosphere that are above the average brightness of the quiet chromosphere, and below the brightness cutoff for plage.

\subsection{Standard activity proxies}

An activity proxy is an observable that aims to measure a given property of an active region (AR) or that aims to be at least proportional to it. Usually, most activity proxies were dedicated to be a tracer of the total filling factor $f_{\text{tot}}(t)$ of active regions.

In photometry, both spots and faculae have been widely investigated with results sometimes in contradiction such as for faculae \citep{Chapman(1982),Unruh(1999),Foukal(1990),Berger(2007),Toriumi(2020)}. For starspots, results are more established and their modulation has been largely investigated thanks to large photometric surveys \citep{Messina(2002),Mirtorabi(2003),Lanza(2009),Bonomo(2012),Nielsen(2013),Roettenbacher(2013),Lanzafame(2018),Breton(2021)}. However, as faculae are dimmer in photometry, yet dominate in RVs due to their larger filling factor \citep{Meunier(2015)}, stable photometry is not sufficient to forecast suitable RV standard stars \citep{Bastien(2014)}. Furthermore, bright and dark regions can cancel out in photometry, whereas their effects add up in RVs. 

High resolution spectra have also been used to investigate active regions. Indeed, reaching an RV precision better than 1 \ms usually requires high resolution spectra ($\mathcal{R}>100\,000$), from fiber-feed spectrographs. Most of them working in the visible due to the richness of atomic transitions in the blue part of the spectrum for G to K dwarfs stars and the absence of strong tellurics bands. Because of the conservation of the etendue \citep{Schroeder(2000)}, small fibers are needed to avoid the development of large optical elements such as the grating in high-resolution instruments, which would be too expensive. Consequently, the stellar point-spread function (PSF) on sky is larger than the fiber sky-projected diameter and part of the stellar light is not captured \citep{Pepe(2021)}, leading to absolute flux variations as large as 50\% induced by seeing variations, guiding errors or change in weather conditions. As a consequence, high-resolution spectra are usually not photometrically calibrated, which prevents the photometric signatures of ARs to be measured. Even if projects of diffraction-limited spectrographs\footnote{aiming to photometric accurate spectra} using adaptive optics are being investigated \citep{Mello(2018),Jovanovic(2016),Bechter(2018),Bechter(2019)}, none of them are currently in use. 

From spectra, the extraction of information on faculae/plages and spots is more difficult. Due to the color dependency of spots \citep{Desort(2007),Carleo(2020),Kossakowski(2022),Miyakawa(2021)}, some authors tend to use chromatic indices such as differential RVs \citep{Zechmeister(2020)}, whereas other have tried to measure the magnetic field via Zeeman broadening \citep{Robinson(1980),Berdyugina(2003a),Anderson(2010),Reiners(2013), Lienhard(2023)} or molecular bands \citep{Afram(2015),Oneal(1997),Oneal(1998),Oneal(2001),Oneal(2004)}. 

In general, the whole spectrum tends to be affected by stellar activity. For example, other stellar lines distortions such as filling of the line core has been reported many times \citep{Stenflo(1977),Basri(1989),Brandt(1990),Malanushenko(2004),Jaime13,Druett2017,Thompson(2017),2021Vissers,Cretignier(2021),Druett2022,Pietrow2022a}. More generally, using the varying sensitivity of different stellar lines has been proposed as a way to reduce degeneracies in models. As an example, combining low and high temperature sensitive lines can recover the effective temperature of stars  \citep{Gray(1988),Gray(1991),Caccin(2002),Teixeira(2016),Joao2018} and this idea has been naturally extrapolated for active regions as well  \citep{Chapman(2001),Catalano(2002),Penza(2004),Frasca(2005),Oneal(2006)}. Using lines formed at different depth in the stellar atmosphere also demonstrated interesting insights on the effect of stellar activity on the convective motions \citep{Cretignier(2020a),Moulla(2023)}.

A simple element that would help to disentangle AR contributions would be to know the spectrum of plages/faculae and spots across the solar disk $I_{\text{AR}}(\lambda,\mu)$ over a large bandpass and at high-resolution, the variable $\mu=\cos(\theta)$ representing the heliocentric angle and $\theta$ the angle between the observer and the normal to the surface vector ($\mu=1$ at stellar center and $\mu=0$ at the stellar limb).

While such studies have been conducted for the quiet Sun in recent years \citep{ramelli17,ramelli19,Pietrow2023,Ellwarth23}, there is still a lack of information for ARs center-to-limb variation (CLV). Such products do not exist or only for peculiar lines \citep{Avrett(2015),Rajhans(2023)} and projects are in preparation, such as a solar telescope on ESPRESSO\footnote{\url{https://poet.iastro.pt}}. A sunspot atlas spectrum can be found in \citet{Wallace(2005)} and is already used to simulate stellar activity \citep{Dumusque(2014)}, but such a spectrum was only taken at a single $\mu$ angle and similar observations do not exist for facula. 

Even if the full spectrum is sometimes studied, researchers tend rather to investigate the chromospheric signatures of the ARs. The hot plage canopy is visible in most chromospheric diagnostics such as the \ion{Ca}{II} \citep{Shapiro(2014),Milbourne(2019),Pietrow2020,Morosin2020}, Magnesium \citep{Sasso(2017),Li2023L}, and \Halpha  \citep{SocasNavarro(2004),Frasca(2008),Flores(2016),Flores(2018)}. Note that, in the context of RV, the question of good activity proxies is still an open question for which the answer may depends on the stellar spectral type \citep{Meunier(2009),Robertson(2016),Maldonado(2019),Schofer(2019), Lafarga(2021)} and recent studies tend to show that the best activity proxy for solar-type star is the unsigned magnetic field \citep{Milbourne(2019),Haywood(2022)}.

\begin{figure}
	
	\centering
	\includegraphics[width=8.5cm]{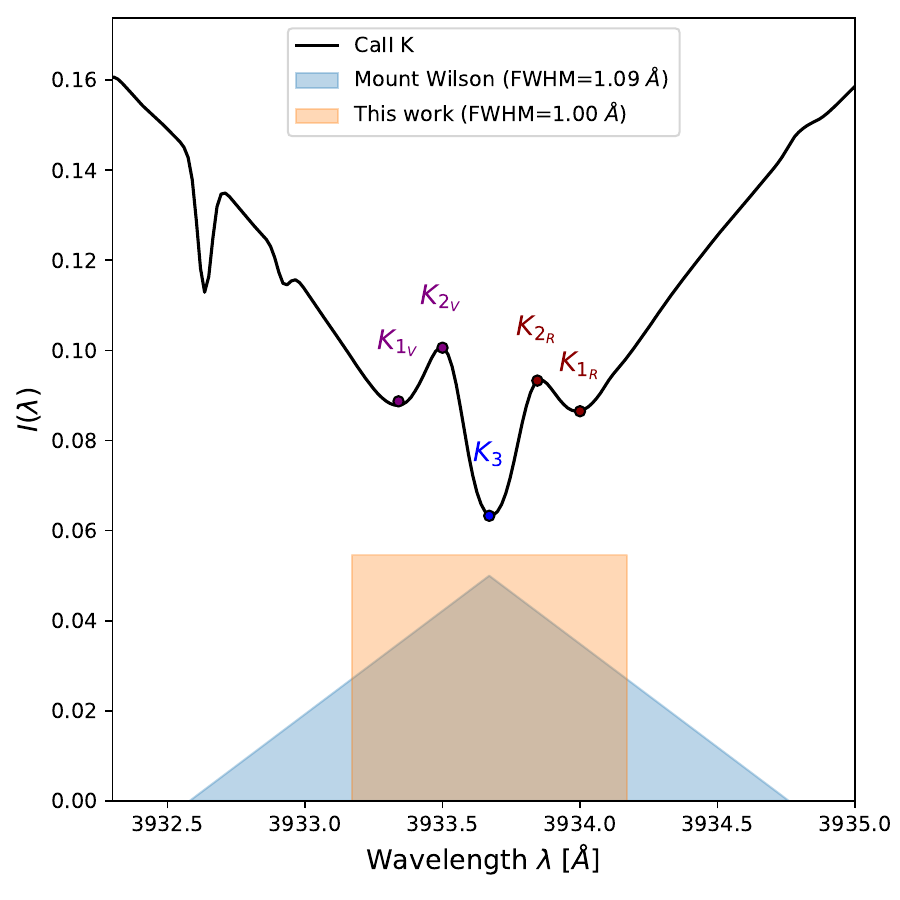}
	\caption{Illustration of the Ca II K line profile obtained by the ISS spectrograph ($\mathcal{R}\sim 300'000$). The peculiar K1 minima, K2 maxima and K3 minimum are indicated. The Mount Wilson triangular bandpass, from which the Mount-Wilson \Sidx ($S_{\text{MW}})$ is derived, is also indicated at the bottom superposed on the bandpass used in this work to derive \Sidx.}
	\label{FigWilson}

\end{figure}

\subsection{The physics of the Ca II H\&K lines}

The most widely used chromospheric lines are the Ca II H \& K  lines. They are amongst the opaquest lines in the visual spectrum and, like the \Mghk lines, they typically have wide damping wings and a central reversal in their cores. The wings of the lines sample the higher photosphere up to the temperature minimum, and can be modelled in LTE \citep{Rutten04}. The line cores have a characteristic reversed shape (see Fig.\ref{FigWilson}) that is caused by the rise in temperature in the chromosphere above the temperature minimum. Afterwards the line decouples from the Planck function and an absorbing core forms \citep{Vernazza(1981),Leenaarts2013,Bjorgen2018}. The H1/K1 minima correlate with the temperature at the higher photosphere, also known as the temperature minimum. However, it has been shown that this value is underestimated due to the line being already partially decoupled from the Planck function \citep{Shine1975,Bjorgen2018}. The H2/K2 peaks form in the lower chromosphere with the R peaks forming lower than the V peaks. Their brightness temperature correlates with the gas temperature in the lower chromosphere, but again should not be taken as an absolute value. The peaks, along with the H1/K1 minima, move apart when observed further away from disk center, and tend to disappear altogether when averaged over large areas \citep{Wilson(1957),White(1978),Druzhinin87,Pietrow2023}. The H3/K3 minimum forms in the higher chromosphere and can be used to measure the vertical velocity at this height.

The strong variation in line shape of Ca II H \& K as a result of chromospheric activity \citep[e.g.][]{Jaime13,NagaVarun2018} combined with broad sensitivity across various heights in the solar atmosphere, makes these lines an excellent activity probe. Other peculiar distortions of the lines were reported such as the broadening of the line as measured by distance of the K2 maxima \citep{Wilson(1957)} or line asymmetry \citep{White(1978)}. Some authors have tried constructing different metrics from the lines, and attempting to interpret them based on physical solar quantities \citep{Pasquini(1992),Livingston(2007),Bertello(2012),Dineva(2022)}, but no clear relations  have been found.

Usually, the study of the Ca II H \& K lines is mainly performed via a bandpass integration in the core of the lines, that people called \Sidx metrics. The most widely used convention is the Mount Wilson \Sidx ($S_{MW}$) that is characterized by a triangular band pass integration \citep{Duncan(1991)}. 

In this work, we performed a wide investigation on the solar Ca II H \& K lines, to better understand how ARs modify their line profiles. We used the Sun as a starting point, since we can test it against resolved observations. We then used the knowledge obtained on the Sun to better characterize the stellar activity on $\alpha$ Cen B in a subsequent paper II \citep{Cretignier(2024)}. 

The paper analyses the data coming from four solar instruments: the Meudon spectroheliograph, the SDO AIA1700 filtergrams, the disk-integrated Ca II~H\&K high resolution spectra from ISS and HARPS-N solar telescope. The structure of the paper is divided in two main parts. The first part (Sect.~\ref{sec:meudon}) focuses on the analysis of low resolution spectra obtained from spatially resolved observations of the meudon spectroheliographs. The second part of the paper (Sect.~\ref{sec:secSDO_HARPS}) focuses on the combination of disk-integrated high resolution spectra with spatially resolved observation. Finally, we conclude in Sect.~\ref{sec:conclusion}. 

\section{Spatially resolved spectra from the Meudon spectroheliograph}
\label{sec:meudon}

The Meudon specroheliograph is a low-resolution ($R\sim40\,000$) slit spectrograph located at the Meudon observatory in Paris that takes daily observations of the Sun with three different filters (Ca II K, Ca II H, \Halpha). The instrument was upgraded in 2018 \citep{Malherbe(2019)} and can be used to study a vast diversity of solar active events \citep{Malherbe(2023)}. Contrary to several other Sun-as-a-star instruments that work with disk-integrated spectra or restricted field of view, the full solar surface is scanned by the slit spectrograph and a spectrum is extracted producing therefore a datacube $I(\lambda,x,y)$. Such datasets are crucial to better constrain stellar activity since they directly lead to the intensity profile $I(\lambda,\mu)$. We recall\footnote{see e.g \citet{Thompson(2006)} for a paper about change of units in solar images} that if $x$ and $y$ are the sky coordinates with their origin at the solar disk centre and given in fractions of the total radius, the center-to-limb angle $\mu$ is given by: $\mu=\sqrt{1-x^2-y^2}$. 

The section is divided as follows: the processing of the Meudon spectroheliogram datacubes is described in Sect.~\ref{sec:processingMeudon}. The analysis of peculiar events is discussed in Sect.~\ref{sec:meudon_analysis}. We then extracted the emission profile per unit area of plages, spots and of the network from Meudon datacubes in Sect.~\ref{sec:meudon_clv}, before to use numerical simulations to recreate the disk-integrated spectra of plages and spots in Sect.~\ref{sec:meudon_soap}.

\subsection{Datacubes processing}
\label{sec:processingMeudon}

The images have sizes of 2048 $\times$ 2048 pixels, whereas the wavelength axis is sampled by 98 and 94 data points for the Ca II K and Ca II H respectively. The wavelength step is 0.093 \ang{}, equivalent to a spectral resolution of $R\sim40\,000$. The stellar radius is about 880 pixels on average\footnote{Due to the eccentricity of the Earth's orbit.} leading to an angular resolution of 1.1", providing slightly more than $\pi\times 880^2 \simeq$ 2 million spectra per observation. One pixel covering roughly 0.41 MSH. 

\renewcommand{\arraystretch}{2.0}
\begin{table}
\caption{Table of the dataset observations downloaded from Meudon spectroheliograph. Six events of active regions crossing the solar disk were selected. A spot dominated event in September 2017, a plage event in May 2018 and 4 events in 2022, the most active solar activity period. A quiet period in 2020 was also selected for sanity check of segmentation algorithms. The first day ($T_1$) and the last day ($T_2$) of each period as well as the number of datacubes ($\# D_3$) extracted for each line is displayed. The total is displayed in the last row. Each datacube contains roughly 2\,000k of spectra. The number of spectra extracted after the segmentation of the datacubes are also indicated. Both the number of spectra for spots ($\#_s$) and plages ($\#_p$) are given in thousands of units (k).}
\label{Table1}
\centering
\begin{tabular}{ccccc}
\hline\hline
$T_1$ [date] & $T_2$ [date] & \#$D_3$(K/H) & $\#_s$(K/H) & $\#_p$(K/H)\\
\hline

01-09-2017 & 07-09-2017 & 6/6 & 37/37 & 122/96 \\
24-05-2018 & 31-05-2018 & 9/10 & 0/0 & 63/61 \\
21-02-2020 & 29-02-2020 & 6/6 & 0/0 & 10/21 \\
18-04-2022 & 27-04-2022 & 10/10 & 48/51 & 709/665 \\
15-05-2022 & 24-05-2022 & 12/15 & 30/48 & 1176/1295 \\
18-06-2022 & 24-06-2022 & 6/6 & 12/15 & 289/259 \\
06-07-2022 & 18-07-2022 & 24/24 & 93/116 & 1255/820 \\

\hline
 - & - & 73/77 & 221/267 & 3624/3216\\

\hline
\end{tabular}
\end{table}

All the observations are publicly available and can be obtained via a web page\footnote{\url{https://bass2000.obspm.fr/home.php}} query.
Due to the size of a single datacube ($\sim$ 2048 $\times$ 2048 $\times$ 100), only 6 events of ARs crossing the stellar disk were analysed. In total, 73 and 77 epochs were hence obtained for the Ca II K and Ca II H lines respectively. Among the 6 events, we selected an event dominated by a spot in September 2017, a single plage event in May 2018 and four periods with large number of ARs in 2022. The datasets in 2017 and 2018 were taken at the minimum of the solar cycle, but represent ARs that are typical in size. The 2022 probes a mid-activity solar level. We also extracted datacubes for a period when no clear AR was visible in 2020. This dataset will be used as a quality control for the segmentation algorithm. A summary of the observations can be found in Table.\ref{Table1}. Even if the number of epochs analysed may seem low, it already represents a total of 51 million of spectra to analyse.

\subsubsection{Quiet Sun CLV correction}

All datacubes were first normalised by the center-to-limb variation of the quiet Sun. Such normalisation, also known as the third solar spectrum \citep[SS3,][]{Stenflo2015,ramelli19}, are common way to analyse the residual intensities \citep{Harvey(1999)}. The CLV law was fit independently for each wavelength element, producing as well CLV datacubes. The solar surface was first binned into a 15 $\times$ 15 grid where the median intensity inside each cell was computed in order to remove the signatures of active regions that will never dominate in area compared to the quiet Sun.

The CLV was approximated by a 5th order polynomial following \citet{Neckel1984} and \citet{Neckel1994}. In order to obtain the coefficients of the CLV relation, we simultaneously fit a basis of 2D polynomial function of degree 10 in $x$ and $y$ directions of the images:

\begin{equation}
\label{eq:3}
\begin{array}{rccccl}
I(\lambda,x,y) & = & P_{10}(\lambda,x,y) & + & \text{CLV}_5(\lambda,x,y) \\
& = & \displaystyle\sum_{i,j=0}^{i+j\le 10} C_{i,j} (\lambda)\cdot x^i \cdot y^j & + & \displaystyle\sum_{k=0}^5 c_k(\lambda) \cdot \mu^{k} 
\end{array}
\end{equation}


The first term was necessary to correct for instrumental systematics\footnote{see also \citet{Shen(2018)} for similar systematics and correction using Zernike polynomials}, while the the second is the CLV relation of the quiet Sun. Note that since CLV is described by radial polynomials, we removed the components of the 2D polynomial basis that were already present in the CLV relation\footnote{which is the case when $i=0$ and $j$ is a even number, or $j=0$ and $i$ is even.}.

Even if dividing out by CLV profiles of the quiet Sun allows to precisely study the residuals, a first issue was related to the absolute flux level of the CLV profiles. These are caused by variable observational conditions that affect the absolute flux level, and the difficulty of setting the continuum level of the lines due to the restricted bandpass of the instrument. An inaccurate flux level will prevent the comparison of the line profiles obtained by different instruments at later stage due to different flux units. To fix it, we used as a reference profile the HARPS-N master quiet spectrum (see Sect.~\ref{sec:harpn}), since HARPS-N has a much wider bandpass that allows to fit a more reliable and accurate continuum. 

The HARPS-N master quiet spectrum was degraded to the Meudon spectral resolution in order to produce a reference spectrum $I_{\text{ref}}(\lambda)$, while each CLV law was disk-integrated to produce a master quiet disk-averaged stellar line $I_{\Sigma}(\lambda)$. In order to determine the proper scaling, we initially tried to fit a simple linear or quadratic relation with a free offset between the two profiles, but doing so often led to non-physical units, such as negative flux values close to the limb. In order to avoid this issue, we removed the free offset in the model and imposed a boundary condition on the fit, namely that the minimum of the CLV datacubes\footnote{which is expected to be at the limb of the core of the line: $I_{\text{min}} \simeq \text{CLV}_5(\lambda_c,\mu=0)$} $I_{\text{min}}$ goes to 0. In practice, such conditioned change of units is obtained by fitting the quadratic relation between the master quiet spectrum of both instruments: $I_{\text{ref}}(\lambda) = a\cdot (I_{\Sigma}(\lambda)-I_{\text{min}}) + b\cdot (I_{\Sigma}(\lambda)-I_{\text{min}})^2$. The parameters $a$ and $b$, and the constant $I_{\text{min}}$ can then be used to scale the Meudon flux units toward HARPS-N flux units.

\subsubsection{Correction of residual systematics}

\begin{figure*}
	
	\centering
	\includegraphics[width=18cm]{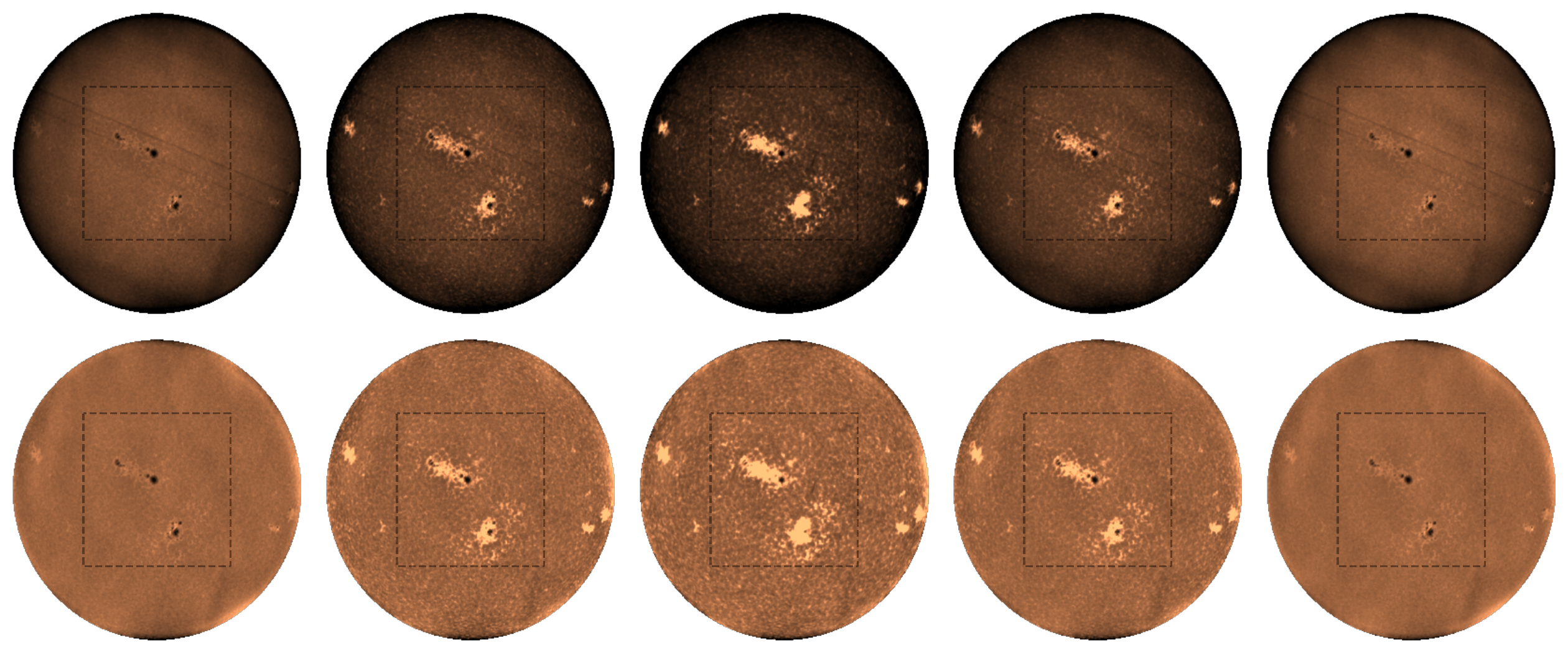}
	\caption{Representation of a subpart of the Meudon spectroheliogram datacubes (5 of the 96 wavelengths) for an observation taken on 4 September 2017. \textbf{Top}: Raw datacubes. The images of the Sun at five different wavelength around the line center $\lambda_c$ of Ca II K are plotted (see Fig.\ref{FigMeudon2}). From left to right: $\lambda_c-0.6$, $\lambda_c-0.3$, $\lambda_c$, $\lambda_c+0.3$ and $\lambda_c+0.6$ \ang{}. A zoom on the central part of the Sun, containing most of the active regions, is displayed in Fig.\ref{FigMeudon2}. \textbf{Bottom}: Datacubes residual intensities after fitting out the quiet CLV and correcting from instrumental systematics.}
	\label{FigMeudon1}

	\centering
	\includegraphics[width=18cm]{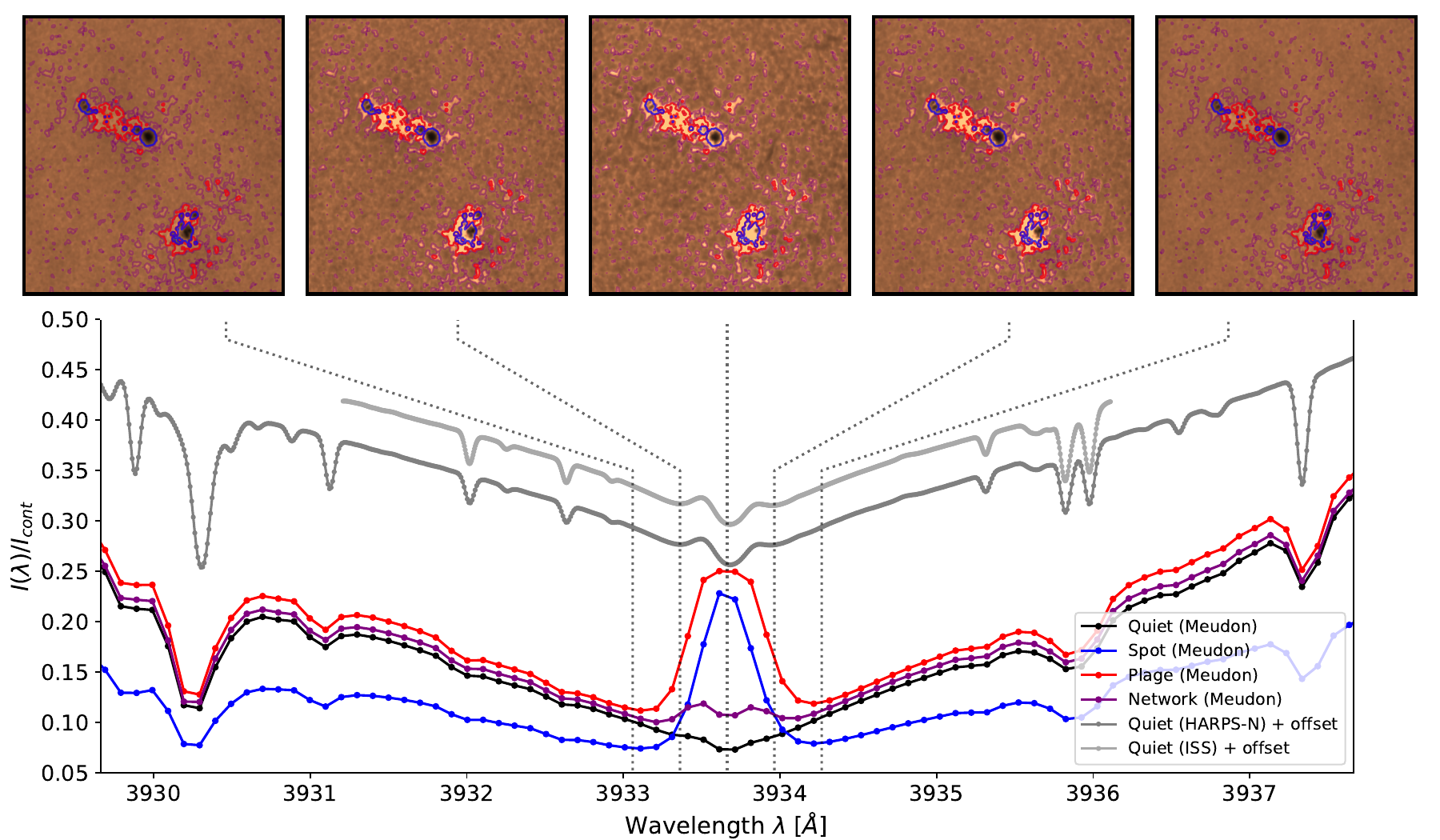}
	\caption{Extracted spectra from the Meudon spectroheliogram for 4 September 2017. \textbf{Top panel:} Segmentation maps for spots (blue), plages (red) and network (purple) is represented for the central part of the Sun (see dashed square in Fig.\ref{FigMeudon1}). The wavelengths displayed are the same as in Fig.\ref{FigMeudon1}. \textbf{Bottom panel:} Average spectrum $I_{\text{AR}}(\lambda,\mu\simeq1)$ obtained from the segmented regions. The quiet spectra from Meudon ($R\sim40\,000$), HARPS-N ($R\sim120\,000$) and ISS ($R\sim300\,000$) are displayed as comparison respectively in black, dark grey and light grey. HARPS-N and ISS spectra are shown with arbitrary offsets.}
	\label{FigMeudon2}
 
\end{figure*}

After such processing, several artefacts in the form of diagonal lines can be observed on the images (already visible on the raw datacubes as displayed in the top row of Fig.\ref{FigMeudon1}). Those artefacts are likely created by scattered light in the spectrograph or changes in weather conditions during the scanning by the slit spectrograph (a scan being typically 1 min long). To complicate matters, the angles of those rays change from one day to the next. However, the angle remains identical for all the wavelengths. For that reason, the mean along the wavelength axis was taken in order to produce a high signal-to-noise ratio master image of the Sun $I_{\text{master}}(x,y)$ used to search for the angle of the rays.

The algorithm used to detect the rays is a "rotating-stacking algorithm" similar to the one developed in \citet{Cretignier(2021)} to detect the ghosts on the HARPS detector. The algorithm consists in computing the median of the pixel intensity for a given angle direction $\theta$, before computing the standard deviation of the distribution. The variance of the distribution will be the smallest once pixels are summed along the best angle, while a flat distribution is expected when summing along a random angle. The optimal angle $\theta_{\text{max}}$ is found with a grid search between -$90^{\circ}$ and $90^{\circ}$ with steps of $0.25^\circ$. Once the optimal angle $\theta_{\text{max}}$ is found, the median along the angle is used to correct the image. The correction is performed for each wavelength independently. We will refer hereafter to these images as the \textit{flat corrected images}. 

An example of the datacubes, after that the CLV of the quiet Sun has been fit out and straight rays corrected, is shown in the bottom row of Fig.\ref{FigMeudon1}. Some minor residuals remains, but they are not a matter of concern, as they will be diluted when spectra from a given $\mu$ angle will be stacked at later stage (see Sect.~\ref{sec:meudon_analysis}), taking advantage of the significant number of epochs processed and the number of ARs.

\subsubsection{Identification of ARs}

Having corrected for the CLV of the quiet Sun $I_{\text{quiet}}(\lambda,\mu)$, we segmented the flat corrected images, before extracting the relative intensity profiles of the AR compared to the quiet chromosphere, namely $I_{\text{AR}}(\lambda,\mu)/I_{\text{quiet}}(\lambda,\mu)$. The results of the segmentation are displayed in Fig.\ref{FigMeudon2}. The cutoff values were set by eye as good trade-off values. The thresholds were defined as follows: we considered a spot as any pixel with a flux intensity on the first wavelength element (located on the wing of the Ca II H\&K lines at $\lambda=3964.097$\ang{} and $\lambda=3929.292$\ang{}) lower than 75\% the quiet photosphere, which usually selects the umbrae and penumbrae of the spots. Note that the spatial resolution of the Meudon spectroheliograph does not allow an easy identification of both components and disentangling umbrae from penumbrae on stellar spectra is unlikely at the moment. On the other hand, we considered as a plage any pixel brighter than 225\% the solar quiet chromosphere in the core of the Ca II lines and brighter than 95\% the quiet photosphere using the first wavelength element. The last condition was mandatory since, as showed in Sect.~\ref{sec:meudon_analysis}, spots also present a clear emission in the core of the line which is also visible in Fig.\ref{FigMeudon2} for the most southern spot in the image. The 30\% of pixels closest to the poles were excluded since ARs are not expected above latitudes $l\simeq 45^\circ$ according to the butterfly diagram \citep{Yabar2015}. No active region at limb angle smaller than $\mu<0.30$ were attempted to be detected due to the large noise close to the limb enhanced by the border instability in the CLV fit. 

We tested the segmentation on the 2020 quiet dataset to assess the level of contamination expected. No region was flagged as spotted; however, we observe that some pixels were classified as plages likely due to stochastic noise in the core of the Calcium lines or network pixels. The average contamination per frame is however lower than 0.001\% of the stellar surface and are therefore negligible validating our choice of the threshold for both types of ARs. 

We also used the 2020 dataset to extract the network intensity spectra, since no other activity components were expected during this period of time. The network was detected by a brighter pixel condition using the average of the 5 wavelength elements the closest to the line center. The lower boundary segmentation criterion was however slightly more difficult to set, compared to plage for instance, since this component is closer to the noise precision of the datacubes. Indeed, this threshold helps to disentangle real signals from noise, which can not be set with a fixed threshold since the value will depend on the S/N of the observations.

We used the property that no relative intensity smaller than 1 is expected from any activity components during this period of time and such values are only produced by the stochastic noise in the images. We therefore compute the half $\sigma$-width of the distribution computed as the median $(I_{50})$ minus the $16^{\rm th}$ percentile ($I_{16}$) and used the same value projected on the positive side multiplied by 3 in order to use an equivalence to a $3\sigma$ definition. Any pixel brighter than  $I/I_{\text{quiet}}>I_{\text{min}}=I_{50}+3\cdot (I_{50}-I_{16})$ and not already classified as a plage was categorised as network.

Once images were segmented, the spectra from the 1) quiet Sun 2) plages 3) spots and 4) the network were extracted. The segmentation resulted in a large dataset of more than $\sim$250\,000 spectra for spots, $\sim$350\,000 spectra for the network and $\sim$3\,000\,000 spectra for plages. Ultimately, all the spectra were corrected for the solar rotation using a solid body rotation model $v_{\text{eq}}\cdot \sin i$ with an equatorial velocity $v_{\text{eq}}=2$\kms{}. The inclination of the Sun observed from the Earth was obtained from the JPL Horizons Ephemerids \citep{Giorgini(1996)}. Deviations from the solid body model (such as differential rotation) were neglected given the spectral resolution of the instrument.   

\subsection{Analysis of the intensity profiles $I_{\text{AR}}(\lambda,\mu)$ for peculiar events}
\label{sec:meudon_analysis}

We first investigated the spectra coming from Meudon spectroheliograph in order to see if substantial variations are observed as a function of $\mu$ angle along half a rotational phase ($\sim$ 13 days) of the Sun. We analysed two isolated AR crossing events: a spot-dominated and a plage-dominated event that took place in 2017 and 2018 respectively. We show the results obtained with the Ca II H line as there were more observations available of these events when compared the Ca II K line, however the results of the latter are very similar. The 2020 dataset was used to study the spectra of the network. 

\subsubsection{The 2020 quiet Sun}
\label{sec:meudon_quiet}

\begin{figure}
	
	\centering
	\includegraphics[width=8.5cm]{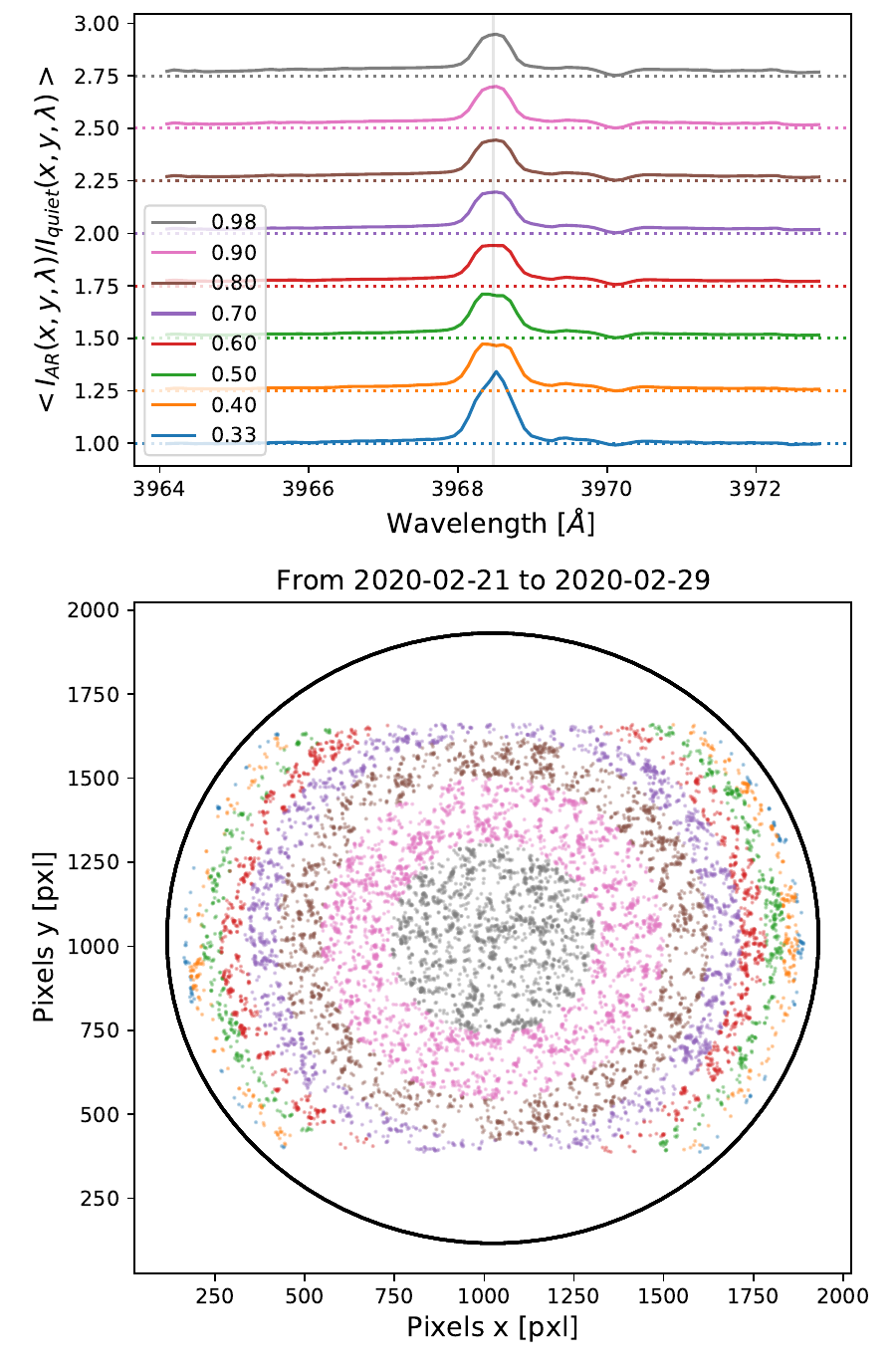}
	\caption{Extraction of the profiles on the Ca II H line for the quiet phase in 2020. The color highlights different $\mu$ bin angle and values are reported in the legend. Note that all the observations of the 2020 dataset have been agglomerated together. 
    The profile of the network does not significantly change from center-to-limb and is described by a core 20\% brighter than the quiet solar chromosphere. The small deep on the right side of the core emission is the Balmer $H_{\epsilon}$ line.}
	\label{FigPhaseN}

\end{figure}

We used the 2020 solar dataset to extract the signal from the network, because no other ARs were visible at that time. Since it is difficult to follow a single network region during a rotational phase, we displayed in Fig.\ref{FigPhaseN} the superposition of all the frames used in the 2020 dataset. The filling factor for the network was in average around 2.7\% which is slightly to large compared to the SDO value around 1.0\% (see Sect.~\ref{sec:sdo}) and may highlight that the segmentation also picked up spurious pixels. The implications of this will be discussed later. 

We chose to highlight the CLV by binning the network in different $\mu$ angle rings, which shows how the $\mu$ angle changes across the surface. This figure gives a better appreciation of how close to the solar limb are $\mu<0.30$ spectra and hence why they are so difficult to extract.  

The network has a weak core emission and is roughly 20\% brighter than the quiet solar chromosphere. We can also observe that the profiles becomes slightly broader closer to the limb, but such effect will be confirmed later (see Sect.\ref{sec:meudon_clv}). Interestingly enough, we also observe a deep at 3970\ang{}, which is the location of the $H_{\varepsilon}$ Balmer line. 

\subsubsection{The 2018 plage event}
\label{sec:meudon_plage}

\begin{figure}
	
	\centering
	\includegraphics[width=8.5cm]{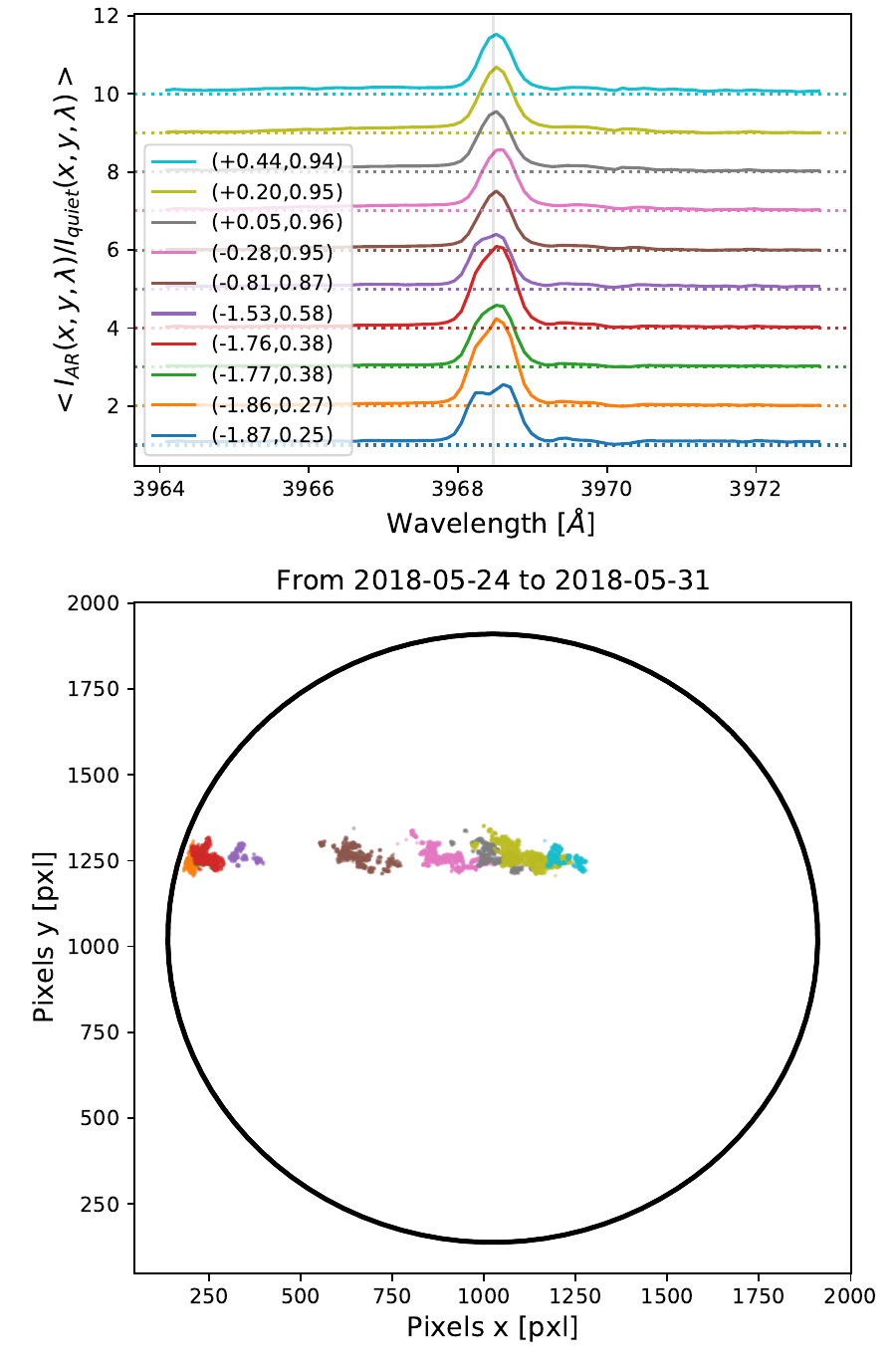}
	\caption{Similar to Fig.\ref{FigPhaseN} for the $30$ May 2018 plage-dominated crossing event. The solar rotation vector is oriented to the top. Only spectra at the AR expected latitude and longitude were extracted. \textbf{Bottom}: Segmentation map for the plages, each color represents an observation taken on a different day. \textbf{Top}: Mean relative intensity profiles extracted from the segmentation maps presented below, using ther same color codes as the bottom panel. An arbitrary offset value was applied for each curve, but the quiet chromospheric level is highlighted for comparison (dotted lines). The average location of the active region ($v,\mu$) is written in the legend. The velocity $v$ is in \kms.}
	\label{FigPhaseP}

\end{figure}

We analysed the series of datacubes taken during the May 2018 plage-dominated event, which first appeared on the eastern limb on 27 May 2018. This region was given the NOAA designation of 12712 and had a filling factor of $f\simeq 1\%$ at disk center. 
In Fig.\ref{FigPhaseP}, we show the extracted intensity profiles of the plage at 10 different locations on the solar disk. Instead of displaying the average location of the AR in the ($x$, $y$) sky-coordinates, we used the ($\nu$, $\mu$) coordinates system, with $\nu$ being the local radial velocity due to the stellar rotation ranging from $-2$ to $2$ \kms from left to right, and $\mu$ the cosine of the heliocentric angle $\theta$. To avoid contamination from other ARs, only spectra around the expected longitude and latitude of the AR were extracted.

We can observe that the relative intensity of the plage spectra with the quiet chromosphere does not change significantly from the limb to the solar center. The classical double peak signature of the lines is not visible, except for $\mu<0.3$. The reason for this is the relatively low spectral resolution of the instrument. Because the separation of the double H2 peaks increases towards the limb \citep{Smith(1960),Bjorgen(2018),Pietrow2023a}, this effect can be detected only for the smallest $\mu$ values. 

The present example shows that, at least for that specific AR, no strong variation is observed from one exposure to the next day and the profile remains coherent across the rotational phase. Small variations are still observed, but those are small compared to the main line profile. The intensity profile of a plage is slightly brighter in the wings, whereas the core is made of a single core emission twice as bright as the quiet chromosphere. This "universality" of the intensity Ca II H \& K  plage profile will be confirmed later when a larger dataset of ARs will be analysed (see Sect.~\ref{sec:meudon_clv}). 

\begin{figure}
	
	\centering
	\includegraphics[width=8.5cm]{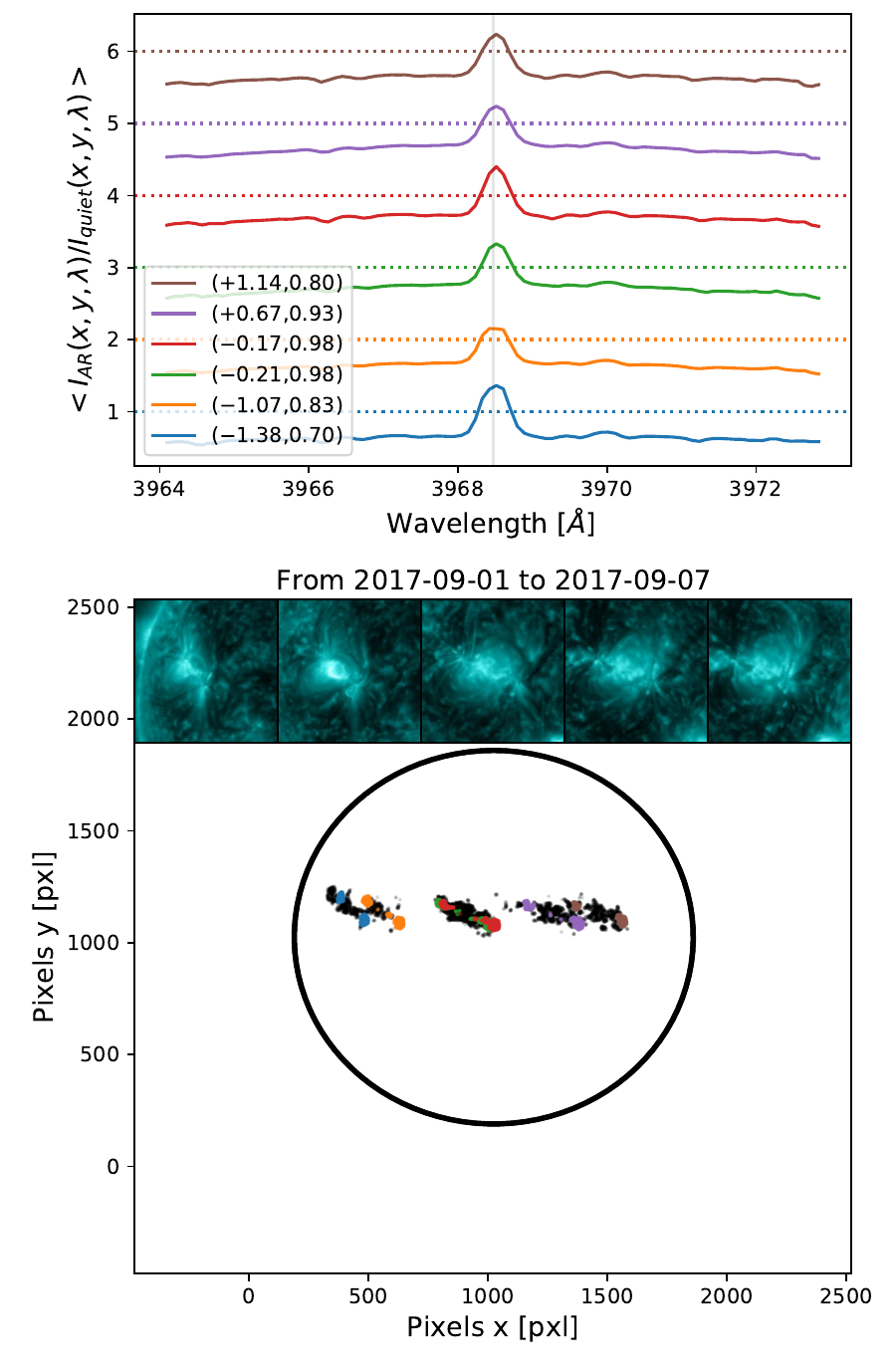}
	\caption{Same as Fig.\ref{FigPhaseP} for the spot-dominated crossing event, on $4^{}$ September 2017. Only the stable sunspots from the northern hemisphere were extracted. Segmentation maps of plage were also indicated (grey area). The profiles are characterized, as expected, by dark wings, but also a clear core emission. Depending on the configuration of plage and spots (see main text), an enhancement of the core emission can be seen, as displayed for the south hemisphere in Fig.\ref{FigPhaseS2}. Inner panels represent the SDO AI131 filtergrams probing the corona.}
	\label{FigPhaseS1}

\end{figure}

\subsubsection{The 2017 spot event}
\label{sec:meudon_spot}

 After investigating the behaviour of a plage and of the network, we investigated the spot event that started on 30 August 2017 when the AR appeared and was assigned the NOAA designation of 12674. This region comprised out of two large spots that were interconnected by a plage region. Additionally, a smaller spot (NOAA 12673) was visible on the southern hemisphere during this crossing. The spots on the northern hemisphere remained largely unchanged during their crossing, while the southern spots evolved rapidly once they crossed the meridian.

\begin{figure}
	
	\centering
	\includegraphics[width=8.5cm]{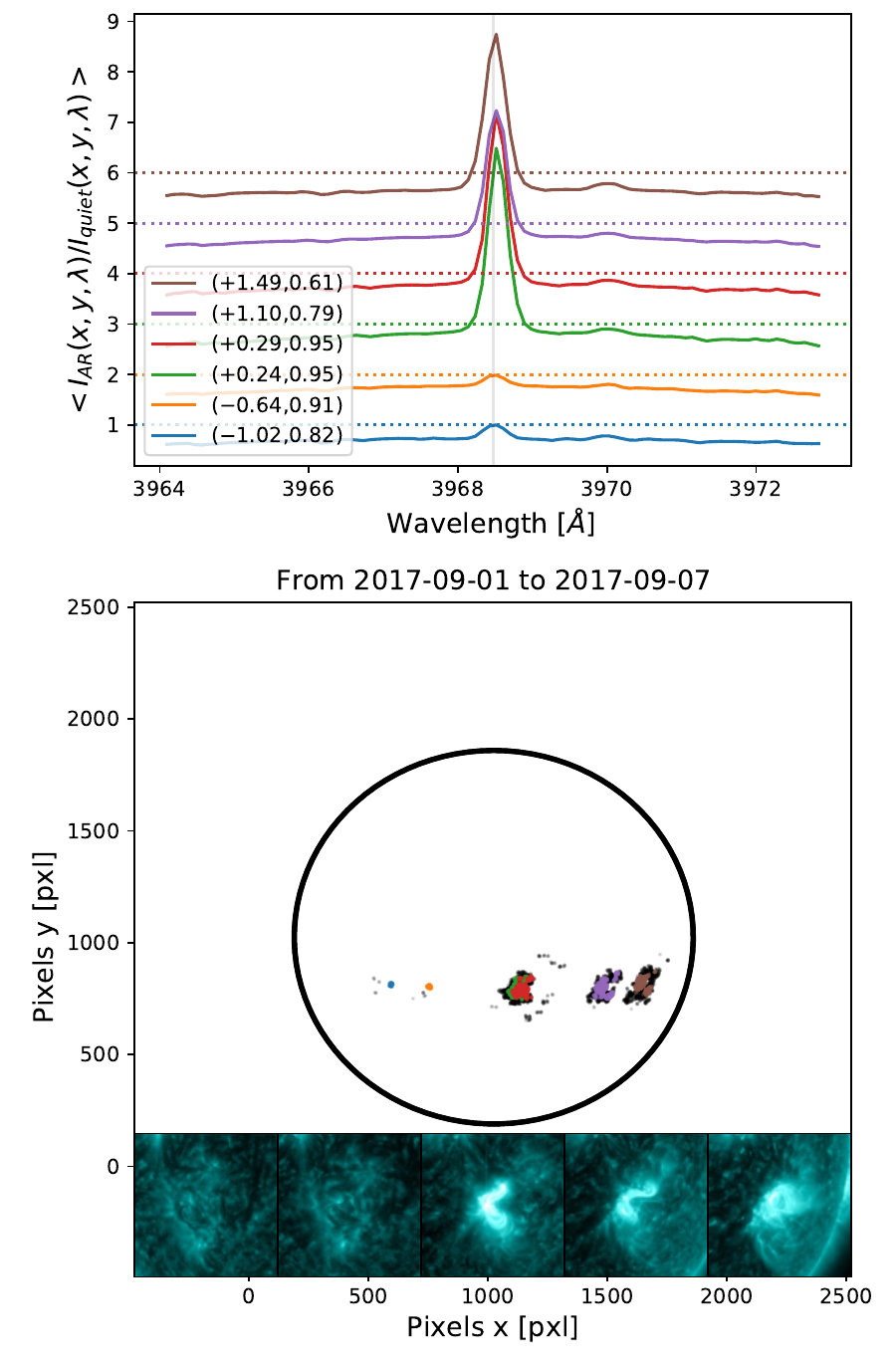}
	\caption{Same as Fig.\ref{FigPhaseS1} but extracting only the emerging sunspots from the southern hemisphere. A rapid evolution of the sunspots (and surrounding plage) is observable as soon as the solar meridian is crossed. The spectra from the pores before the sudden emergence are similar to those in Fig.\ref{FigPhaseS1}. However, once the meridian crossed, a huge flux excess becomes visible in the core. Its origin seems related to a peculiar magnetic field configuration as visible from the AI131 filtergrams.}
	\label{FigPhaseS2}

\end{figure}

The emission spectra as a function of the solar rotational phase are drawn in Fig.\ref{FigPhaseS1} and Fig.\ref{FigPhaseS2} for the northern and southern spot regions respectively. The stable sunspots regions from the North show a time-independent profile with darker photospheric wings and a core emission, similarly to the results obtained for plage, although less bright. The main feature to notice is the darker photospheric profile, which is expected because of the segmentation criterion. It can be seen by eye that the recovered spot spectra are narrower than those recovered for the plage, a behavior which we will discuss further in  Sect.~\ref{sec:meudon_clv}.

\begin{figure*}
	
	\centering
	\includegraphics[width=18cm]{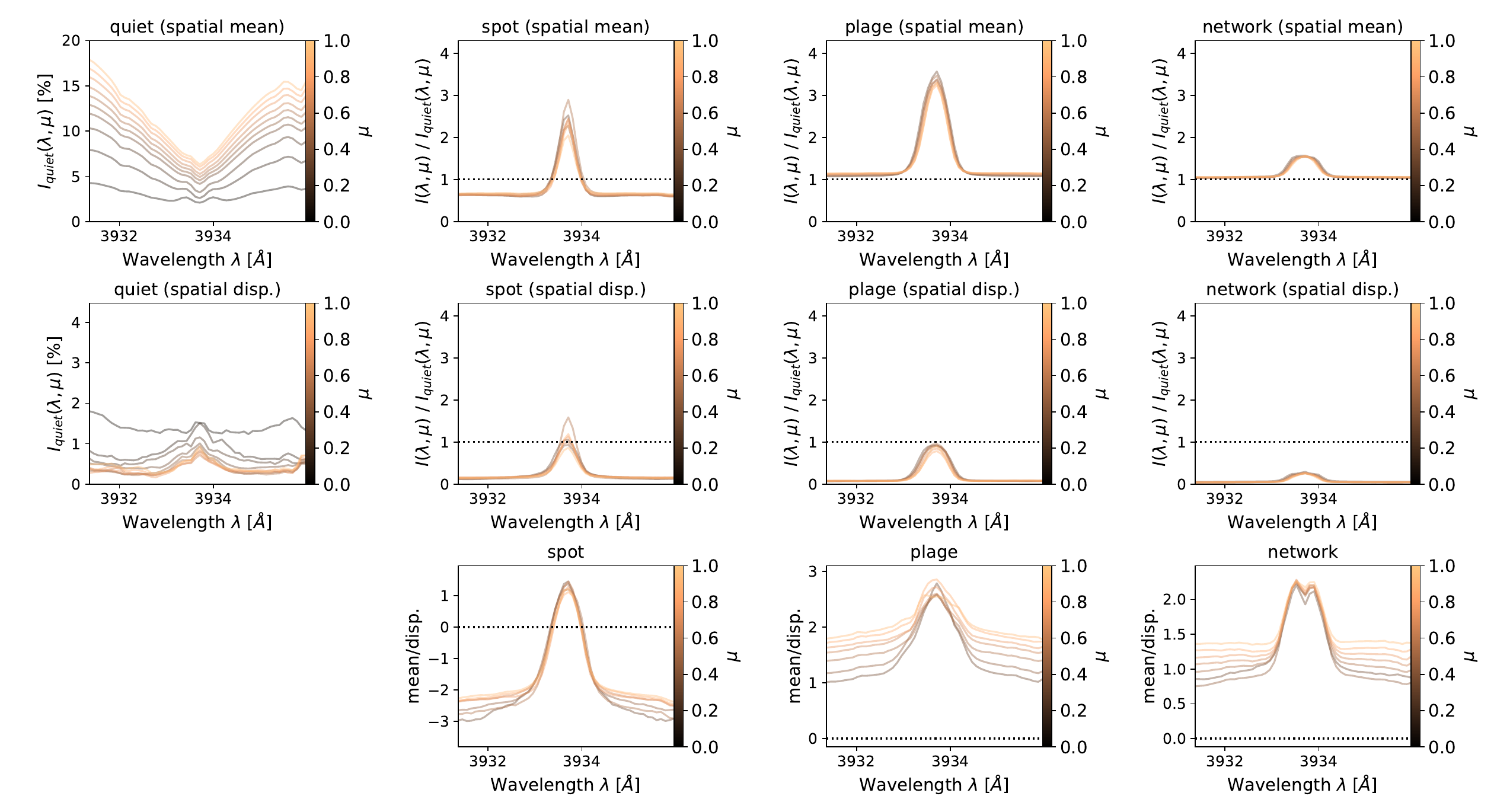}
	\caption{Stacked profiles of the quiet spectrum $I_{\text{quiet}}(\lambda,\mu)$ (left panel) and relative active spectra $I_{\text{AR}}(\lambda,\mu)/I_{\text{quiet}}(\lambda,\mu)$ for seven bins of $\mu$ angle between $\mu=1.0$ and $\mu=0.3$ (color code). For the quiet Sun, profiles up to $\mu=0.1$ are shown. Both the spatial average (first row), spatial dispersion (second row) and their ratio (third row) are displayed. A null spatial dispersion implies universal profiles (see main text). Different regions are displayed such as spots (second column), plage (third columns) and the network (fourth columns). Spots are darker in the wings but present an emission in their core. The plage emission profiles are always brighter, but present a similar core emission as spots. The network exhibits a weak emission profile slightly brighter in the core of the line. The result of the Gaussian fit on the intensity line profiles for the different active regions is displayed in Fig.\ref{FigMeudon4}.}
	\label{FigMeudon3}
 
\end{figure*}
 
While the northern regions show a time-independent behaviour, things are radically different for southern hemispheric regions. Indeed, before the AR crossed the meridian, the intensity profiles of the pores were almost identical to those of the northern hemisphere AR. However, once it crossed the meridian, we observed a rapid evolution of the AR. Note that this AR will produce one of the largest, and best studied flare recorded during the cycle on the 6th of September 2017 \citep[e.g.][]{Yang2017,Hou2018, Inoue_2018, Romano2018, Wang2018, Verma2018, Inoue_2021,Zou2019,Zou_2020,2021Vissers,Pietrow23flare}. That is however not included in our Meudon dataset.

A strong emission in the core of the lines is visible on the evolving spot (also visible in the middle panel of Fig.\ref{FigMeudon2}). It is unclear what is causing such different spectra. Is that spectrum a specific behaviour of spots in evolution or could it be contamination from adjacent plage that "covers" the sunspot?

Unlike the AR seen on the northern hemisphere, 
(classified as a dipole $\beta$ configuration\footnote{See \citet{Hale1919} and \citet{Nikbakhsh(2019)} for a detail description of the classification and the modified Zürich/McIntosh \citep{McIntosh1990} systems. We refer the reader to the SDIC page for an overview of the classification systems. \url{https://www.sidc.be/educational/classification.php}}), we can see that the spots in the southern region are surrounded by plages (classified as a complexe $\beta \gamma \delta$ configuration), which means that the resulting magnetic field strength and vector will be very different. This was shown by \citet{Anan21}, who found that the plage magnetic field in the high chromosphere tends to be horizontal, while \citet{Pietrow2020} and \citet{Morosin2020} found a mostly vertical field for a plage in between two pores. Note that the latter studies were done using the \ion{Ca}{II}~8542 \ang{} line, which forms lower down than the \ion{He}{I}~10830 \ang{} line studied by the former \citep[e.g.][]{Jaime2019}. As the wings are darker than the quiet chromosphere, we are still observing a sunspot, but the 3D magnetic field geometry is likely different. 

This was confirmed by looking at other AIA filtergrams that probe higher atmospheric layers such as the corona. We displayed the 131 \ang{} filter in the inner panels of the Fig.\ref{FigPhaseS1} and Fig.\ref{FigPhaseS2}. Looking those filters, the southern region also shows a clear anomalous bright excess, that may indicate that the magnetic field loops were contained at a different height compared to the northern hemisphere.  

\subsection{Analysis of the intensity spectrum CLV of active regions}
\label{sec:meudon_clv}

We extended the sample size by merging all the datasets coming from different epochs, and by binning the spectra in 7 bins from $\mu=0.3$ to $\mu=1$. Doing so assumes that no time-dependency exists in the spectrum itself, or to say it differently that the spectrum of the network, spots and plages are "universal". This hypothesis is therefore the same as assuming similar physical conditions (in temperature, velocity flows, magnetic field strength and direction) for all ARs of that type, which is clearly not true, but we believe that this is still a valid way of representing the average behavior of these regions.

We represent in the top row of Fig.\ref{FigMeudon3} the spatial average of the spectra for different $\mu$ angles\footnote{The spatial average profiles are provided in machine readable format in the online supplementary material on CDS}, as well as the spatial dispersion in the bottom row. The spatial dispersion is given by the standard deviation among the different spectra averaged. The ratio between the spatial average and spatial dispersion can be understood as a sort of "universality" measurement. Absolute ratio larger than 1 giving a rough delimitation for the "universality" region. 

At first glance, there are already clear differences between the spots and plage, with plage (third column) being broader and brighter at disk center, which is in line with the observations of \citep{Smith(1960)}. The profile for the spots (second column) is thinner and similar to the one published for the MgII lines in \citet{Avrett(2015)}. The network (fourth column) as a limited intensity roughly ten times smaller than plages.

\begin{figure*}
	
	\centering
	\includegraphics[width=18cm]{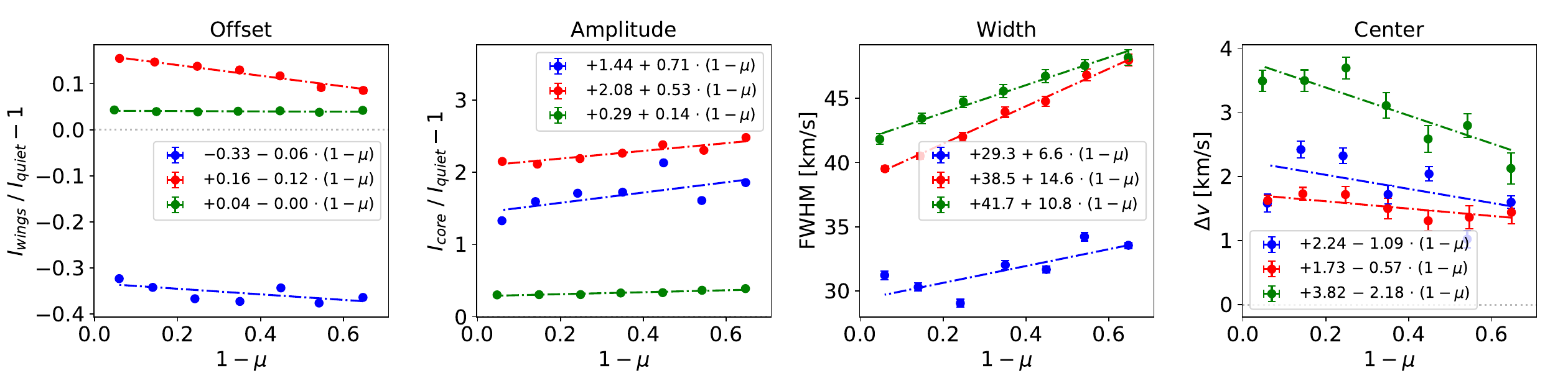}
	\caption{Parameters of the Gaussian fit on the active spectra of a spot (blue), plage (red) and network (green) as a function of the $1-\mu$ angle (see Fig.\ref{FigMeudon3}). The parameter of the Gaussian function is indicated in the title of each subplot. For each parameter, a linear relation as a function of $1-\mu$ angle (star center on the left, limb on the right) was fit and the relation displayed in the legend. The CLV relation for plage show rather a fix intensity emission profile in agreement with the universal emission plage profile \citep{Oranje(1983b)}. A notable broadening of 14.6 \kms from center-to-limb for plages can be observed already reported in \citep{Engvold(1966),Engvold(1967)}. For spots, no trend with $\mu$ angle can be clearly extracted even if the mean of the parameters strongly differ from the plage emission profile. }
	\label{FigMeudon4}

\end{figure*}

The spatial dispersion of the quiet Sun follows a constant level of 0.5\% of noise that is likely the superposition of stellar signals, such as granulation and p-modes oscillations \citep{Pietrow2023}, with the limited algorithmic precision of the CLV fit. This is confirmed by the two largest dispersions ($\sim$$1.5 \%$) observed for $\mu<0.3$, which shows the difficulty of obtaining stable fits close to the edges of the solar disk. A small excess is visible in the core of the line, which may be explained by other structures too small to be spatially resolved ($<0.40$ MSH).

A weak spatial dispersion is observed for plages, meaning that the plage profile seems "universal" as already suggested in \citet{Oranje(1983b)}. The standard deviation of the profile across the surface (0.8 at the line core) being typically 4.5 times smaller than the mean profile (that peaks at 3.5).  

In contrary to the plage, spots present more diverse behaviour. The "universality" of spots is limited to their darker photospheric wings, whereas more diverse behaviours are observed in the core of the lines (1.0 of standard deviation against a mean profile at 2.5). This was already visible in Fig.\ref{FigMeudon2}, where the bottom spot presented a clear emission, whereas the one at the top was still partially dark. We showed in Sect.~\ref{sec:meudon_spot} that this behaviour was likely related to a geometrical effect due to the peculiar configuration with adjacent plages. 

Further possible explanations include that spots with different umbra versus penumbra ratio exhibit different chromospheric signatures, since both components have rather different profiles (see Fig.2 of \citet{Yoon(1995)}). Umbrae typically represents between 15-30\% of the sunspot size \citep{Tlatov(2014)}. Also, because the cores of the \ion{Ca}{II} lines probe the upper part of the chromosphere \citep{Vernazza(1981),Bjorgen2018}, we should consider that spots are partially covered by the surrounding fibrils and other structures (e.g. Fig. 1 of \citet{Cauzi2012}, Fig. 1 of \citet{Yadav2022}, Fig. 1 of \citet{Pietrow2022a}, and Fig. 3 of \citet{Morosin2022}).

In order to extract the $\mu$ dependency of the AR profiles, we fit a Gaussian to each emission profile\footnote{Note that we also tried to extract other parameters of the profiles (such as distribution moments) in order to confirm that the trends observed were real and not induced by the choice of the Gaussian parametrisation.}. The parameters for the Gaussian fit, namely the offset, amplitude, full-width at half maximum (FWHM) and center are displayed for the Ca II K (Fig.\ref{FigMeudon4}) and Ca II H (Fig.\ref{FigMeudon5}) lines respectively. The offset parameter can be understood as the photospheric brightness, while the amplitude is associated with the chromospheric brightness. A similar behaviour is observed for both lines and we will only describe the behaviour of the Ca II K line. A linear relation as a function of $\mu$ was fit for each Gaussian parameter to highlight any linear CLV of the active spectra. 

The main conclusions drawn from these figures are that plage profiles as a function of $\mu$ see 1) their photospheric component becoming dimmer, 2) their chromopsheric component becoming brighter, 3) their profile becoming significantly broader. The broadening of the line is similar to the one observed on the quiet Sun by \citet{Pietrow2023}, meaning that this behavior might be intrinsic to the line itself and how it changes from the center to the limb. The clear broadening of the lines detected towards the limb is also in agreement with old solar physics studies \citep{Engvold(1966),Engvold(1967)}. Note that similar broadening toward the limb has also been detected for photospheric lines \citep{Bruls(2004)}. However, the darkening of the photospheric component (the offset parameter) towards the limb contradicts the idea of faculae becoming brighter towards the limb \citep{Unruh(1999)}, but it is hard to compare the photospheric level in the wings of the Ca II H \& K  to the one observed lower in the solar atmosphere by G-band continuum images. 
 
For spots, there seems to be some trends also, but the dispersion is much larger than for plage. This is notably visible on the amplitude parameter of the Gaussian that shows strong jitter, that is seemingly uncorrelated with the $\mu$ angle. Both activity profiles are red shifted by $\sim$~1.5 \kms{} but we do not know if the Meudon spectroheliograph is stabilised and accurate in RVs. This is particularly true given that the network seems redshifted by 3 \kms{}, while such shift is not confirmed with the Ca II H line (see appendix Fig.\ref{FigMeudon5}). We do however note that the trends are very similar in Ca II H and Ca II K, suggesting that the spread that we see is tied to physical processes rather than instrumental ones.

\subsection{Simulating disk-integrated spectra from active intensity profiles}
\label{sec:meudon_soap}

The last sections focused on disk resolved observations where ARs could be picked out by simple thresholding techniques, leading to $I_{\text{AR}}(\lambda,\mu)$. However, this is of course not possible with disk-integrated stellar spectra and a natural question that raises is how those spectra would look in disk-integrated observations?
We therefore simulated the behaviour of the Ca II K line using the activity profiles obtained in the previous Sect.~\ref{sec:meudon_clv}. Several codes exist today to recreate light curves photometry, stellar spectra, CCF or RV time series based on an AR configuration. As examples, we can cite SOAP \citep{Boisse(2012),Dumusque(2014)} that has recently be upgraded to produce spectra using a GPU \citep{Zhao(2023)}, but also StarSim \citep{Herrero(2016)}, and Starry \citep{Luger(2019)}. Creating a spectrum is really straightforward once $I_{\text{quiet}}(\lambda,\mu)$ and $I_{\text{AR}}(\lambda,\mu)$ are known. The main issue is to precisely obtain such spectra across the stellar disk, on a long band pass and at high-resolution. Such products have only been obtained for the quiet Sun recently \citep{Lohner(2019)}.

We created a solar disk image in a 880 $\times$ 880 grid. The $\mu_i$ angle of each cell are known and each cell contains a spectrum $I_{\text{quiet}}(\lambda,\mu_i)$ or $I_{\text{AR}}(\lambda,\mu_i)$. For the ARs spectra, we used the spatial averaged profiles (top middle and right panel in Fig.\ref{FigMeudon4}) rather than the analytical Gaussian fit approximation in Fig.\ref{FigMeudon4}. To extract an analytical approximation, we fit a polynomial function of degree 4 on each wavelength independently $P_{\text{AR}}(\lambda,\mu)=\sum_{i=0}^4 A_{i}(\lambda) \cdot \mu^i$. 
Note that those parameters are in fact the ratio profiles of the AR with the quiet Sun: $I_{\text{AR}}(\lambda,\mu)/I_{\text{quiet}}(\lambda,\mu)$ and therefore need to be rescaled afterwards. To derive $I_{\text{quiet}}(\lambda,\mu)$, a similar polynomial function $I_{\text{quiet}}(\lambda,\mu)=\sum_{i=0}^4 Q _{i}(\lambda) \cdot \mu^i$ was used on the profiles of the  quiet Sun (top left panel in Fig.\ref{FigMeudon3}). All the profiles are then shifted to reproduce the $\pm 2$ \kms solar rotation across the stellar disk. Differential rotation was neglected for simplicity in the toy model. Once all the spectra are known across the stellar disk, an AR covering $\mu_j$ locations can be placed and the integral over the surface performed to simulate a disk-integrated Sun-as-a-star observation: 

\begin{equation}
I_{\text{obs}}(\lambda) = \sum_{i} I_{\text{quiet}}(\lambda,\mu_i) + \sum_{j\neq i} I_{\text{quiet}}(\lambda,\mu_j) \cdot P_{\text{AR}}(\lambda,\mu_j)
\end{equation}

Even if this $I_{\text{obs}}(\lambda)$ is the natural product obtained by several codes, this intensity spectrum is not the one as measured by high-resolution spectra. Indeed, because high-resolution spectra are not accurate in flux, researchers rather study their ratio or difference with a reference spectrum after the continuum of both spectra have been adjusted to match. We reproduced that behaviour, using as reference the spectrum obtained by integrating the full quiet Sun: $I_{\text{quiet}}(\lambda)$. The continuum of the observed spectrum $I_{\text{obs}}(\lambda)$ was adjusted to the continuum level of $I_{\text{quiet}}(\lambda)$ by computing the scaling factor $C$ between both spectra outside the core of the line, where the core was defined as $\pm1$ \ang{} around the line center. Such renormalisation tends to match the photospheric flux level between the reference spectrum and the observations. Once the observed spectra scaled, we analysed the difference of both:

\begin{equation}
\Delta I(\lambda) = C \times I_{\text{obs}}(\lambda) - I_{\text{quiet}}(\lambda) 
\end{equation}

\begin{figure*}
	
	\centering
	\includegraphics[width=18cm]{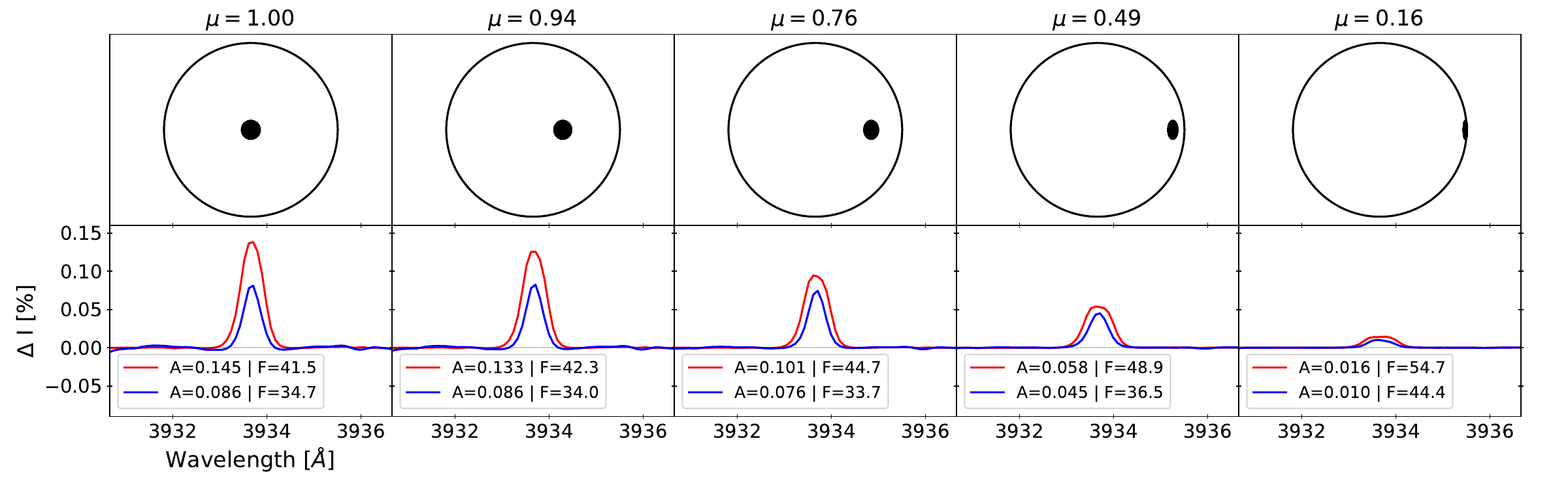}
	\caption{Recreated disk-integrated spectra of the Ca II K line (bottom panels) from the polynomial CLV relation fit on profiles in Fig.\ref{FigMeudon3}. We simulated the effect of a 1\% filling factor AR at the equator at five different rotational phases (top panels) equidistant by $20^\circ$ ($\sim$1.5 days). The spectra are normalised by the photospheric flux level, as would high-resolution spectra observations be (see plain text), explaining why the spot does not look darker and only exhibits core emission. A clear brightening of plages (red curve) and spot (blue curve) occurs from limb to star center, mainly due to the increased apparent size. The amplitude of the peak ($A$) as well as the FWHM ($F$) in \kms{} are indicated in the legend of each subplot. Note that the last phase ($\mu=0.16$) is an extrapolation of the polynomial CLV, since outside the fitting domain ($\mu<0.30$).}
	\label{FigMeudon6}
 
	\centering
	\includegraphics[width=18cm]{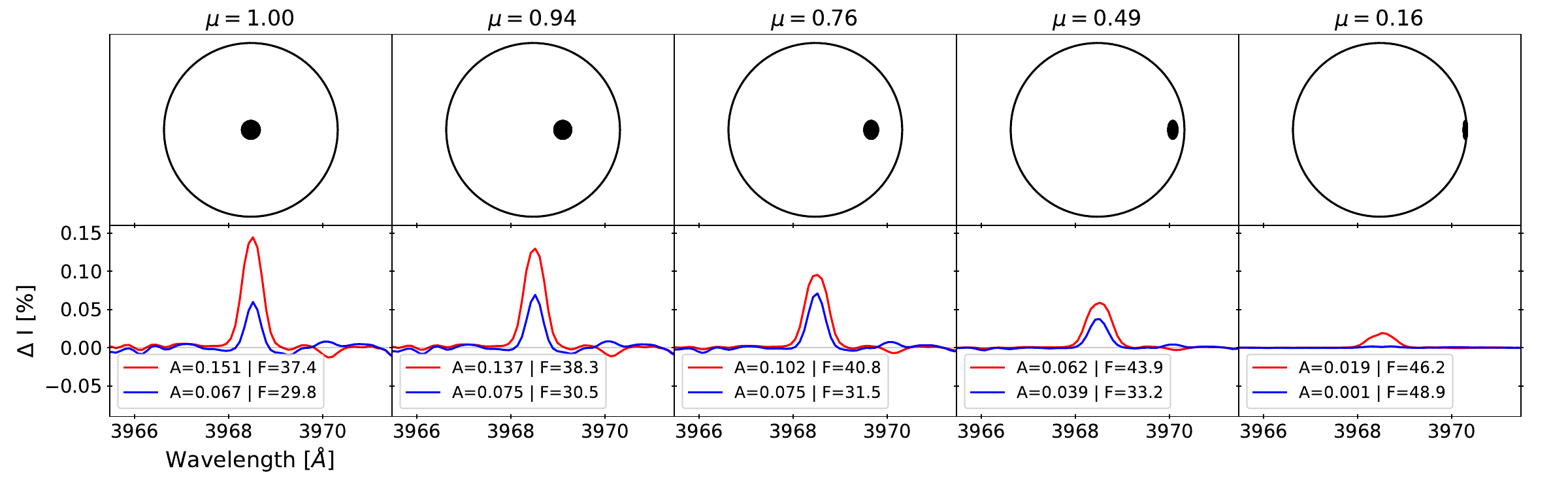}
	\caption{Same as Fig.\ref{FigMeudon6} for the Ca II H line. A similar behaviour is observed, a different behaviour of the Balmer $H_{\epsilon}$ at $\lambda=3970$ \ang{} is also visible. The line becoming either deeper or shallower depending if the AR is a spot or a faculae respectively. A behaviour already noticeable in \citet{Martinez(1990)}.}
	\label{FigMeudon7}

	\centering
	\includegraphics[width=18cm]{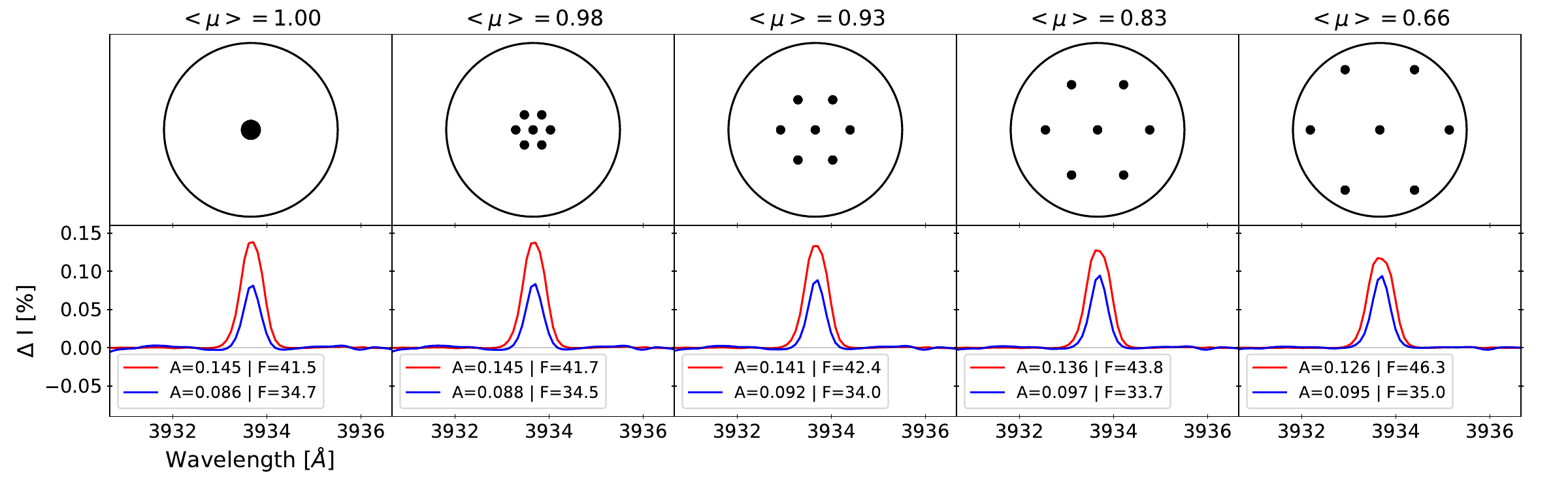}
	\caption{Same as Fig.\ref{FigMeudon6} for different $\mu$-distributions of ARs. By construction, the filling factor is fixed to $f=1\%$ in all the cases. The stellar rotation was removed to focus on the CLV contribution. If spots do not show clear effect, the plages show on the $<\mu>=0.66$ spread configuration a weakening and broadening of 10\% compared to the $<\mu>=1$ packed configuration.}
	\label{FigMeudon8}
 
\end{figure*}

We displayed in Fig.\ref{FigMeudon6} the series of profiles obtained for an AR going from the solar center to the limb with a filling factor of $f=1\%$. We did not study the network component, since this component has a weak rotational modulation. We note that spots and plage both have near identical profiles now that the photospheric flux has been renormalised. The main difference between the profile is a slight difference of amplitude and of width, already visible from the Gaussian parameters in Fig.\ref{FigMeudon4}, the spot profile being $\sim$25\% thinner than the plage profile and 40\% less bright. We notice that both ARs are less bright on the limb compared to the center, because of the $\mu(t)$ projection effect of the AR. Such effect likely shows that integrating the flux in the core of the line is a good proxy for their filling factor $f_{\text{AR}}(t)$. Such relation between the core integration and the plage filling factor have already been established \citep{Skumanich(1975),Ortiz(2005)}, but we highlight that the authors never tried to split the contribution from plage and spots in the Ca II H \& K  lines, and that good correlation were found mainly because the plage filling factor dominates the spot filling factor by one order of magnitude \citep{Chapman(2001)}. Such a splitting of both contributions will be done later (see  Sect.~\ref{sec:iss_analysis} and Sect.~\ref{sec:harpn_analysis}).

The intensity at center being 0.15\% and 0.09\%, for a $f=1$\% plage and spot AR respectively. Detecting such effect on stellar observation is not straightforward since the signal-to-noise ratio in the extreme blue is often limited. For a plage, the signal is easier to detect since plages can easily cover up to 10\% of the solar surface but spots rarely exceed 1\% in size (see Sect.~\ref{sec:sdo}). 

We performed the same reconstruction for the Ca II H line in Fig.\ref{FigMeudon7}. A similar behaviour is observed that highlights once again the similarity of both Ca II H \& K lines. However, an interesting extra feature is visible on the right side of the line, which is the behaviour of the Balmer line $H_{\epsilon}$ at $\lambda$=3970 \ang{}. As visible, the line is becoming deeper for a plage, but shallower for a spot; a behaviour already visible in \citet{Martinez(1990)}. This difference theoretically opens a way to lift the degeneracy between the spot and plage contributions. However, its amplitude does not exceed 0.01\% for a filling factor $f=1\%$ which is extremely challenging to measure in stellar observations. The further information brought by the $H_{\epsilon}$ Balmer line was also raised for solar activity studies recently \citep{Krikova(2023)}.

We pushed further the investigations by simulating the behaviour of the Ca II K line for different configurations of ARs. In the previous simulation, the decrease of the line intensity was dominated by the size projection of the ARs that decrease linearly from center-to-limb. In order to remove the projection dependency and focus on the CLV effect, we simulated five different distributions of ARs, that are all configured with a filling factor of $f=1$\%. 

The simulations are created using seven small circular\footnote{we used circular shape on the projected surface for simplicity and since the intrinsic shape is irrelevant for the final integrated profile.} ARs, for which the six outermost ARs are gradually moved toward the limb. We kept an AR at center since it is likely to have an AR at $\mu>0.90$ when randomly observing a star due to geometrical consideration. The result of this simulation is showed in Fig.\ref{FigMeudon8}. For this peculiar simulation, we also removed the rotational velocity to avoid any broadening from that component. As visible, once the filling factor is fixed, the peak intensity is now far less affected compared to Fig.\ref{FigMeudon6}. The profile of spots is almost constant, indicating a low CLV effect. For plage, when the six ARs are moved to $\mu=0.60$ (a disk-averaged value of $<\mu>\,=0.66$) the profile becomes broader by 11\% and the amplitude smaller by 15\%.

The present example shows that the CLV relation has a weak, but measurable effect, in disk-integrated spectra in particular for the plages. Therefore, different configurations lead to slightly different integrated profiles. However, it is unclear if disk-integrated spectra could be inverted to infer the distribution of ARs due to degenaracies between spots or plages ratio and their locations.

\section{Combining spatially resolved observations with high-resolution disk-integrated spectra}
\label{sec:secSDO_HARPS}

A strong advantage of the Meudon spectroheliograms is that the individual spectra coming from different active regions can directly be obtained from the spatially resolved surface elements of the datacubes. However, such advantage is at the expense of the spectral resolution which is therefore limited. In this section, we solved for the issue by combining disk-integrated high-resolution spectra from ISS and HARPS-N spectrograph, with spatially resolved information from SDO. We used the SDO public database to segment the solar surface and extract the filling factor of the different active regions. The choice of using SDO filtergrams, rather than the Meudon spectroheliograms is justified by the more extended and better temporal sampling of the SDO space mission compared to Meudon spectroheliograms. 

The section is divided as follows: we first extracted the filling factor time series of the different activity components using SDO filtergrams in Sect.~\ref{sec:sdo}. We combined these filling factors time series with the ISS spectra in Sect.~\ref{sec:ISS} in order to derive the emission profile of the ARs per unit of filling factor. The same analysis is conducted in Sect.\ref{sec:harpn} for HARPS-N. The profiles obtained between the instruments are commented in Sect.\ref{sec:comparison}. Finally, an estimation of the contribution of the different types of ARs in an \Sidx-like metric is investigated in Sect.\ref{sec:sindex}.

\begin{figure*}
	
	\centering
	\includegraphics[width=18cm]{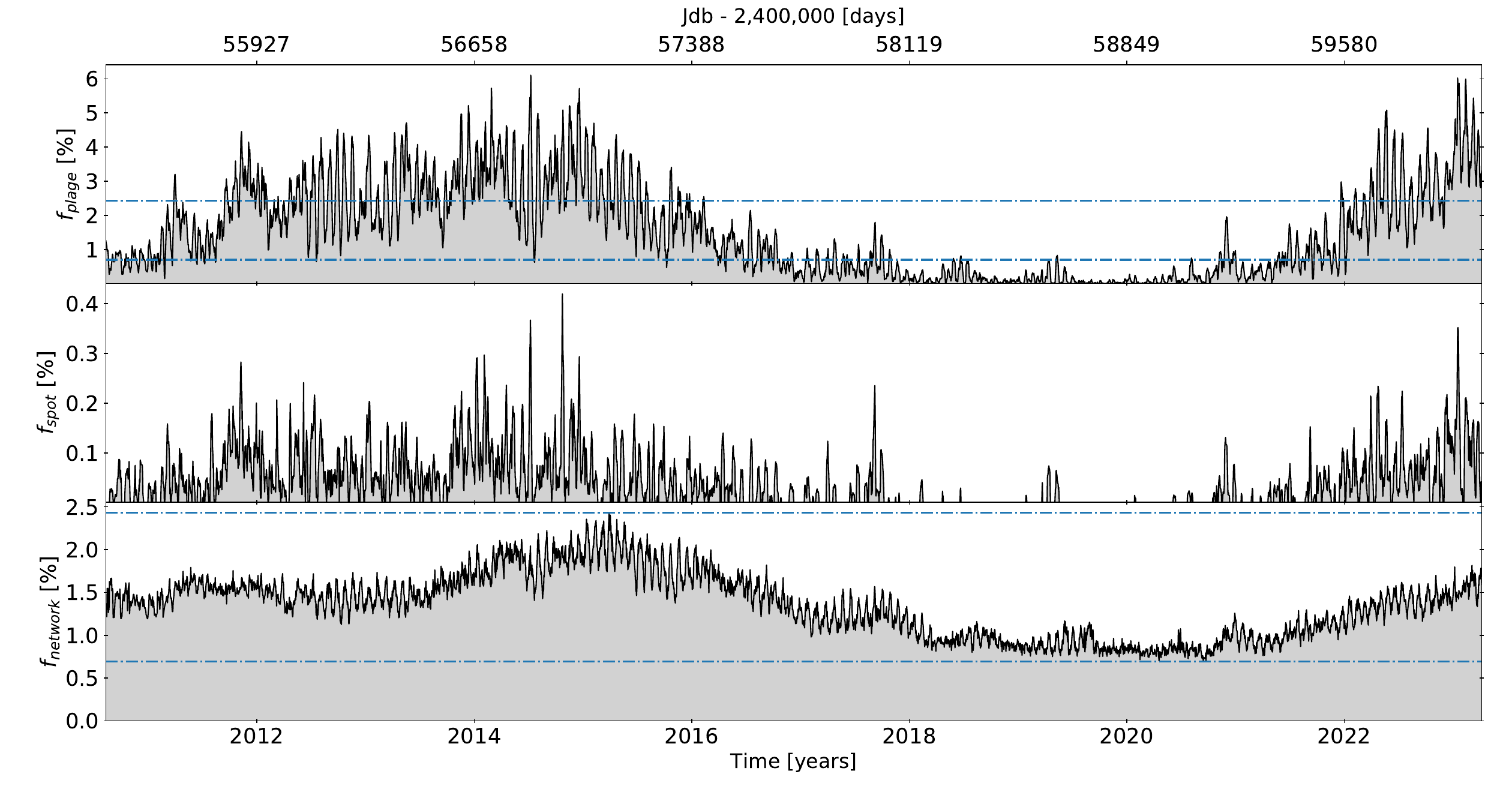}
	\caption{Extraction of the filling factor time series for plages, spots and network by segmentation of the AIA1700 SDO filtergrams. The dataset begins from August 2010 to March 2023 and probes the full $24^{\rm th}$ magnetic cycle as well as the beginning of the $25^{\rm th}$ solar cycle. \textbf{Top}: Plage filling factor time series $f_{\text{plage}}(t)$. The maximum and minimum values of the network filling factor are indicated with the blue dashed lines. \textbf{Middle}: Spot filling factor time series $f_{\text{spot}}(t)$. \textbf{Bottom}: Network filling factor time series $f_{\text{network}}(t)$. Unlike the  two other components, the network time series has been cleaned from discontinuities and outliers.}
	\label{FigSDO1}
 
\end{figure*}

\subsection{Fillling factor of plages, spots and network from SDO images}
\label{sec:sdo}
The Solar Dynamics Observatory \citep[SDO,][]{Pesnell2012} is a solar observer launched in February 2010. The satellite is equipped with the Atmospheric Imaging Assembly \citep[AIA,][]{Lemen2012} producing filtergrams in several bandpasses \citep{Title(2006)}, probing different layers of the solar atmosphere. Among the SDO/AIA filters, AIA1700 measures the ultraviolet flux continuum at 1700\ang{}, which probes the temperature minimum, located between the photosphere and the lower chromosphere from $4500$K to $5000$K \citep{Simoes2019}. More information on the filters can be found on the official NASA web site\footnote{\url{https://www.nasa.gov/content/goddard/sdo-aia-1700-angstrom/}}. An advantage of this filter is that both bright plages and dark spots are visible and can therefore be segmented from the same image (see Fig.\ref{FigSDO2}). The same can be done with the wavelengths close the K1/H1 minima in the Meudon spectroheliogram datacubes (see the second and fourth panel in Fig.\ref{FigMeudon2}).

We downloaded 4518 images from the public archive\footnote{\url{https://sdo.gsfc.nasa.gov/data/aiahmi/}}, aiming for a cadence of 1 image per day on a baseline of 13 years, between 2010 and 2023. In this work, we focus on detecting plages and spots that constitutes large ARs. Because of this, we restricted ourselves to the 512 $\times$ 512 pixel filtegrams that are faster to process than the raw 4048 $\times$ 4048 images. The stellar radius is on average about 198 pixels, hence the spatial resolution is around 5'' arcsec (or 8 MSH per pixel), which prevents the detection of small structures (such as pores) but which is sufficient for plages and spots that are larger than 20 MSH \citep{Tlatov(2014)}. All the selected images passed a visual inspection to reject anomalous filtergrams, which occurs when the exposure began with improper setting of the alignment of the filter or when the filtergrams were not properly centered by the SDO data extraction pipelines. 

Each SDO image is processed in a similar way as the Meudon spectroheliogram datacubes to remove the CLV signature of the quiet Sun (see Sect.~\ref{sec:meudon}). A difficulty consists in selecting a good brightness threshold for the \textit{flat corrected images} to disentangle plages from the enhanced network \citep{Dineva(2022)}. We selected a conservative value that was performing a satisfactory result visually. The thresholds brightness for plage and spots were set to brighter and darker than 25\% the quiet Sun values respectively. Again, we did not attempt to detect any ARs at limb angle smaller than $\mu<0.3$ due to the large amount of noise close to the limb. The network time series suffered from discontinuities and clear outliers, mainly induced by issues during the CLV fit that bias the $3\sigma$ threshold to large value (therefore producing small filling factor values). Such time-domain signatures are unexpected, since a smooth component is visible everywhere else on the time series. We corrected the time series from the discontinuities by "sticking" back the different part of the signal and cleared the outliers using the average of the next and previous day.  

The time series for the filling factor of plages, spots and network are shown in Fig.\ref{FigSDO1}. Our result is similar to the one obtained by \citet{Denker2019} and \cite{Milbourne(2021)}. The $24^{\rm th}$ magnetic cycle is clearly identified as well as the start of the $25^{\rm th}$ cycle. Cycle $24^{\rm th}$ was the quietest cycle in the past century \citep{Love(2021)} with a 100-day smoothed filling factor of 4\% at maximum for plages. The rotational modulation due to plages is clearly visible on top of this. The threshold used for the plage segmentation seems adequate since their filling factor value converges toward 0 at the minimum of the solar cycle, whereas the network saturates to its basal minimal intensity around 1\%, as in \citet{Milbourne(2021)}.

On average, spots are roughly ten times smaller than plages, in agreement with \citet{Chapman(2001)}. They were almost absent from the $24^{\rm th}$ cycle, except for a large sunspot event in 2015. Due to the dataset used, the spatial resolution is unable to detect pores or spots < 8 MSH. However, most of the sunspots are larger than 100 MSH \citep{Tlatov(2014),Nagovitsyn(2016)} and are therefore confidently detected. The magnetic cycle is far less clear with spots, which is in contradiction with several studies presenting sunspot number \citep{Love(2021)}. It may indicate that the number of spots shows a clear magnetic cycle, but the total area does not, which seems to be a peculiarity of the 24th solar cycle\footnote{See the sunspot area in \url{https://lasp.colorado.edu/lisird/data/sfo_sunspot_indices/}}.  

\begin{figure*}
	
	\centering
	\includegraphics[width=18cm]{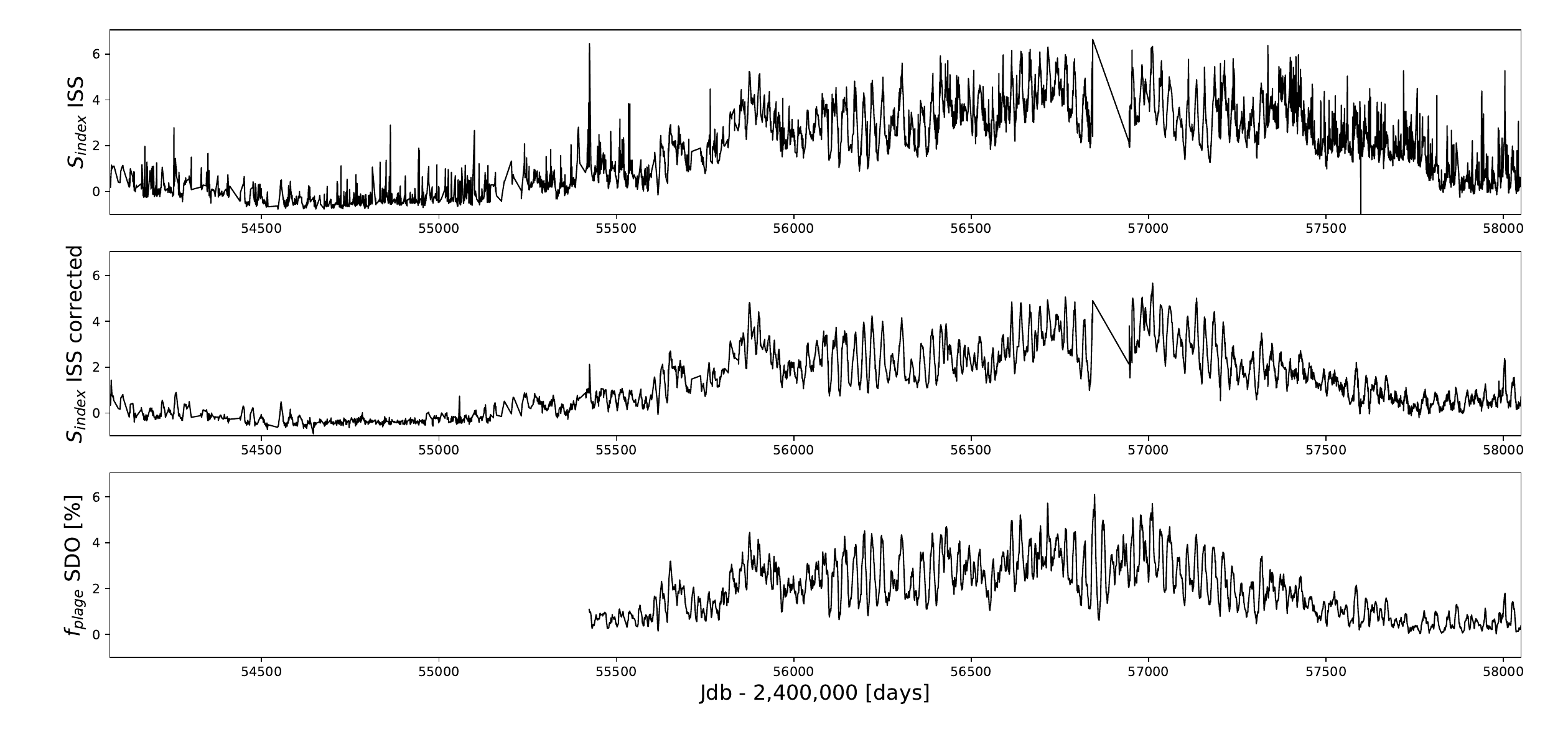}
	\caption{Core integration of the Ca II K lines obtained from the uncorrected and corrected ISS spectra. The \Sidx metrics have been rescaled in unit to provide the closest 1:1 relation with the SDO plages filling factor of Fig.\ref{FigSDO1}. \textbf{Top panel}: Core integration with a rectangular 1\ang{} band pass in order to obtain a \Sidx metric from the uncorrected ISS spectra. \textbf{Middle panel}: Same as top panel, but from the ISS corrected spectra. \textbf{Bottom panel}: SDO plage filling factor $f_{\text{plage}}(t)$ from Fig.\ref{FigSDO1}. The Pearson correlation coefficient $\mathcal{R}$ between plage filling factor and the core integrated emission was boosted from $\mathcal{R}=0.69$ to $\mathcal{R}=0.81$. The effect was even more impressive for the Ca II H line slightly more affected, with a correlation increased from $\mathcal{R}=0.51$ to $\mathcal{R}=0.80$.}
	\label{FigISS_sindex}

\end{figure*}

While the modulation of the signals is similar, we noted a difference in the absolute filling factor values for spots in comparison with \citet{Milbourne(2021)}. For instance, looking their Fig.2, we can see that their spot filling factor rises to $\sim$ 0.75\% for the 2015 event, while ours remains at $0.40$ \%. A difference of filling factor could be due to different threshold brightness (which are always arbitrary to some extent), but the reason for the discrepancy here is likely physical. The difference in filling factor mainly arises because we observe a different solar atmospheric layer. Indeed, while we used the AIA1700 filter that probes the chromosphere, \citet{Milbourne(2021)} used HMI that mainly probes the photosphere. In order to confirm this hypothesis, we compared for a given date the spot filling factor obtained between HMI and AIA1700. Defining sunspots as any region darker than 90\% the quiet photosphere for HMI, we found that spot filling factor was roughly twice as large in HMI as in AIA1700. Furthermore, 80\% of the regions missing from our segmentation of AIA1700 are brighter than 75\% the quiet solar photosphere in HMI, which indicates that our SDO segmentation likely misses the penumbrae, but catches the umbrae of sunspots.   

Note that because the present paper aims to study the behaviour of the Ca II H \& K  lines that are formed in the chromosphere, we are not particularly interested in physical quantities measured in the photosphere, justifying the use of AIA1700 rather than HMI for this work.

\subsection{Combining the spatially resolved SDO filtergrams with ISS disk-integrated spectra}
\label{sec:ISS}

The Integrated Sunlight Spectrometer (ISS) of the Synoptic Optical Long-term Investigations of the Sun \citep[SOLIS,][]{Keller2003}
that was installed at the National Solar Observatory has taken daily disk-integrated images of the Sun at extremely high spectral resolution ($\mathcal{R}$ $\sim$ 300\,000) for almost a decade, from 2007 to 2017. Public spectra are available and can be downloaded from an archive\footnote{\url{https://solis.nso.edu/pubkeep/}} . Several lines are available with ISS, but in this paper, we focus on the Ca II H \& K  lines. Standard products are the daily stacked observations with a S/N in the continuum of the CaII lines of S/N$\sim800$.

\subsubsection{Processing of the disk-integrated ISS spectra}
\label{sec:ISS2}

Despite the impressive baseline and spectral resolution of the dataset, the spectrograph suffered from clear scattered light or background correction issues, which can be highlighted when computing the integral of the core emission. We show in Fig.\ref{FigISS_sindex} (top panel), the time series of the rectangular core integration of the Ca II K lines. Strong peak excursions are visible at different times (see e.g BJD=$2\,455\,450$) and a clear degradation of the instrument is visible toward the end of its life.

A simple way to correct for background scattering consists in fitting a parabolic relation using the wings of the Ca II lines that probe the photosphere. Doing so will absorb any real signature coming from that solar layer, but these are assumed to be small compared to the chromopheric signatures. Moreover, such photospheric normalisation are standard when studying spectra time series where residuals spectra are compared to a reference spectrum. We defined the reference spectrum $I_{\text{ref}}(\lambda)$ as the median of the 10\% quietest spectra. An example of the line profile correction is displayed in Appendix.\ref{appendix:b} (Fig.\ref{FigISS_correction}). The parabolic fit is performed on the wings selecting the wavelength range where $I_{\text{ref}}(\lambda)>0.10$, which excludes the core of the lines.

Similarly to the Meudon spectra (see Sect.~\ref{sec:meudon}), another issue was related to the lack of reliable continuum level due to the restrictive bandpass. We therefore also calibrated the ISS flux units to match the HARPS-N flux units by fitting a linear relation between both master spectra. The calibration leads to a relation: $f_{\text{HARPS-N}} = 0.741 \times f_{\text{ISS}} -0.005$ and was used to convert the flux units of all the ISS spectra. 

We show the previous rectangular core integration obtained with the clean spectra in middle panel of Fig.\ref{FigISS_sindex}. By comparing the two time series with the plage filling factor of SDO (see Sect.~\ref{sec:sdo}), we observed that the \Sidx obtained with the clean spectra was now strongly correlated with the SDO $f_{\text{plage}}(t)$ time series ($\mathcal{R}\simeq0.80$ in average for both lines), while a moderate correlation ($\mathcal{R}\simeq0.60$) was obtained with the uncorrected spectra.

\begin{figure*}
	
	\centering
	\includegraphics[width=18cm]{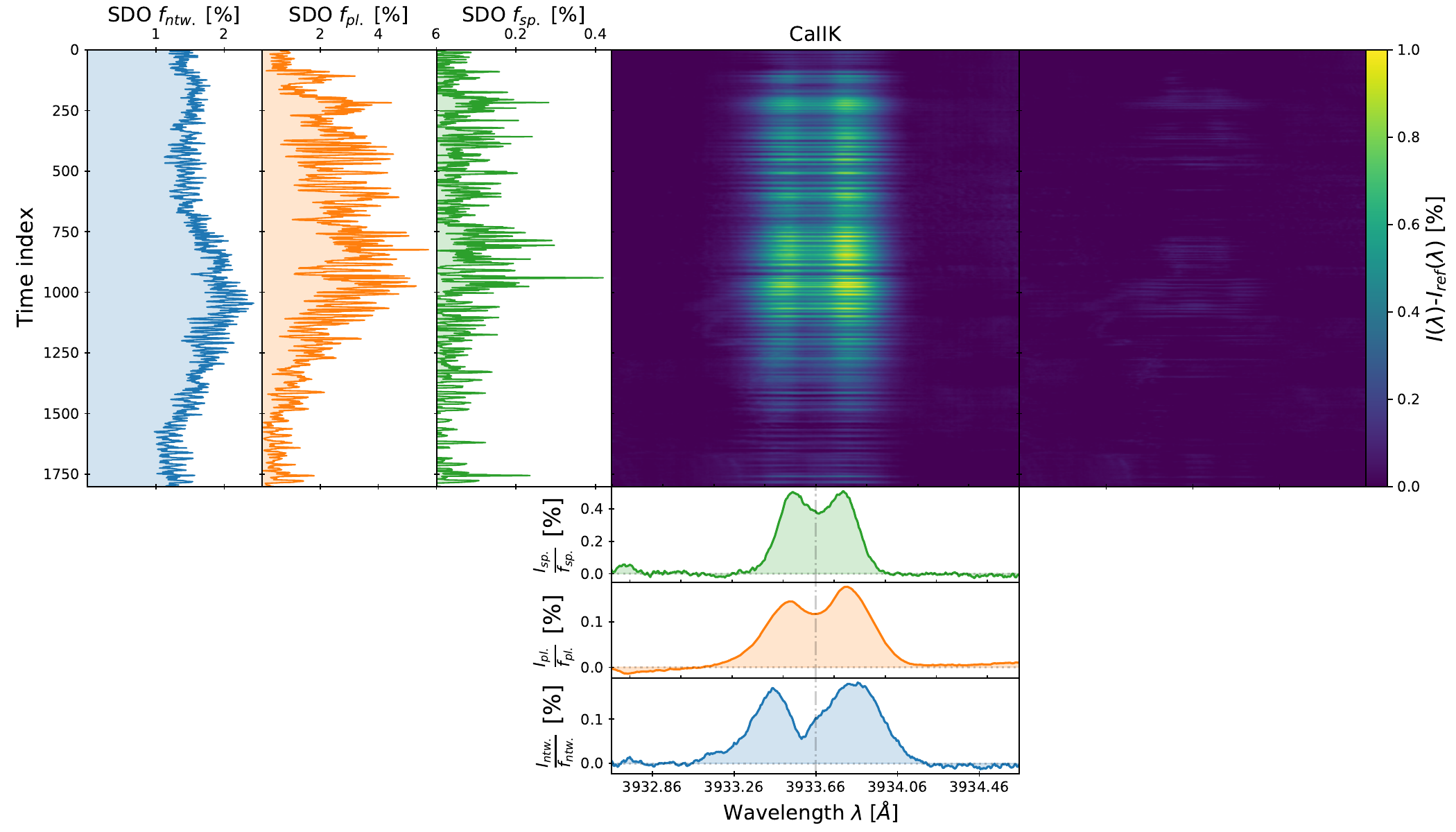}
	\caption{Extraction of the intensity emission profiles per unit of area from spots $I_{\text{spot}}(\lambda)$, plages $I_{\text{plage}}(\lambda)$ and the network $I_{\text{ntw}}(\lambda)$ using the ISS spectra from 2010-09-01 to 2017-10-23. In order to extract the profiles, the ISS spectra time series (colormap) was fit with a multi-linear model of SDO spot (green), plage (orange) and network (blue) filling factor according to Eq.\ref{EqFund2}, plus a free offset parameter. The SDO time series are the same as the one presented in Fig.\ref{FigSDO1} except that only intersection of SDO and ISS observations were kept. The residuals spectra are displayed on the right. Some rotational behaviour are still visible in residuals and are either due to non-universality or CLV such as the broadening of the plage emission spectrum mentioned in Sect.~\ref{sec:meudon_analysis}.}
	\label{FigComp1}

\end{figure*}

\subsubsection{Extraction of the ISS activity emission profiles per unit of area}
\label{sec:iss_analysis}

The Meudon spectroheliogram showed that spots and plages present different emission profiles. Unfortunately, the moderated spectral resolution of the instrument prevents the extraction of intensity profile for high-resolution spectrographs. Since the goal is to obtain the emission intensity profile per unit of area, we can wonder if SDO filtergrams and ISS disk-integrated spectra could not be combined in a joint analysis. A solution for it consists in fitting the filling factors of plage and spots obtained from SDO onto the ISS spectra time series to recover the emission profile per unit of area.

We first selected all the days where simultaneous SDO and ISS spectra were available, that corresponds to 1617 days between September 2010 and October 2017. Because we are interested in rotational timescale modulation, a precise match of the observations is not required, and we considered 1-day as the maximum time difference acceptable. We then modeled the spectra time series $\delta(\lambda,t) = I(\lambda,t)-I_{\text{ref}}(\lambda)$ of ISS, with $I_{\text{ref}}(\lambda)$ the median of the 1\% spectra with the smallest integrated flux emission\footnote{Because of the reference choice, these filling factor values have to be understood as relative values compared to the mean reference epochs level $\Delta f_X(t)=f_X(t)-f_X(t_{\text{ref}})$}. The residual spectra time series was fit by a three-component model described as the superposition of the spots $f_{\text{spot}}(t)$, plage $f_{\text{plage}}(t)$ and network $f_{\text{ntwk}}(t)$ SDO filling factors: 

\begin{equation}
\label{EqFund2}
\delta(\lambda,t) = I_{\text{spot}}(\lambda)\cdot f_{\text{spot}}(t)+I_{\text{plage}}(\lambda)\cdot f_{\text{plage}}(t)+I_{\text{ntwk}}(\lambda)\cdot f_{\text{ntwk}}(t)
\end{equation}

Such a model assumes only three components, which seems to be relatively true for the Ca II H \& K lines based on Meudon spectroheliogram, but which could be wrong for other lines. For example, the $H_{\alpha}$ line would require another component for the filaments that are 
darker compared to the quiet chromosphere  \citep[e.g.][]{Kuckein2016,Zhu(2019)}. Furthermore, this model assumes that the AR profiles do not change over time, which is only partially true as seen in Sect.~\ref{sec:meudon_analysis} since ARs evolve over time and due to the CLV. However, these are secondary effects and we are more interested to derive the average behaviour of the ARs.

On top of that, it is important to note that, for a target with an inclination of 90 degrees (equator-on), any AR will spend $75\%$ of its time at $\mu>0.5$, and $90\%$ at $\mu>0.30$ due to the spherical geometry of stars \citep[e.g. see Fig. 2 of][]{Pietrow2023b}. This means that for most of the time AR are visible, the Gaussian parameters of the ARs change by less than 15\%. 

The decomposition to recover $I_{\text{spot}}(\lambda)$, $I_{\text{plage}}(\lambda)$ and $I_{\text{ntwk}}(\lambda)$ is displayed for the Ca II K line in Fig.\ref{FigComp1} and for the Ca II H line in the appendix (Fig.\ref{FigComp2}). Similar intensity spectra are obtained for both lines. The emission profiles are compatible with those published in literature. For example, the double-peak of the plage emission profile is almost identical to the one published by \citet{Oranje(1983b)} with K2R higher than K2V, whereas the spot emission profile is compatible with the one published by \citet{Yoon(1995)}. A thinner emission profile was also observed in penumbrae (see Fig.1 of \citet{Sheminova(2005)}). Finally, the network spectrum shows a broader peak separation of the K2 maxima compared to plage with core depletion. So far as we know, we are not aware of any studies focusing on the profiles of the network in the Ca II H \& K lines. 

The similarity of the spot and plage profiles is a huge concern since it means that most of the band pass to disentangle both contributions come from the wavelength range cover by the wings of their profiles, which has strong implication for the minimum S/N needed to extract both components. The same conclusion holds for the plage and network components. The peak separation between the K2 maxima is gradually increasing from spots, plage to network with separation of 17.5 \kms{}, 21.0 \kms{} and 30.0 \kms{} respectively.

Looking at those three intensity profiles, we can easily understand that the different parametrization of the lines profiles (such as distance between the K2 maxima, K1 minima or similar handmade metrics) had poor chance to disentangle the three signatures and their investigations has lead so far to none meaningful result \citep{Keil(1998),Livingston(2007),Bertello(2011),Dineva(2022)}. The time series extracted from them were certainly a complex mixture of spot, plage and network signals strongly affected by their respective ratio area. 

In a similar fashion, integration with bandpasses \citep{Gomes(2018)} will extract with difficulty the three contributions that are strongly similar. The \Sidx and its triangular integration was likely the weighted average of all the contributions which is a meaningful metric if the chromospheric flux excess, independently on the nature of the ARs, want to be measured (for instance to measure the global stellar activity level). However, such metric has a limited interest for RV correction for which each activity component is known to contribute in a different way \citep{Milbourne(2021)}. 

\begin{figure*}
	
	\centering
	\includegraphics[width=18cm]{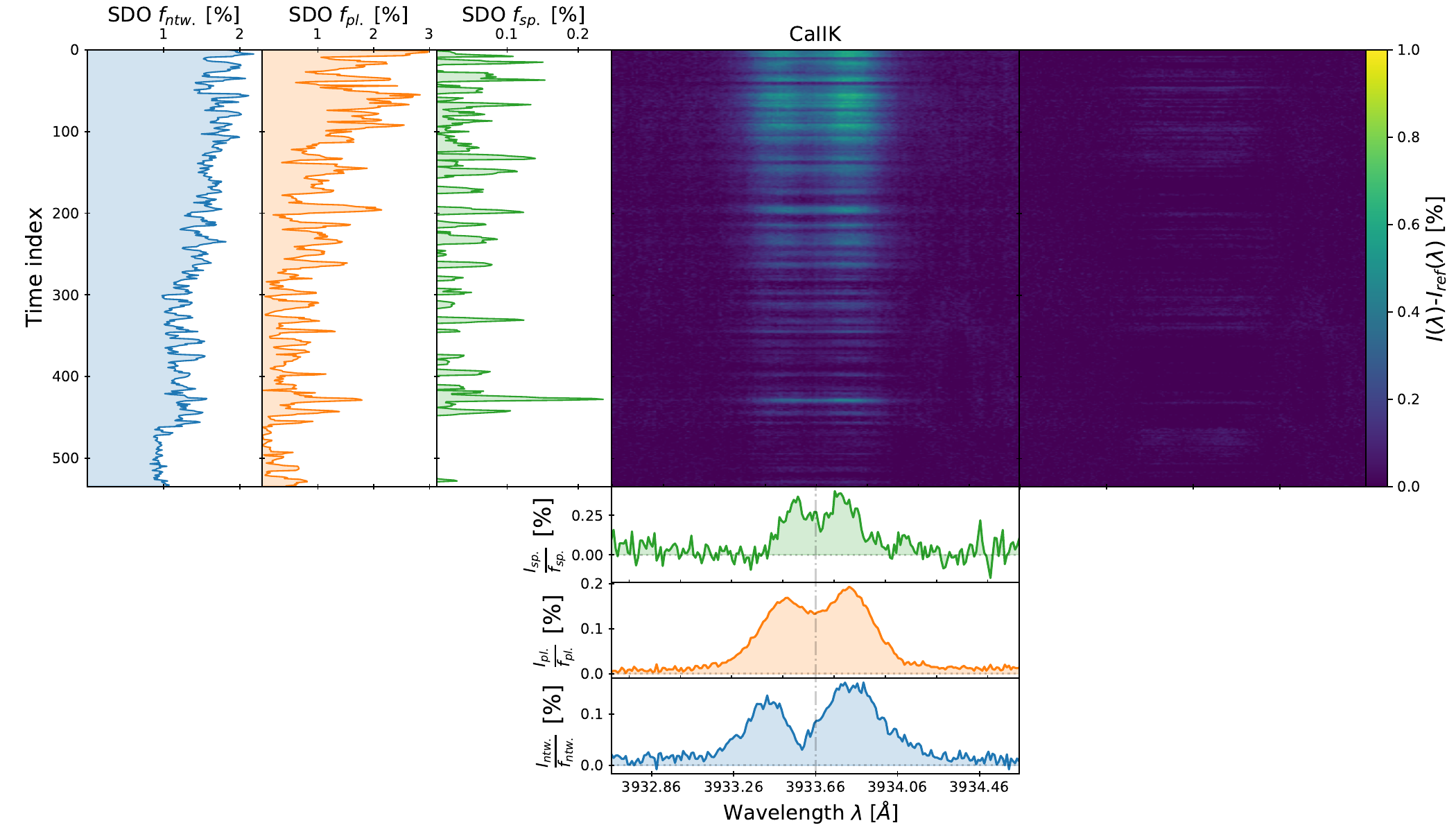}
	\caption{Same as Fig.\ref{FigComp1} for the HARPS-N solar spectra. The double peak broad plage emission profile is recovered as well as its broader equivalence for the network. A noisier spot emission profile is recovered, where the larger noise can be explained both by the lower S/N of HARPS-N in the extreme blue (S/N$_{4000}\sim$ 450) and by larger degeneracy in the Eq.\ref{EqFund2}, since HARPS-N has mainly probed the minimum of the solar magnetic cycle from 20 July 2015 (equivalent to time index 1123 in Fig.\ref{FigComp1}) to 28 June 2018, and only a few spots events have been measured so far.}
	\label{FigComp3}

\end{figure*}

\subsection{Combining the spatially resolved SDO filtergrams with HARPS-N disk-integrated spectra}
\label{sec:harpn}

HARPS-N is a high-resolution spectrograph ($\mathcal{R}\sim$ 110\,000) installed on the TNG at La Palma. A solar telescope has been installed in 2015 \citep{Dumusque(2015)} and is observing continuously the Sun with 5-min exposure time to mitigate the p-mode. A public data release was presented in \citet{Dumusque(2021)} that has been used in several occasions for different stellar activity studies \citep{Milbourne(2021),Collier(2021),Langellier(2021),Beurs(2022),Haywood(2022),Camacho(2022),Hara(2023),Lienhard(2023),Moulla(2022),Moulla(2023)}.

\subsubsection{Processing of the disk-integrated HARPS-N spectra}
\label{sec:harpn_processing}

We processed 535 daily stacked publicly available HARPS-N spectra from 20 July 2015 to 28 June 2018 \citep{Dumusque(2021)}. We analysed the 1D order-merged spectra produced by the official DRS. As usual, with high-resolution spectrographs, spectra are not photometrically calibrated due to differential extinction, filters in the instrument, change in weather condition but also since the fiber sky-projected size is usually smaller than the on-site seeing value \citep{Pepe(2021)}. As a consequence, part of the light is not captured inducing large variations in the continuum from one exposure to the next. The spectra were continuum normalised with RASSINE \citep{Cretignier(2020b)}. The code\footnote{A Python code publicly available on GitHub: \url{https://github.com/MichaelCretignier/Rassine_public}} uses the concept of alpha shapes to perform a filtering between the stellar lines and the upper envelop of the continuum. As part of the standard RASSINE process, all the spectra were evenly resampled on a 0.01 \ang{} wavelength grid by a linear interpolation.

Unlike the ISS, the background scattered light is highly limited in HARPS-N due to the intrinsic RV precision that the instrument is aiming for. Unfortunately, the Ca II H \& K  lines are contaminated by two ghosts \citep{Dumusque(2021)} which are spurious reflections of the physical echelle-order inside the spectrograph. In the official DRS, the \Sidx is corrected for the ghost effect, but such correction is not performed at the spectrum level and the 1D spectrum are therefore still contaminated. We post-processed the spectra using YARARA \citep{Cretignier(2021)} since the pipeline possesses a recipe to mitigate the ghost contamination using Principal Component Analysis (PCA). Since YARARA also corrects for stellar activity, we disabled the stellar activity correction. As part of the standard post-processing, spectra are daily stacked in order to mitigate short timescale phenomena and in order to boost the S/N. Its value in the continuum of the CaII lines is in average around S/N$\sim450$.

\subsubsection{Extraction of the HARPS-N activity emission profiles per unit of area}
\label{sec:harpn_analysis}

We performed the same analysis as in the previous Sect.~\ref{sec:iss_analysis}, but combining SDO filling factors with HARPS-N solar observations. Because of the ghosts falling on both \ion{Ca}{II} lines \citep{Dumusque(2021)}, only the Ca II K line, slightly less contaminated, was analysed and the Ca II H line was dropped. Compared to the ISS dataset, less observations are available and most of them probe the minimum of the solar cycle. Hence, we only fit for a linear trend ($N=1$) for the polynomial detrending.

The solution for the Eq.\ref{EqFund2} is plotted in Fig.\ref{FigComp3}. It leads to emission profiles similar to the one obtained by ISS in the previous section (see Sect.~\ref{sec:iss_analysis}). The plage profile once again shows the characteristic double peak emission with a brighter K2R than K2V maximum. The network profile is also similar to the ISS one. The amplitude of the spot profile is smaller compared to ISS and may be induced by the non-universality of the spot chromospheric emission profiles, already detected with Meudon (see Sect.~\ref{sec:meudon_analysis}). 

We also observed that the extracted emission profile of spots is noisier than the one obtained with ISS, even if a similar thin emission profile with double peaks is visible. The noisier profile can be due to the lower S/N observations (S/N$\sim 450$) or to the number of observations four times less numerous, but also to larger degeneracies in the model. Indeed, contrary to the ISS observations that were probing the full $24^{\rm th}$ solar cycle, HARPS-N has so far mainly probed the minimum part of the solar cycle for which less spot events are visible. Therefore, the current dataset is more subject to degeneracy and spurious correlations. 

\subsection{Comparison between the instruments}
\label{sec:comparison}
We now discuss and compare the different profiles obtained by the different instruments. 

A good agreement was found between both ISS and HARPS-N spectrograph, ISS producing smoother profiles as a result of the larger S/N. Globally the profiles are in agreement with intensity profile per unit of filling factor $I_X$ around 0.15\% for plage and network. A small disagreement is detected for the intensity of spots with intensity of 0.25\% for HARPS-N against 0.40\% for ISS. Such a discrepancy may be explained by the results of Meudon that the spot signature likely depends on the magnetic field geometry and is therefore non-universal. 

When comparing Meudon with the two other instruments, results are more difficult to interpret. It should be reminded that the scaling of the flux units, on the HARPS-N reference, was more difficult to obtain on the Meudon datacubes compared to the ISS disk-integrated spectra. Hence, the flux units of Meudon are less reliable in absolute. 

Quantitatively, all the previous activity profiles are recovered. For instance, a thinner/larger intensity profile of spot/network are obtained compared to plage. In Sect.\ref{sec:meudon_soap}, we recovered similarly $0.15\pm0.05$\% for the plage intensity profile per unit of filling factor. However, the spot profile intensity was about $0.09\pm0.06$\% which is far weaker than the value obtained with the previous instruments. However, we note than only a few spots events have been measured with Meudon, and therefore the statistical size of the sample may be too small to produce reliable average values. 

Finally, whereas the intensity profile for the network was found of a similar level than for the plage on ISS and HARPS-N, results obtained on Meudon were rather providing a contribution 5 times smaller compared to plages (see the amplitude in Fig.\ref{FigMeudon4}). Since the same profiles were obtained with ISS and HARPS-N, the discrepancy with Meudon either occurs because of the SDO segmentation values and/or Meudon segmentations ones. 

\begin{figure*}
	
	\centering
	\includegraphics[width=18cm]{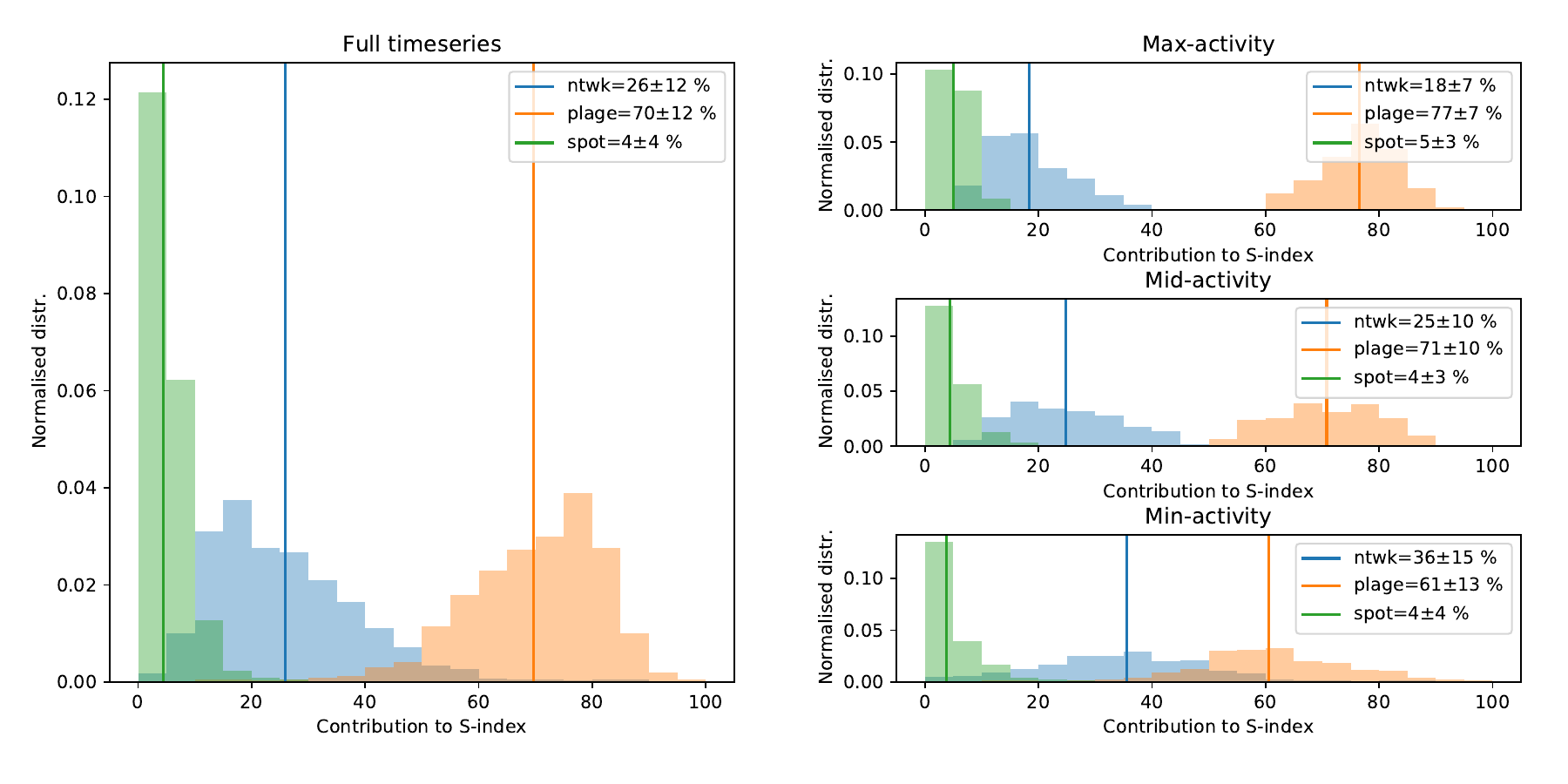}
	\caption{Normalised distribution to the \Sidx contribution of the three activity components: network (blue), plage (orange) and spot (green). Vertical lines highlight the median values of the distributions that are also specified in the legend of the figure with the standard deviation values. The different contributions are obtained by the integration of the different terms in Eq.\ref{EqFund2} fitted on the ISS data ($24^{\rm \text{th}}$ solar cycle) with a 1\ang{} band pass rectangular kernel. The compositions are separately investigated for the full time series (left), and three different activity levels: high (top right), mid (middle right) and low (bottom right). Between low and high activity level, the contribution gradually decreases for the network and increase for plages.}
	\label{FigSindexCompo}

\end{figure*}

We suspect that the Meudon precision was not good enough to perform an accurate segmentation of the network or that an extra activity component contaminates the network segmentation maps. For instance, two datacubes taken a few tens of hours apart were showing filling factor variation larger than 2\% which is too large compared to the temporal variation of the network (see bottom panel of SDO in Fig.\ref{FigSDO1}). This indicates that the Meudon datacubes either contain another bright activity component or noisy pixels. This could also be identified in Fig.\ref{FigMeudon2} where the segmentation seems to pick up the quiet solar chromosphere as well.

On the 2020 dataset, we remarked that the average filling factor was roughly 2.7 times larger than the SDO value in 2020. Hence, it is likely that our network signal is diluted by the same amount due to imperfect segmentation. Therefore, the difference between the plage and network intensity could be a factor of $5/2.7\simeq1.9$, rather than 5. 

Another observation to raise is the difficulty to disentangle the network component from plage using a single filter for the segmentation. Trivially, since we defined in the present work plages as region brighter than a given threshold $I>I_{\text{min}}$, increasing or decreasing the threshold value $I_{\text{min}}$ will naturally convert some network regions as plages or reversely. If plage can easily be identified as concentrated or packed ARs, the network is made of small-scale structures more diffuse.

Another major difference between the Meudon and SDO dataset is also naturally the dates of the observations. For Meudon, we only used the 2020 dataset to form the network sample because this latter was free of any ARs. However, it could be possible that the network component itself is reduced in intensity at the solar minima compared to its average magnetic cycle value. To check this eventuality, we compared the network spectra obtained from the April 2022 datacubes with those of the 2020 dataset. We found that, on this dataset, the intensity of the network was on average twice larger compared to its intensity in 2020. Using again that the signal is diluted by 2.7 would produce a profile $2 \times 2.7= 5.4$ larger, reducing the tension with the extracted network amplitude from HARPS-N and ISS data. 

Note that since the network component seems to evolve with time (or has similarly a large variance spatially), it violates the hypothesis of the model used in Eq.\ref{EqFund2} and therefore the result on ISS/HARPS-N may also be subject to bias. As a consequence it is likely that the network intensity is likely between the lower value of the Meudon instrument ($I_{\text{plage}}/I_{\text{ntwk}}\sim 1.9$) and ISS values ($I_{\text{plage}}/I_{\text{ntwk}}\sim 1$). We note that \citet{Worden1998} derived a factor 1.5 for the contrast ratio between plage and enhanced network in the core of the CaII K line during the $22^{\text{th}}$ solar cycle.

\subsection{Disentangling the S-index composition}
\label{sec:sindex}

To conclude the analysis, we determined the contribution of the different components to the final \Sidx value. To do so, we extracted the intensity map of each contribution ($I_X \cdot f_X$) based on the Eq.\ref{EqFund2} for the ISS decomposition presented in Sect.\ref{sec:iss_analysis}. Each map was then integrated with a rectangular bandpass of $1\ang{}$ to mimic an \Sidx like metric. We display in Fig.\ref{FigSindexCompo} the contribution of all the components on the full cycle as well as at three different activity level obtained by splitting the \Sidx distribution at $33^{\rm \text{th}}$ and $66^{\rm \text{th}}$ percentiles. We found out that along the $24^{\rm \text{th}}$ magnetic cycle, the \Sidx is composed at $70\pm12$~\% by plage intensity, $26 \pm 12$~\% by the network and $4\pm4$~\% by spots. Because those values are the average value along the cycle and since the mixture ratio of these three components evolves with time, it is possible to find some time epochs where the network contribution to the \Sidx raises up to 50\%. For instance, at the minimum of the solar magnetic cycle, we found a contribution of $61\pm13$~\% by plage intensity, $36 \pm 15$~\% by the network and $4\pm4$~\% by spots. In comparison, at the maximum of the magnetic cycle, the distributions are $77\pm7$~\% of plage intensity, $18 \pm 7$~\% of network and $5\pm3$~\% of spots. Such a change in the composition of the \Sidx between high and low activity level would drastically impact its ability to correct the RVs. 

Because the \Sidx is likely a weighted mixture coming both from plage and network, and since only\footnote{This is in contradiction with solar physics studies that also showed the peculiar convective velocities observed in the network component \citep{Buehler2019}} plage have been found to have a significant effect on RVs  \citep{Milbourne(2021)}, it could explain why the \Sidx has been previously classified as an imperfect activity proxy for RV correction \citep{Haywood(2016)}. Furthermore, other studies also demonstrated that a departure from a strict linear relation may exist between chromospheric emission and RVs \citep{Meunier(2019)}. It should be reminded that RVs are mainly photospheric by nature, and no 1:1 relation exist between the photosphere and the chromosphere. For this reason, photospheric activity proxies may be more powerful \citep{Cretignier(2022)}.

However, more complex extraction of the plage signal from the Ca II H \& K lines could already lead to a better RV correction. This will be shown in our next paper II \citep{Cretignier(2024)}. Theoretically, it is possible to elaborate a complex bandpass to extract the different contributions, but none of them will outperforms the extraction by the real profiles as the one presented in Fig.\ref{FigComp1}. 

\section{Conclusions}

\label{sec:conclusion}

We showed in this paper the different profiles expected for the activity components given by the Ca II H \& K lines. Those results will be used as Rosetta stones for our next paper II \citep{Cretignier(2024)}.

We first processed the $I(\lambda,x,y)$ Meudon datacubes to show that mainly three distinct profiles exists for spots, plages and the network. An universal profile is found for plage that shows a minor CLV such as a broadening toward the limb. The network intensity was found lower by one order of magnitude, but that is due partially to a bias that diluted its signal. For spots, an emission profile $\sim$25\% thinner than for plages is measured. This result was detected on three independent solar instruments. Furthermore, the spot emission profile seems non-universal, which may be since umbrae and penumbrae both exhibit different intensity profiles and so different ratio area may lead to different profiles on the Meudon spectra. Also, investigating a sunspot along a rotational phase, a peculiar brightness excess was detected when the AR emerged, that may be related to a peculiar configuration of spots surrounded by plages or by the peculiar height where the magnetic fields were confined. 

By disk integrating the Meudon spectra and normalising by the photospheric flux, we recovered a broad plage emission profile with a narrower emission profile for spots. The plage emission profile also shows a clear and measurable CLV effect that is visible by the broadening of the line.  

By combining SDO filtergrams with ISS disk-integrated spectra of Ca II H \& K lines, we extracted the intensity profile per unit area based on a three-component model and showed that, once again, a broad emission profile for plage is detected and a thinner emission profiles for spots, while a broader profile is obtained for the network compared to plage. Similar results were obtained with HARPS-N even though noisier profiles were recovered. An intensity profile for plage $I_{\text{plage}}=0.15$\% per percent of filling factor was detected independently for the three instruments. However, different values were obtained for spots and for the network. For this latter, a relative intensity $I_{\text{plage}}/I_{\text{ntwk}}$ between 1 and 2 was found depending on the instrument used, in agreement with previous estimation \citep{Worden1998}. 

Finally, we estimated that the popular \Sidx contained in average for the $24^{\rm th}$ solar cycle: $70\pm12\%$ of the plage intensity, $26\pm12\%$ of the network intensity and $4\pm4 \%$ of the spots intensity. If the spot contribution is almost negligible, the network contribution is not. This result may explain why this activity proxy is only partially good for RV correction. Our results for the sunspot contribution are in line with \citet{Sowmya(2023)} who showed that sunspots do not contribute in the \Sidx, except for very active stars.

The present work could be improved by investigating other strong chromospheric lines in the visible such as $H_{\alpha}$ and $H_{\beta}$ that could help to lift the degeneracies between the activity profiles components. Also, the Meudon datacubes appears to contain a wealth of information that we only partially explored. A better categorisation of the ARs along the full Meudon database could be interesting to conduct to better understand the deep relations between Ca II H \& K lines and ARs properties (that we assumed time-independent during this work). Finally, CLV of the profiles could also be extracted using the disk-integrated spectra, if filling factor time series in different $\mu$ bin angle were fit.

In our next paper II \citep{Cretignier(2024)}, we will apply and extrapolate the information acquired on the Sun to better constrain the stellar activity on $\alpha$ Cen B. In particular, we will show that, in the context of RVs, better activity proxies can be extracted.

\section*{Acknowledgements}

We acknowledge Paris Observatory for the use of spectroheliograms. Meudon spectroheliograph data are courtesy of the solar and BASS2000 teams, as part of operational services of Paris Observatory. M.C. acknowledges the SNSF support under grant P500PT\_211024. AP was supported by the 824135 (SOLARNET -- Integrating High Resolution Solar Physics) grant. SA acknowledges funding from the European Research Council (ERC) under the European Union’s Horizon 2020 research and innovation programme (Grant agreement No. 865624)

This research has made use of NASA's Astrophysics Data System (ADS) bibliographic services. 
We acknowledge the community efforts devoted to the development of the following open-source packages that were used in this work: numpy (\href{http:\\numpy.org}{numpy.org}), matplotlib (\href{http:\\matplotlib.org}{matplotlib.org}), and astropy (\href{http:\\astropy.org}{astropy.org}).

\section*{Data Availability}

The data underlying this article are publicly available. The datasets were derived from sources in the public domain: Meudon spectroheliograms (\url{https://bass2000.obspm.fr/home.php}), SDO images (\url{https://sdo.gsfc.nasa.gov/data/aiahmi/}), ISS spectra (\url{https://solis.nso.edu/pubkeep/}) and HARPS-N spectra (\url{https://dace.unige.ch/pythonAPI/?tutorialId=22}).



\bibliographystyle{mnras}
\bibliography{Master} 

\begin{thebibliography}{}
\makeatletter
\relax
\def\mn@urlcharsother{\let\do\@makeother \do\$\do\&\do\#\do\^\do\_\do\%\do\~}
\def\mn@doi{\begingroup\mn@urlcharsother \@ifnextchar [ {\mn@doi@}
  {\mn@doi@[]}}
\def\mn@doi@[#1]#2{\def\@tempa{#1}\ifx\@tempa\@empty \href
  {http://dx.doi.org/#2} {doi:#2}\else \href {http://dx.doi.org/#2} {#1}\fi
  \endgroup}
\def\mn@eprint#1#2{\mn@eprint@#1:#2::\@nil}
\def\mn@eprint@arXiv#1{\href {http://arxiv.org/abs/#1} {{\tt arXiv:#1}}}
\def\mn@eprint@dblp#1{\href {http://dblp.uni-trier.de/rec/bibtex/#1.xml}
  {dblp:#1}}
\def\mn@eprint@#1:#2:#3:#4\@nil{\def\@tempa {#1}\def\@tempb {#2}\def\@tempc
  {#3}\ifx \@tempc \@empty \let \@tempc \@tempb \let \@tempb \@tempa \fi \ifx
  \@tempb \@empty \def\@tempb {arXiv}\fi \@ifundefined
  {mn@eprint@\@tempb}{\@tempb:\@tempc}{\expandafter \expandafter \csname
  mn@eprint@\@tempb\endcsname \expandafter{\@tempc}}}

\bibitem[\protect\citeauthoryear{{Afram} \& {Berdyugina}}{{Afram} \&
  {Berdyugina}}{2015}]{Afram(2015)}
{Afram} N.,  {Berdyugina} S.~V.,  2015, \mn@doi [\aap]
  {10.1051/0004-6361/201425314}, \href
  {https://ui.adsabs.harvard.edu/abs/2015A&A...576A..34A} {576, A34}

\bibitem[\protect\citeauthoryear{{Al Moulla}, {Dumusque}, {Cretignier}, {Zhao}
  \& {Valenti}}{{Al Moulla} et~al.}{2022}]{Moulla(2022)}
{Al Moulla} K.,  {Dumusque} X.,  {Cretignier} M.,  {Zhao} Y.,   {Valenti}
  J.~A.,  2022, arXiv e-prints, \href
  {https://ui.adsabs.harvard.edu/abs/2022arXiv220507047A} {p. arXiv:2205.07047}

\bibitem[\protect\citeauthoryear{{Al Moulla}, {Dumusque}, {Figueira}, {Lo
  Curto}, {Santos}  \& {Wildi}}{{Al Moulla} et~al.}{2023}]{Moulla(2023)}
{Al Moulla} K.,  {Dumusque} X.,  {Figueira} P.,  {Lo Curto} G.,  {Santos}
  N.~C.,   {Wildi} F.,  2023, \mn@doi [\aap] {10.1051/0004-6361/202244663},
  \href {https://ui.adsabs.harvard.edu/abs/2023A&A...669A..39A} {669, A39}

\bibitem[\protect\citeauthoryear{{Anan} et~al.,}{{Anan} et~al.}{2021}]{Anan21}
{Anan} T.,  et~al., 2021, \mn@doi [\apj] {10.3847/1538-4357/ac1b9c}, \href
  {https://ui.adsabs.harvard.edu/abs/2021ApJ...921...39A} {921, 39}

\bibitem[\protect\citeauthoryear{{Anderson}, {Reiners}  \&
  {Solanki}}{{Anderson} et~al.}{2010}]{Anderson(2010)}
{Anderson} R.~I.,  {Reiners} A.,   {Solanki} S.~K.,  2010, \mn@doi [aap]
  {10.1051/0004-6361/201014769}, \href
  {https://ui.adsabs.harvard.edu/abs/2010A&A...522A..81A} {522, A81}

\bibitem[\protect\citeauthoryear{{Avrett}, {Tian}, {Landi}, {Curdt}  \&
  {W{\"u}lser}}{{Avrett} et~al.}{2015}]{Avrett(2015)}
{Avrett} E.,  {Tian} H.,  {Landi} E.,  {Curdt} W.,   {W{\"u}lser} J.~P.,  2015,
  \mn@doi [\apj] {10.1088/0004-637X/811/2/87}, \href
  {https://ui.adsabs.harvard.edu/abs/2015ApJ...811...87A} {811, 87}

\bibitem[\protect\citeauthoryear{{Basri}, {Wilcots}  \& {Stout}}{{Basri}
  et~al.}{1989}]{Basri(1989)}
{Basri} G.,  {Wilcots} E.,   {Stout} N.,  1989, \mn@doi [pasp]
  {10.1086/132464}, \href
  {https://ui.adsabs.harvard.edu/abs/1989PASP..101..528B} {101, 528}

\bibitem[\protect\citeauthoryear{{Bastien} et~al.,}{{Bastien}
  et~al.}{2014}]{Bastien(2014)}
{Bastien} F.~A.,  et~al., 2014, \mn@doi [aj] {10.1088/0004-6256/147/2/29},
  \href {https://ui.adsabs.harvard.edu/abs/2014AJ....147...29B} {147, 29}

\bibitem[\protect\citeauthoryear{{Bechter}, {Bechter}, {Crepp}, {King}  \&
  {Crass}}{{Bechter} et~al.}{2018}]{Bechter(2018)}
{Bechter} A.~J.,  {Bechter} E.~B.,  {Crepp} J.~R.,  {King} D.,   {Crass} J.,
  2018, in {Evans} C.~J.,  {Simard} L.,   {Takami} H.,  eds,  Society of
  Photo-Optical Instrumentation Engineers (SPIE) Conference Series Vol. 10702,
  Ground-based and Airborne Instrumentation for Astronomy VII. p. 107026T
  (\mn@eprint {arXiv} {1812.02704}), \mn@doi{10.1117/12.2313658}

\bibitem[\protect\citeauthoryear{{Bechter}, {Bechter}, {Crepp}, {Crass}  \&
  {King}}{{Bechter} et~al.}{2019}]{Bechter(2019)}
{Bechter} E.~B.,  {Bechter} A.~J.,  {Crepp} J.~R.,  {Crass} J.,   {King} D.,
  2019, \mn@doi [pasp] {10.1088/1538-3873/aaf278}, \href
  {https://ui.adsabs.harvard.edu/abs/2019PASP..131b4504B} {131, 024504}

\bibitem[\protect\citeauthoryear{{Beck} \& {Chapman}}{{Beck} \&
  {Chapman}}{1993}]{Beck(1993)}
{Beck} J.~G.,  {Chapman} G.~A.,  1993, \mn@doi [solphys] {10.1007/BF00662169},
  \href {https://ui.adsabs.harvard.edu/abs/1993SoPh..146...49B} {146, 49}

\bibitem[\protect\citeauthoryear{{Berdyugina}, {Solanki}  \&
  {Frutiger}}{{Berdyugina} et~al.}{2003}]{Berdyugina(2003a)}
{Berdyugina} S.~V.,  {Solanki} S.~K.,   {Frutiger} C.,  2003, \mn@doi [aap]
  {10.1051/0004-6361:20031473}, \href
  {https://ui.adsabs.harvard.edu/abs/2003A%26A...412..513B} {412, 513}

\bibitem[\protect\citeauthoryear{{Berger}, {Rouppe van der Voort}  \&
  {L{\"o}fdahl}}{{Berger} et~al.}{2007}]{Berger(2007)}
{Berger} T.~E.,  {Rouppe van der Voort} L.,   {L{\"o}fdahl} M.,  2007, \mn@doi
  [apj] {10.1086/517502}, \href
  {https://ui.adsabs.harvard.edu/abs/2007ApJ...661.1272B} {661, 1272}

\bibitem[\protect\citeauthoryear{{Bertello}, {Pevtsov}, {Harvey}  \&
  {Toussaint}}{{Bertello} et~al.}{2011}]{Bertello(2011)}
{Bertello} L.,  {Pevtsov} A.~A.,  {Harvey} J.~W.,   {Toussaint} R.~M.,  2011,
  \mn@doi [solphys] {10.1007/s11207-011-9820-8}, \href
  {https://ui.adsabs.harvard.edu/abs/2011SoPh..272..229B} {272, 229}

\bibitem[\protect\citeauthoryear{{Bertello}, {Pevtsov}  \&
  {Pietarila}}{{Bertello} et~al.}{2012}]{Bertello(2012)}
{Bertello} L.,  {Pevtsov} A.~A.,   {Pietarila} A.,  2012, \mn@doi [apj]
  {10.1088/0004-637X/761/1/11}, \href
  {https://ui.adsabs.harvard.edu/abs/2012ApJ...761...11B} {761, 11}

\bibitem[\protect\citeauthoryear{{Bj{\o}rgen}, {Sukhorukov}, {Leenaarts},
  {Carlsson}, {de la Cruz Rodr{\'\i}guez}, {Scharmer}  \&
  {Hansteen}}{{Bj{\o}rgen} et~al.}{2018a}]{Bjorgen2018}
{Bj{\o}rgen} J.~P.,  {Sukhorukov} A.~V.,  {Leenaarts} J.,  {Carlsson} M.,  {de
  la Cruz Rodr{\'\i}guez} J.,  {Scharmer} G.~B.,   {Hansteen} V.~H.,  2018a,
  \mn@doi [\aap] {10.1051/0004-6361/201731926}, \href
  {https://ui.adsabs.harvard.edu/abs/2018A&A...611A..62B} {611, A62}

\bibitem[\protect\citeauthoryear{{Bj{\o}rgen}, {Sukhorukov}, {Leenaarts},
  {Carlsson}, {de la Cruz Rodr{\'\i}guez}, {Scharmer}  \&
  {Hansteen}}{{Bj{\o}rgen} et~al.}{2018b}]{Bjorgen(2018)}
{Bj{\o}rgen} J.~P.,  {Sukhorukov} A.~V.,  {Leenaarts} J.,  {Carlsson} M.,  {de
  la Cruz Rodr{\'\i}guez} J.,  {Scharmer} G.~B.,   {Hansteen} V.~H.,  2018b,
  \mn@doi [aap] {10.1051/0004-6361/201731926}, \href
  {https://ui.adsabs.harvard.edu/abs/2018A&A...611A..62B} {611, A62}

\bibitem[\protect\citeauthoryear{{Boisse}, {Bonfils}  \& {Santos}}{{Boisse}
  et~al.}{2012}]{Boisse(2012)}
{Boisse} I.,  {Bonfils} X.,   {Santos} N.~C.,  2012, \mn@doi [aap]
  {10.1051/0004-6361/201219115}, \href
  {https://ui.adsabs.harvard.edu/abs/2012A&A...545A.109B} {545, A109}

\bibitem[\protect\citeauthoryear{{Bonomo} \& {Lanza}}{{Bonomo} \&
  {Lanza}}{2012}]{Bonomo(2012)}
{Bonomo} A.~S.,  {Lanza} A.~F.,  2012, \mn@doi [aap]
  {10.1051/0004-6361/201219999}, \href
  {https://ui.adsabs.harvard.edu/abs/2012A&A...547A..37B} {547, A37}

\bibitem[\protect\citeauthoryear{{Borrero} \& {Ichimoto}}{{Borrero} \&
  {Ichimoto}}{2011}]{Borrero(2011)}
{Borrero} J.~M.,  {Ichimoto} K.,  2011, \mn@doi [Living Reviews in Solar
  Physics] {10.12942/lrsp-2011-4}, \href
  {https://ui.adsabs.harvard.edu/abs/2011LRSP....8....4B} {8, 4}

\bibitem[\protect\citeauthoryear{{Brandt} \& {Solanki}}{{Brandt} \&
  {Solanki}}{1990}]{Brandt(1990)}
{Brandt} P.~N.,  {Solanki} S.~K.,  1990, aap, \href
  {https://ui.adsabs.harvard.edu/abs/1990A&A...231..221B} {231, 221}

\bibitem[\protect\citeauthoryear{{Brants} \& {Zwaan}}{{Brants} \&
  {Zwaan}}{1982}]{Brants(1982)}
{Brants} J.~J.,  {Zwaan} C.,  1982, \mn@doi [solphys] {10.1007/BF00147972},
  \href {https://ui.adsabs.harvard.edu/abs/1982SoPh...80..251B} {80, 251}

\bibitem[\protect\citeauthoryear{{Breton}, {Santos}, {Bugnet}, {Mathur},
  {Garc{\'\i}a}  \& {Pall{\'e}}}{{Breton} et~al.}{2021}]{Breton(2021)}
{Breton} S.~N.,  {Santos} A.~R.~G.,  {Bugnet} L.,  {Mathur} S.,  {Garc{\'\i}a}
  R.~A.,   {Pall{\'e}} P.~L.,  2021, \mn@doi [aap]
  {10.1051/0004-6361/202039947}, \href
  {https://ui.adsabs.harvard.edu/abs/2021A&A...647A.125B} {647, A125}

\bibitem[\protect\citeauthoryear{{Bruls} \& {Solanki}}{{Bruls} \&
  {Solanki}}{2004}]{Bruls(2004)}
{Bruls} J.~H.~M.~J.,  {Solanki} S.~K.,  2004, \mn@doi [\aap]
  {10.1051/0004-6361:20041311}, \href
  {https://ui.adsabs.harvard.edu/abs/2004A&A...427..735B} {427, 735}

\bibitem[\protect\citeauthoryear{{Buehler}, {Lagg}, {Solanki}  \& {van
  Noort}}{{Buehler} et~al.}{2015}]{Buehler2015}
{Buehler} D.,  {Lagg} A.,  {Solanki} S.~K.,   {van Noort} M.,  2015, \mn@doi
  [\aap] {10.1051/0004-6361/201424970}, \href
  {https://ui.adsabs.harvard.edu/abs/2015A&A...576A..27B} {576, A27}

\bibitem[\protect\citeauthoryear{{Buehler}, {Lagg}, {van Noort}  \&
  {Solanki}}{{Buehler} et~al.}{2019}]{Buehler2019}
{Buehler} D.,  {Lagg} A.,  {van Noort} M.,   {Solanki} S.~K.,  2019, \mn@doi
  [\aap] {10.1051/0004-6361/201833585}, \href
  {https://ui.adsabs.harvard.edu/abs/2019A&A...630A..86B} {630, A86}

\bibitem[\protect\citeauthoryear{{Caccin}, {Penza}  \& {Gomez}}{{Caccin}
  et~al.}{2002}]{Caccin(2002)}
{Caccin} B.,  {Penza} V.,   {Gomez} M.~T.,  2002, \mn@doi [aap]
  {10.1051/0004-6361:20020217}, \href
  {https://ui.adsabs.harvard.edu/abs/2002A&A...386..286C} {386, 286}

\bibitem[\protect\citeauthoryear{{Camacho}, {Faria}  \& {Viana}}{{Camacho}
  et~al.}{2022}]{Camacho(2022)}
{Camacho} J.~D.,  {Faria} J.~P.,   {Viana} P.~T.~P.,  2022, arXiv e-prints,
  \href {https://ui.adsabs.harvard.edu/abs/2022arXiv220506627C} {p.
  arXiv:2205.06627}

\bibitem[\protect\citeauthoryear{{Carleo} et~al.,}{{Carleo}
  et~al.}{2020}]{Carleo(2020)}
{Carleo} I.,  et~al., 2020, \mn@doi [aap] {10.1051/0004-6361/201937369}, \href
  {https://ui.adsabs.harvard.edu/abs/2020A&A...638A...5C} {638, A5}

\bibitem[\protect\citeauthoryear{{Carlsson}, {De Pontieu}  \&
  {Hansteen}}{{Carlsson} et~al.}{2019}]{Carlsson2019}
{Carlsson} M.,  {De Pontieu} B.,   {Hansteen} V.~H.,  2019, \mn@doi [\araa]
  {10.1146/annurev-astro-081817-052044}, \href
  {https://ui.adsabs.harvard.edu/abs/2019ARA&A..57..189C} {57, 189}

\bibitem[\protect\citeauthoryear{{Catalano}, {Biazzo}, {Frasca}  \&
  {Marilli}}{{Catalano} et~al.}{2002}]{Catalano(2002)}
{Catalano} S.,  {Biazzo} K.,  {Frasca} A.,   {Marilli} E.,  2002, \mn@doi [aap]
  {10.1051/0004-6361:20021223}, \href
  {https://ui.adsabs.harvard.edu/abs/2002A&A...394.1009C} {394, 1009}

\bibitem[\protect\citeauthoryear{{Cauzzi} \& {Reardon}}{{Cauzzi} \&
  {Reardon}}{2012}]{Cauzi2012}
{Cauzzi} G.,  {Reardon} K.,  2012, IAU Special Session, \href
  {https://ui.adsabs.harvard.edu/abs/2012IAUSS...6E.511C} {6, E5.11}

\bibitem[\protect\citeauthoryear{{Chapman} \& {Klabunde}}{{Chapman} \&
  {Klabunde}}{1982}]{Chapman(1982)}
{Chapman} G.~A.,  {Klabunde} D.~P.,  1982, \mn@doi [\apj] {10.1086/160349},
  \href {https://ui.adsabs.harvard.edu/abs/1982ApJ...261..387C} {261, 387}

\bibitem[\protect\citeauthoryear{{Chapman}, {Cookson}, {Dobias}  \&
  {Walton}}{{Chapman} et~al.}{2001}]{Chapman(2001)}
{Chapman} G.~A.,  {Cookson} A.~M.,  {Dobias} J.~J.,   {Walton} S.~R.,  2001,
  \mn@doi [apj] {10.1086/321466}, \href
  {http://adsabs.harvard.edu/abs/2001ApJ...555..462C} {555, 462}

\bibitem[\protect\citeauthoryear{{Chowdhury}, {Belur}, {Bertello}  \&
  {Pevtsov}}{{Chowdhury} et~al.}{2022}]{Chowdhury2022}
{Chowdhury} P.,  {Belur} R.,  {Bertello} L.,   {Pevtsov} A.~A.,  2022, \mn@doi
  [\apj] {10.3847/1538-4357/ac3983}, \href
  {https://ui.adsabs.harvard.edu/abs/2022ApJ...925...81C} {925, 81}

\bibitem[\protect\citeauthoryear{{Collier Cameron} et~al.,}{{Collier Cameron}
  et~al.}{2021}]{Collier(2021)}
{Collier Cameron} A.,  et~al., 2021, \mn@doi [mnras] {10.1093/mnras/stab1323},
  \href {https://ui.adsabs.harvard.edu/abs/2021MNRAS.505.1699C} {505, 1699}

\bibitem[\protect\citeauthoryear{{Cretignier}, {Dumusque}, {Allart}, {Pepe}  \&
  {Lovis}}{{Cretignier} et~al.}{2020a}]{Cretignier(2020a)}
{Cretignier} M.,  {Dumusque} X.,  {Allart} R.,  {Pepe} F.,   {Lovis} C.,
  2020a, \mn@doi [aap] {10.1051/0004-6361/201936548}, \href
  {https://ui.adsabs.harvard.edu/abs/2020A&A...633A..76C} {633, A76}

\bibitem[\protect\citeauthoryear{{Cretignier}, {Francfort}, {Dumusque},
  {Allart}  \& {Pepe}}{{Cretignier} et~al.}{2020b}]{Cretignier(2020b)}
{Cretignier} M.,  {Francfort} J.,  {Dumusque} X.,  {Allart} R.,   {Pepe} F.,
  2020b, \mn@doi [aap] {10.1051/0004-6361/202037722}, \href
  {https://ui.adsabs.harvard.edu/abs/2020A&A...640A..42C} {640, A42}

\bibitem[\protect\citeauthoryear{{Cretignier}, {Dumusque}, {Hara}  \&
  {Pepe}}{{Cretignier} et~al.}{2021}]{Cretignier(2021)}
{Cretignier} M.,  {Dumusque} X.,  {Hara} N.~C.,   {Pepe} F.,  2021, \mn@doi
  [aap] {10.1051/0004-6361/202140986}, \href
  {https://ui.adsabs.harvard.edu/abs/2021A&A...653A..43C} {653, A43}

\bibitem[\protect\citeauthoryear{{Cretignier}, {Dumusque}  \&
  {Pepe}}{{Cretignier} et~al.}{2022}]{Cretignier(2022)}
{Cretignier} M.,  {Dumusque} X.,   {Pepe} F.,  2022, \mn@doi [aap]
  {10.1051/0004-6361/202142435}, \href
  {https://ui.adsabs.harvard.edu/abs/2022A&A...659A..68C} {659, A68}

\bibitem[\protect\citeauthoryear{{Cretignier}, {Hara}, {Pietrow}, {Zhao}  \&
  {Aigrain}}{{Cretignier} et~al.}{prep}]{Cretignier(2024)}
{Cretignier} M.,  {Hara} N.,  {Pietrow} A.,  {Zhao} Y.,   {Aigrain} S.,
  in\,prep., \mnras

\bibitem[\protect\citeauthoryear{{Denker} \& {Verma}}{{Denker} \&
  {Verma}}{2019}]{Denker2019}
{Denker} C.,  {Verma} M.,  2019, \mn@doi [\solphys]
  {10.1007/s11207-019-1459-x}, \href
  {https://ui.adsabs.harvard.edu/abs/2019SoPh..294...71D} {294, 71}

\bibitem[\protect\citeauthoryear{{Desort}, {Lagrange}, {Galland}, {Udry}  \&
  {Mayor}}{{Desort} et~al.}{2007}]{Desort(2007)}
{Desort} M.,  {Lagrange} A.~M.,  {Galland} F.,  {Udry} S.,   {Mayor} M.,  2007,
  \mn@doi [aap] {10.1051/0004-6361:20078144}, \href
  {https://ui.adsabs.harvard.edu/abs/2007A&A...473..983D} {473, 983}

\bibitem[\protect\citeauthoryear{{Dineva}, {Pearson}, {Ilyin}, {Verma},
  {Diercke}, {Strassmeier}  \& {Denker}}{{Dineva} et~al.}{2022}]{Dineva(2022)}
{Dineva} E.,  {Pearson} J.,  {Ilyin} I.,  {Verma} M.,  {Diercke} A.,
  {Strassmeier} K.~G.,   {Denker} C.,  2022, \mn@doi [Astronomische
  Nachrichten] {10.1002/asna.20223996}, \href
  {https://ui.adsabs.harvard.edu/abs/2022AN....34323996D} {343, e23996}

\bibitem[\protect\citeauthoryear{{Druett}, {Scullion}, {Zharkova}, {Matthews},
  {Zharkov}  \& {Rouppe van der Voort}}{{Druett} et~al.}{2017}]{Druett2017}
{Druett} M.,  {Scullion} E.,  {Zharkova} V.,  {Matthews} S.,  {Zharkov} S.,
  {Rouppe van der Voort} L.,  2017, \mn@doi [Nature Communications]
  {10.1038/ncomms15905}, \href
  {https://ui.adsabs.harvard.edu/abs/2017NatCo...815905D} {8, 15905}

\bibitem[\protect\citeauthoryear{{Druett}, {Pietrow}, {Vissers}, {Robustini}
  \& {Calvo}}{{Druett} et~al.}{2022}]{Druett2022}
{Druett} M.~K.,  {Pietrow} A. G.~M.,  {Vissers} G. J.~M.,  {Robustini} C.,
  {Calvo} F.,  2022, \mn@doi [RAS Techniques and Instruments]
  {10.1093/rasti/rzac003}, \href
  {https://ui.adsabs.harvard.edu/abs/2022RASTI...1...29D} {1, 29}

\bibitem[\protect\citeauthoryear{{Druzhinin}, {Pevtsov}  \&
  {Teplitskaja}}{{Druzhinin} et~al.}{1987}]{Druzhinin87}
{Druzhinin} S.~A.,  {Pevtsov} A.~A.,   {Teplitskaja} R.~B.,  1987,
  Astronomicheskij Tsirkulyar, \href
  {https://ui.adsabs.harvard.edu/abs/1987ATsir1512....5D} {1512, 5}

\bibitem[\protect\citeauthoryear{{Dumusque}, {Boisse}  \& {Santos}}{{Dumusque}
  et~al.}{2014}]{Dumusque(2014)}
{Dumusque} X.,  {Boisse} I.,   {Santos} N.~C.,  2014, \mn@doi [apj]
  {10.1088/0004-637X/796/2/132}, \href
  {http://adsabs.harvard.edu/abs/2014ApJ...796..132D} {796, 132}

\bibitem[\protect\citeauthoryear{{Dumusque} et~al.,}{{Dumusque}
  et~al.}{2015}]{Dumusque(2015)}
{Dumusque} X.,  et~al., 2015, \mn@doi [apjl] {10.1088/2041-8205/814/2/L21},
  \href {http://adsabs.harvard.edu/abs/2015ApJ...814L..21D} {814, L21}

\bibitem[\protect\citeauthoryear{{Dumusque} et~al.,}{{Dumusque}
  et~al.}{2021}]{Dumusque(2021)}
{Dumusque} X.,  et~al., 2021, \mn@doi [aap] {10.1051/0004-6361/202039350},
  \href {https://ui.adsabs.harvard.edu/abs/2021A&A...648A.103D} {648, A103}

\bibitem[\protect\citeauthoryear{{Duncan} et~al.,}{{Duncan}
  et~al.}{1991}]{Duncan(1991)}
{Duncan} D.~K.,  et~al., 1991, \mn@doi [\apjs] {10.1086/191572}, \href
  {https://ui.adsabs.harvard.edu/abs/1991ApJS...76..383D} {76, 383}

\bibitem[\protect\citeauthoryear{{Ellwarth}, {Sch{\"a}fer}, {Reiners}  \&
  {Zechmeister}}{{Ellwarth} et~al.}{2023}]{Ellwarth23}
{Ellwarth} M.,  {Sch{\"a}fer} S.,  {Reiners} A.,   {Zechmeister} M.,  2023,
  \mn@doi [arXiv e-prints] {10.48550/arXiv.2303.08205}, \href
  {https://ui.adsabs.harvard.edu/abs/2023arXiv230308205E} {p. arXiv:2303.08205}

\bibitem[\protect\citeauthoryear{{Engvold}}{{Engvold}}{1966}]{Engvold(1966)}
{Engvold} O.,  1966, Astrophysica Norvegica, \href
  {https://ui.adsabs.harvard.edu/abs/1966ApNr...10..101E} {10, 101}

\bibitem[\protect\citeauthoryear{{Engvold}}{{Engvold}}{1967}]{Engvold(1967)}
{Engvold} O.,  1967, \mn@doi [\solphys] {10.1007/BF00155926}, \href
  {https://ui.adsabs.harvard.edu/abs/1967SoPh....2..234E} {2, 234}

\bibitem[\protect\citeauthoryear{{Flores}, {Gonz{\'a}lez}, {Jaque Arancibia},
  {Buccino}  \& {Saffe}}{{Flores} et~al.}{2016}]{Flores(2016)}
{Flores} M.,  {Gonz{\'a}lez} J.~F.,  {Jaque Arancibia} M.,  {Buccino} A.,
  {Saffe} C.,  2016, \mn@doi [aap] {10.1051/0004-6361/201628145}, \href
  {https://ui.adsabs.harvard.edu/abs/2016A&A...589A.135F} {589, A135}

\bibitem[\protect\citeauthoryear{{Flores}, {Gonz{\'a}lez}, {Jaque Arancibia},
  {Saffe}, {Buccino}, {L{\'o}pez}, {Iba{\~n}ez Bustos}  \&
  {Miquelarena}}{{Flores} et~al.}{2018}]{Flores(2018)}
{Flores} M.,  {Gonz{\'a}lez} J.~F.,  {Jaque Arancibia} M.,  {Saffe} C.,
  {Buccino} A.,  {L{\'o}pez} F.~M.,  {Iba{\~n}ez Bustos} R.~V.,   {Miquelarena}
  P.,  2018, \mn@doi [aap] {10.1051/0004-6361/201833330}, \href
  {https://ui.adsabs.harvard.edu/abs/2018A&A...620A..34F} {620, A34}

\bibitem[\protect\citeauthoryear{{Foukal}}{{Foukal}}{2013}]{Foukal2003}
{Foukal} P.,  2013, {Solar astrophysics, 3rd, Revised Edition}

\bibitem[\protect\citeauthoryear{{Foukal}, {Little}, {Graves}, {Rabin}  \&
  {Lynch}}{{Foukal} et~al.}{1990}]{Foukal(1990)}
{Foukal} P.,  {Little} R.,  {Graves} J.,  {Rabin} D.,   {Lynch} D.,  1990,
  \mn@doi [apj] {10.1086/168660}, \href
  {https://ui.adsabs.harvard.edu/abs/1990ApJ...353..712F} {353, 712}

\bibitem[\protect\citeauthoryear{{Frasca}, {Biazzo}, {Catalano}, {Marilli},
  {Messina}  \& {Rodon{\`o}}}{{Frasca} et~al.}{2005}]{Frasca(2005)}
{Frasca} A.,  {Biazzo} K.,  {Catalano} S.,  {Marilli} E.,  {Messina} S.,
  {Rodon{\`o}} M.,  2005, \mn@doi [aap] {10.1051/0004-6361:20041373}, \href
  {https://ui.adsabs.harvard.edu/abs/2005A&A...432..647F} {432, 647}

\bibitem[\protect\citeauthoryear{{Frasca}, {Biazzo}, {Ta{\c{s}}}, {Evren}  \&
  {Lanzafame}}{{Frasca} et~al.}{2008}]{Frasca(2008)}
{Frasca} A.,  {Biazzo} K.,  {Ta{\c{s}}} G.,  {Evren} S.,   {Lanzafame} A.~C.,
  2008, \mn@doi [aap] {10.1051/0004-6361:20077915}, \href
  {https://ui.adsabs.harvard.edu/abs/2008A&A...479..557F} {479, 557}

\bibitem[\protect\citeauthoryear{{Giorgini} et~al.,}{{Giorgini}
  et~al.}{1996}]{Giorgini(1996)}
{Giorgini} J.~D.,  et~al., 1996, in AAS/Division for Planetary Sciences Meeting
  Abstracts \#28. p. 25.04

\bibitem[\protect\citeauthoryear{{Gomes da Silva}, {Figueira}, {Santos}  \&
  {Faria}}{{Gomes da Silva} et~al.}{2018}]{Gomes(2018)}
{Gomes da Silva} J.,  {Figueira} P.,  {Santos} N.,   {Faria} J.,  2018, \mn@doi
  [The Journal of Open Source Software] {10.21105/joss.00667}, \href
  {https://ui.adsabs.harvard.edu/abs/2018JOSS....3..667G} {3, 667}

\bibitem[\protect\citeauthoryear{{Gray}}{{Gray}}{1988}]{Gray(1988)}
{Gray} D.~F.,  1988, {Lectures on spectral-line analysis: F,G, and K stars}

\bibitem[\protect\citeauthoryear{{Gray} \& {Johanson}}{{Gray} \&
  {Johanson}}{1991}]{Gray(1991)}
{Gray} D.~F.,  {Johanson} H.~L.,  1991, \mn@doi [pasp] {10.1086/132839}, \href
  {http://adsabs.harvard.edu/abs/1991PASP..103..439G} {103, 439}

\bibitem[\protect\citeauthoryear{{Hale}, {Ellerman}, {Nicholson}  \&
  {Joy}}{{Hale} et~al.}{1919}]{Hale1919}
{Hale} G.~E.,  {Ellerman} F.,  {Nicholson} S.~B.,   {Joy} A.~H.,  1919, \mn@doi
  [\apj] {10.1086/142452}, \href
  {https://ui.adsabs.harvard.edu/abs/1919ApJ....49..153H} {49, 153}

\bibitem[\protect\citeauthoryear{{Hara} \& {Delisle}}{{Hara} \&
  {Delisle}}{2023}]{Hara(2023)}
{Hara} N.~C.,  {Delisle} J.-B.,  2023, \mn@doi [arXiv e-prints]
  {10.48550/arXiv.2304.08489}, \href
  {https://ui.adsabs.harvard.edu/abs/2023arXiv230408489H} {p. arXiv:2304.08489}

\bibitem[\protect\citeauthoryear{{Harvey} \& {White}}{{Harvey} \&
  {White}}{1999}]{Harvey(1999)}
{Harvey} K.~L.,  {White} O.~R.,  1999, \mn@doi [\apj] {10.1086/307035}, \href
  {https://ui.adsabs.harvard.edu/abs/1999ApJ...515..812H} {515, 812}

\bibitem[\protect\citeauthoryear{{Haywood} et~al.,}{{Haywood}
  et~al.}{2016}]{Haywood(2016)}
{Haywood} R.~D.,  et~al., 2016, \mn@doi [mnras] {10.1093/mnras/stw187}, \href
  {http://adsabs.harvard.edu/abs/2016MNRAS.457.3637H} {457, 3637}

\bibitem[\protect\citeauthoryear{{Haywood} et~al.,}{{Haywood}
  et~al.}{2022}]{Haywood(2022)}
{Haywood} R.~D.,  et~al., 2022, \mn@doi [\apj] {10.3847/1538-4357/ac7c12},
  \href {https://ui.adsabs.harvard.edu/abs/2022ApJ...935....6H} {935, 6}

\bibitem[\protect\citeauthoryear{{Herrero}, {Ribas}, {Jordi}, {Morales},
  {Perger}  \& {Rosich}}{{Herrero} et~al.}{2016}]{Herrero(2016)}
{Herrero} E.,  {Ribas} I.,  {Jordi} C.,  {Morales} J.~C.,  {Perger} M.,
  {Rosich} A.,  2016, \mn@doi [aap] {10.1051/0004-6361/201425369}, \href
  {https://ui.adsabs.harvard.edu/abs/2016A&A...586A.131H} {586, A131}

\bibitem[\protect\citeauthoryear{{Hou}, {Zhang}, {Li}, {Yang}  \& {Li}}{{Hou}
  et~al.}{2018}]{Hou2018}
{Hou} Y.~J.,  {Zhang} J.,  {Li} T.,  {Yang} S.~H.,   {Li} X.~H.,  2018, \mn@doi
  [\aap] {10.1051/0004-6361/201732530}, \href
  {https://ui.adsabs.harvard.edu/abs/2018A&A...619A.100H} {619, A100}

\bibitem[\protect\citeauthoryear{Inoue \& Bamba}{Inoue \&
  Bamba}{2021}]{Inoue_2021}
Inoue S.,  Bamba Y.,  2021, \mn@doi [The Astrophysical Journal]
  {10.3847/1538-4357/abf835}, 914, 71

\bibitem[\protect\citeauthoryear{Inoue, Shiota, Bamba  \& Park}{Inoue
  et~al.}{2018}]{Inoue_2018}
Inoue S.,  Shiota D.,  Bamba Y.,   Park S.-H.,  2018, \mn@doi [The
  Astrophysical Journal] {10.3847/1538-4357/aae079}, 867, 83

\bibitem[\protect\citeauthoryear{{Jovanovic}, {Schwab}, {Cvetojevic}, {Guyon}
  \& {Martinache}}{{Jovanovic} et~al.}{2016}]{Jovanovic(2016)}
{Jovanovic} N.,  {Schwab} C.,  {Cvetojevic} N.,  {Guyon} O.,   {Martinache} F.,
   2016, \mn@doi [pasp] {10.1088/1538-3873/128/970/121001}, \href
  {https://ui.adsabs.harvard.edu/abs/2016PASP..128l1001J} {128, 121001}

\bibitem[\protect\citeauthoryear{{Keil}, {Henry}  \& {Fleck}}{{Keil}
  et~al.}{1998}]{Keil(1998)}
{Keil} S.~L.,  {Henry} T.~W.,   {Fleck} B.,  1998, in {Balasubramaniam} K.~S.,
  {Harvey} J.,   {Rabin} D.,  eds,  Astronomical Society of the Pacific
  Conference Series Vol. 140, Synoptic Solar Physics. p.~301

\bibitem[\protect\citeauthoryear{{Keller}, {Harvey}  \& {Giampapa}}{{Keller}
  et~al.}{2003}]{Keller2003}
{Keller} C.~U.,  {Harvey} J.~W.,   {Giampapa} M.~S.,  2003, in {Keil} S.~L.,
  {Avakyan} S.~V.,  eds,  Society of Photo-Optical Instrumentation Engineers
  (SPIE) Conference Series Vol. 4853, Innovative Telescopes and Instrumentation
  for Solar Astrophysics. pp 194--204, \mn@doi{10.1117/12.460373}

\bibitem[\protect\citeauthoryear{{Keppens}}{{Keppens}}{2000}]{Keppens00}
{Keppens} R.,  2000, in {Murdin} P.,  ed., , Encyclopedia of Astronomy and
  Astrophysics.
p.~2043, \mn@doi{10.1888/0333750888/2043}

\bibitem[\protect\citeauthoryear{{Kneer}, {Scharmer}, {Mattig}, {Wyller},
  {Artzner}, {Lemaire}  \& {Vial}}{{Kneer} et~al.}{1981}]{Kneer(1981)}
{Kneer} F.,  {Scharmer} G.,  {Mattig} W.,  {Wyller} A.,  {Artzner} G.,
  {Lemaire} P.,   {Vial} J.~C.,  1981, \mn@doi [\solphys] {10.1007/BF00149995},
  \href {https://ui.adsabs.harvard.edu/abs/1981SoPh...69..289K} {69, 289}

\bibitem[\protect\citeauthoryear{{Kossakowski} et~al.,}{{Kossakowski}
  et~al.}{2022}]{Kossakowski(2022)}
{Kossakowski} D.,  et~al., 2022, arXiv e-prints, \href
  {https://ui.adsabs.harvard.edu/abs/2022arXiv220905814K} {p. arXiv:2209.05814}

\bibitem[\protect\citeauthoryear{{Krikova}, {Pereira}  \& {Rouppe van der
  Voort}}{{Krikova} et~al.}{2023}]{Krikova(2023)}
{Krikova} K.,  {Pereira} T.~M.~D.,   {Rouppe van der Voort} L.~H.~M.,  2023,
  \mn@doi [arXiv e-prints] {10.48550/arXiv.2307.11131}, \href
  {https://ui.adsabs.harvard.edu/abs/2023arXiv230711131K} {p. arXiv:2307.11131}

\bibitem[\protect\citeauthoryear{{Kuckein}, {Verma}  \& {Denker}}{{Kuckein}
  et~al.}{2016}]{Kuckein2016}
{Kuckein} C.,  {Verma} M.,   {Denker} C.,  2016, \mn@doi [\aap]
  {10.1051/0004-6361/201526636}, \href
  {https://ui.adsabs.harvard.edu/abs/2016A&A...589A..84K} {589, A84}

\bibitem[\protect\citeauthoryear{{Lafarga} et~al.,}{{Lafarga}
  et~al.}{2021}]{Lafarga(2021)}
{Lafarga} M.,  et~al., 2021, \mn@doi [aap] {10.1051/0004-6361/202140605}, \href
  {https://ui.adsabs.harvard.edu/abs/2021A&A...652A..28L} {652, A28}

\bibitem[\protect\citeauthoryear{{Langellier} et~al.,}{{Langellier}
  et~al.}{2021}]{Langellier(2021)}
{Langellier} N.,  et~al., 2021, \mn@doi [aj] {10.3847/1538-3881/abf1e0}, \href
  {https://ui.adsabs.harvard.edu/abs/2021AJ....161..287L} {161, 287}

\bibitem[\protect\citeauthoryear{{Lanza} et~al.,}{{Lanza}
  et~al.}{2009}]{Lanza(2009)}
{Lanza} A.~F.,  et~al., 2009, \mn@doi [aap] {10.1051/0004-6361:200810591},
  \href {https://ui.adsabs.harvard.edu/abs/2009A&A...493..193L} {493, 193}

\bibitem[\protect\citeauthoryear{{Lanzafame} et~al.,}{{Lanzafame}
  et~al.}{2018}]{Lanzafame(2018)}
{Lanzafame} A.~C.,  et~al., 2018, \mn@doi [\aap] {10.1051/0004-6361/201833334},
  \href {https://ui.adsabs.harvard.edu/abs/2018A&A...616A..16L} {616, A16}

\bibitem[\protect\citeauthoryear{{Leenaarts}, {Pereira}, {Carlsson},
  {Uitenbroek}  \& {De Pontieu}}{{Leenaarts} et~al.}{2013}]{Leenaarts2013}
{Leenaarts} J.,  {Pereira} T.~M.~D.,  {Carlsson} M.,  {Uitenbroek} H.,   {De
  Pontieu} B.,  2013, \mn@doi [\apj] {10.1088/0004-637X/772/2/90}, \href
  {https://ui.adsabs.harvard.edu/abs/2013ApJ...772...90L} {772, 90}

\bibitem[\protect\citeauthoryear{{Lemaire}, {Choucq-Bruston}  \&
  {Vial}}{{Lemaire} et~al.}{1984}]{Lemaire(1984)}
{Lemaire} P.,  {Choucq-Bruston} M.,   {Vial} J.~C.,  1984, \mn@doi [\solphys]
  {10.1007/BF00153785}, \href
  {https://ui.adsabs.harvard.edu/abs/1984SoPh...90...63L} {90, 63}

\bibitem[\protect\citeauthoryear{{Lemen} et~al.,}{{Lemen}
  et~al.}{2012}]{Lemen2012}
{Lemen} J.~R.,  et~al., 2012, \mn@doi [\solphys] {10.1007/s11207-011-9776-8},
  \href {https://ui.adsabs.harvard.edu/abs/2012SoPh..275...17L} {275, 17}

\bibitem[\protect\citeauthoryear{{Li} et~al.,}{{Li} et~al.}{2023}]{Li2023L}
{Li} H.,  et~al., 2023, \mn@doi [arXiv e-prints] {10.48550/arXiv.2301.12792},
  \href {https://ui.adsabs.harvard.edu/abs/2023arXiv230112792L} {p.
  arXiv:2301.12792}

\bibitem[\protect\citeauthoryear{{Lienhard}, {Mortier}, {Cegla}, {Cameron},
  {Klein}  \& {Watson}}{{Lienhard} et~al.}{2023}]{Lienhard(2023)}
{Lienhard} F.,  {Mortier} A.,  {Cegla} H.~M.,  {Cameron} A.~C.,  {Klein} B.,
  {Watson} C.~A.,  2023, \mn@doi [\mnras] {10.1093/mnras/stad1343}, \href
  {https://ui.adsabs.harvard.edu/abs/2023MNRAS.522.5862L} {522, 5862}

\bibitem[\protect\citeauthoryear{{Lites}, {Elmore}, {Seagraves}  \&
  {Skumanich}}{{Lites} et~al.}{1993}]{Lites(1993)}
{Lites} B.~W.,  {Elmore} D.~F.,  {Seagraves} P.,   {Skumanich} A.~P.,  1993,
  \mn@doi [apj] {10.1086/173450}, \href
  {https://ui.adsabs.harvard.edu/abs/1993ApJ...418..928L} {418, 928}

\bibitem[\protect\citeauthoryear{{Livingston}, {Wallace}, {White}  \&
  {Giampapa}}{{Livingston} et~al.}{2007}]{Livingston(2007)}
{Livingston} W.,  {Wallace} L.,  {White} O.~R.,   {Giampapa} M.~S.,  2007,
  \mn@doi [apj] {10.1086/511127}, \href
  {http://adsabs.harvard.edu/abs/2007ApJ...657.1137L} {657, 1137}

\bibitem[\protect\citeauthoryear{{L{\"o}hner-B{\"o}ttcher}, {Schmidt},
  {Schlichenmaier}, {Steinmetz}  \& {Holzwarth}}{{L{\"o}hner-B{\"o}ttcher}
  et~al.}{2019}]{Lohner(2019)}
{L{\"o}hner-B{\"o}ttcher} J.,  {Schmidt} W.,  {Schlichenmaier} R.,  {Steinmetz}
  T.,   {Holzwarth} R.,  2019, \mn@doi [aap] {10.1051/0004-6361/201834925},
  \href {https://ui.adsabs.harvard.edu/abs/2019A&A...624A..57L} {624, A57}

\bibitem[\protect\citeauthoryear{{Love}}{{Love}}{2021}]{Love(2021)}
{Love} J.~J.,  2021, \mn@doi [Space Weather] {10.1029/2020SW002579}, \href
  {https://ui.adsabs.harvard.edu/abs/2021SpWea..1902579L} {19, e02579}

\bibitem[\protect\citeauthoryear{{Luger}, {Agol}, {Foreman-Mackey}, {Fleming},
  {Lustig-Yaeger}  \& {Deitrick}}{{Luger} et~al.}{2019}]{Luger(2019)}
{Luger} R.,  {Agol} E.,  {Foreman-Mackey} D.,  {Fleming} D.~P.,
  {Lustig-Yaeger} J.,   {Deitrick} R.,  2019, \mn@doi [\aj]
  {10.3847/1538-3881/aae8e5}, \href
  {https://ui.adsabs.harvard.edu/abs/2019AJ....157...64L} {157, 64}

\bibitem[\protect\citeauthoryear{{Malanushenko}, {Jones}  \&
  {Livingston}}{{Malanushenko} et~al.}{2004}]{Malanushenko(2004)}
{Malanushenko} O.,  {Jones} H.~P.,   {Livingston} W.,  2004, in {Stepanov}
  A.~V.,  {Benevolenskaya} E.~E.,   {Kosovichev} A.~G.,  eds,  Vol. 223,
  Multi-Wavelength Investigations of Solar Activity. pp 645--646,
  \mn@doi{10.1017/S1743921304007161}

\bibitem[\protect\citeauthoryear{{Maldonado} et~al.,}{{Maldonado}
  et~al.}{2019}]{Maldonado(2019)}
{Maldonado} J.,  et~al., 2019, \mn@doi [aap] {10.1051/0004-6361/201935233},
  \href {https://ui.adsabs.harvard.edu/abs/2019A&A...627A.118M} {627, A118}

\bibitem[\protect\citeauthoryear{{Malherbe} \& {Dalmasse}}{{Malherbe} \&
  {Dalmasse}}{2019}]{Malherbe(2019)}
{Malherbe} J.~M.,  {Dalmasse} K.,  2019, \mn@doi [\solphys]
  {10.1007/s11207-019-1441-7}, \href
  {https://ui.adsabs.harvard.edu/abs/2019SoPh..294...52M} {294, 52}

\bibitem[\protect\citeauthoryear{{Malherbe}, {Bual{\'e}}, {Crussaire}, {Cornu}
  \& {Corbard}}{{Malherbe} et~al.}{2023}]{Malherbe(2023)}
{Malherbe} J.-M.,  {Bual{\'e}} I.,  {Crussaire} D.,  {Cornu} F.,   {Corbard}
  T.,  2023, \mn@doi [Advances in Space Research] {10.1016/j.asr.2022.07.058},
  \href {https://ui.adsabs.harvard.edu/abs/2023AdSpR..71.1922M} {71, 1922}

\bibitem[\protect\citeauthoryear{{Martinez Pillet}, {Garcia Lopez}, {del Toro
  Iniesta}, {Rebolo}, {Vazquez}, {Beckman}  \& {Char}}{{Martinez Pillet}
  et~al.}{1990}]{Martinez(1990)}
{Martinez Pillet} V.,  {Garcia Lopez} R.~J.,  {del Toro Iniesta} J.~C.,
  {Rebolo} R.,  {Vazquez} M.,  {Beckman} J.~E.,   {Char} S.,  1990, \mn@doi
  [\apjl] {10.1086/185832}, \href
  {https://ui.adsabs.harvard.edu/abs/1990ApJ...361L..81M} {361, L81}

\bibitem[\protect\citeauthoryear{{McIntosh}}{{McIntosh}}{1990}]{McIntosh1990}
{McIntosh} P.~S.,  1990, \mn@doi [\solphys] {10.1007/BF00158405}, \href
  {https://ui.adsabs.harvard.edu/abs/1990SoPh..125..251M} {125, 251}

\bibitem[\protect\citeauthoryear{{Mello}, {Bouchez}, {Szentgyorgyi}, {van Dam}
  \& {Lupinari}}{{Mello} et~al.}{2018}]{Mello(2018)}
{Mello} A. J.~T.~S.,  {Bouchez} A.~H.,  {Szentgyorgyi} A.,  {van Dam} M.~A.,
  {Lupinari} H.,  2018, \mn@doi [mnras] {10.1093/mnras/sty2542}, \href
  {https://ui.adsabs.harvard.edu/abs/2018MNRAS.481.3804M} {481, 3804}

\bibitem[\protect\citeauthoryear{{Messina} \& {Guinan}}{{Messina} \&
  {Guinan}}{2002}]{Messina(2002)}
{Messina} S.,  {Guinan} E.~F.,  2002, \mn@doi [aap]
  {10.1051/0004-6361:20021000}, \href
  {https://ui.adsabs.harvard.edu/abs/2002A&A...393..225M} {393, 225}

\bibitem[\protect\citeauthoryear{{Meunier} \& {Delfosse}}{{Meunier} \&
  {Delfosse}}{2009}]{Meunier(2009)}
{Meunier} N.,  {Delfosse} X.,  2009, \mn@doi [aap]
  {10.1051/0004-6361/200911823}, \href
  {https://ui.adsabs.harvard.edu/abs/2009A&A...501.1103M} {501, 1103}

\bibitem[\protect\citeauthoryear{{Meunier}, {Lagrange}, {Borgniet}  \&
  {Rieutord}}{{Meunier} et~al.}{2015}]{Meunier(2015)}
{Meunier} N.,  {Lagrange} A.-M.,  {Borgniet} S.,   {Rieutord} M.,  2015,
  \mn@doi [aap] {10.1051/0004-6361/201525721}, \href
  {http://adsabs.harvard.edu/abs/2015A%26A...583A.118M} {583, A118}

\bibitem[\protect\citeauthoryear{{Meunier}, {Lagrange}  \& {Cuzacq}}{{Meunier}
  et~al.}{2019}]{Meunier(2019)}
{Meunier} N.,  {Lagrange} A.~M.,   {Cuzacq} S.,  2019, \mn@doi [aap]
  {10.1051/0004-6361/201935348}, \href
  {https://ui.adsabs.harvard.edu/abs/2019A&A...632A..81M} {632, A81}

\bibitem[\protect\citeauthoryear{{Meunier}, {Kretzschmar}, {Gravet}, {Mignon}
  \& {Delfosse}}{{Meunier} et~al.}{2022}]{Meunier(2022)}
{Meunier} N.,  {Kretzschmar} M.,  {Gravet} R.,  {Mignon} L.,   {Delfosse} X.,
  2022, \mn@doi [aap] {10.1051/0004-6361/202142120}, \href
  {https://ui.adsabs.harvard.edu/abs/2022A&A...658A..57M} {658, A57}

\bibitem[\protect\citeauthoryear{{Milbourne} et~al.,}{{Milbourne}
  et~al.}{2019}]{Milbourne(2019)}
{Milbourne} T.~W.,  et~al., 2019, \mn@doi [apj] {10.3847/1538-4357/ab064a},
  \href {http://adsabs.harvard.edu/abs/2019ApJ...874..107M} {874, 107}

\bibitem[\protect\citeauthoryear{{Milbourne} et~al.,}{{Milbourne}
  et~al.}{2021}]{Milbourne(2021)}
{Milbourne} T.~W.,  et~al., 2021, \mn@doi [apj] {10.3847/1538-4357/ac1266},
  \href {https://ui.adsabs.harvard.edu/abs/2021ApJ...920...21M} {920, 21}

\bibitem[\protect\citeauthoryear{{Mirtorabi}, {Wasatonic}  \&
  {Guinan}}{{Mirtorabi} et~al.}{2003}]{Mirtorabi(2003)}
{Mirtorabi} M.~T.,  {Wasatonic} R.,   {Guinan} E.~F.,  2003, \mn@doi [\aj]
  {10.1086/368247}, \href
  {https://ui.adsabs.harvard.edu/abs/2003AJ....125.3265M} {125, 3265}

\bibitem[\protect\citeauthoryear{{Miyakawa}, {Hirano}, {Fukui}, {Mann},
  {Gaidos}  \& {Sato}}{{Miyakawa} et~al.}{2021}]{Miyakawa(2021)}
{Miyakawa} K.,  {Hirano} T.,  {Fukui} A.,  {Mann} A.~W.,  {Gaidos} E.,   {Sato}
  B.,  2021, \mn@doi [aj] {10.3847/1538-3881/ac111d}, \href
  {https://ui.adsabs.harvard.edu/abs/2021AJ....162..104M} {162, 104}

\bibitem[\protect\citeauthoryear{{Morosin}, {de la Cruz Rodr{\'\i}guez},
  {Vissers}  \& {Yadav}}{{Morosin} et~al.}{2020}]{Morosin2020}
{Morosin} R.,  {de la Cruz Rodr{\'\i}guez} J.,  {Vissers} G. J.~M.,   {Yadav}
  R.,  2020, \mn@doi [\aap] {10.1051/0004-6361/202038754}, \href
  {https://ui.adsabs.harvard.edu/abs/2020A&A...642A.210M} {642, A210}

\bibitem[\protect\citeauthoryear{{Morosin}, {de la Cruz Rodr{\'\i}guez},
  {D{\'\i}az Baso}  \& {Leenaarts}}{{Morosin} et~al.}{2022}]{Morosin2022}
{Morosin} R.,  {de la Cruz Rodr{\'\i}guez} J.,  {D{\'\i}az Baso} C.~J.,
  {Leenaarts} J.,  2022, \mn@doi [\aap] {10.1051/0004-6361/202243461}, \href
  {https://ui.adsabs.harvard.edu/abs/2022A&A...664A...8M} {664, A8}

\bibitem[\protect\citeauthoryear{{Muller}}{{Muller}}{1983}]{Muller1983}
{Muller} R.,  1983, \mn@doi [\solphys] {10.1007/BF00148262}, \href
  {https://ui.adsabs.harvard.edu/abs/1983SoPh...85..113M} {85, 113}

\bibitem[\protect\citeauthoryear{{Nagovitsyn} \& {Pevtsov}}{{Nagovitsyn} \&
  {Pevtsov}}{2016}]{Nagovitsyn(2016)}
{Nagovitsyn} Y.~A.,  {Pevtsov} A.~A.,  2016, \mn@doi [\apj]
  {10.3847/1538-4357/833/1/94}, \href
  {https://ui.adsabs.harvard.edu/abs/2016ApJ...833...94N} {833, 94}

\bibitem[\protect\citeauthoryear{{Neckel} \& {Labs}}{{Neckel} \&
  {Labs}}{1984}]{Neckel1984}
{Neckel} H.,  {Labs} D.,  1984, \mn@doi [\solphys] {10.1007/BF00173953}, \href
  {https://ui.adsabs.harvard.edu/abs/1984SoPh...90..205N} {90, 205}

\bibitem[\protect\citeauthoryear{{Neckel} \& {Labs}}{{Neckel} \&
  {Labs}}{1994}]{Neckel1994}
{Neckel} H.,  {Labs} D.,  1994, \mn@doi [\solphys] {10.1007/BF00712494}, \href
  {https://ui.adsabs.harvard.edu/abs/1994SoPh..153...91N} {153, 91}

\bibitem[\protect\citeauthoryear{{Nielsen}, {Gizon}, {Schunker}  \&
  {Karoff}}{{Nielsen} et~al.}{2013}]{Nielsen(2013)}
{Nielsen} M.~B.,  {Gizon} L.,  {Schunker} H.,   {Karoff} C.,  2013, \mn@doi
  [\aap] {10.1051/0004-6361/201321912}, \href
  {https://ui.adsabs.harvard.edu/abs/2013A&A...557L..10N} {557, L10}

\bibitem[\protect\citeauthoryear{{Nikbakhsh}, {Tanskanen}, {K{\"a}pyl{\"a}}  \&
  {Hackman}}{{Nikbakhsh} et~al.}{2019}]{Nikbakhsh(2019)}
{Nikbakhsh} S.,  {Tanskanen} E.~I.,  {K{\"a}pyl{\"a}} M.~J.,   {Hackman} T.,
  2019, \mn@doi [\aap] {10.1051/0004-6361/201935486}, \href
  {https://ui.adsabs.harvard.edu/abs/2019A&A...629A..45N} {629, A45}

\bibitem[\protect\citeauthoryear{{O'Neal}}{{O'Neal}}{2006}]{Oneal(2006)}
{O'Neal} D.,  2006, \mn@doi [apj] {10.1086/504318}, \href
  {https://ui.adsabs.harvard.edu/abs/2006ApJ...645..659O} {645, 659}

\bibitem[\protect\citeauthoryear{{O'Neal} \& {Neff}}{{O'Neal} \&
  {Neff}}{1997}]{Oneal(1997)}
{O'Neal} D.,  {Neff} J.~E.,  1997, \mn@doi [\aj] {10.1086/118331}, \href
  {https://ui.adsabs.harvard.edu/abs/1997AJ....113.1129O} {113, 1129}

\bibitem[\protect\citeauthoryear{{O'Neal}, {Neff}  \& {Saar}}{{O'Neal}
  et~al.}{1998}]{Oneal(1998)}
{O'Neal} D.,  {Neff} J.~E.,   {Saar} S.~H.,  1998, \mn@doi [\apj]
  {10.1086/306340}, \href
  {https://ui.adsabs.harvard.edu/abs/1998ApJ...507..919O} {507, 919}

\bibitem[\protect\citeauthoryear{{O'Neal}, {Neff}, {Saar}  \& {Mines}}{{O'Neal}
  et~al.}{2001}]{Oneal(2001)}
{O'Neal} D.,  {Neff} J.~E.,  {Saar} S.~H.,   {Mines} J.~K.,  2001, \mn@doi
  [\aj] {10.1086/323093}, \href
  {https://ui.adsabs.harvard.edu/abs/2001AJ....122.1954O} {122, 1954}

\bibitem[\protect\citeauthoryear{{O'Neal}, {Neff}, {Saar}  \& {Cuntz}}{{O'Neal}
  et~al.}{2004}]{Oneal(2004)}
{O'Neal} D.,  {Neff} J.~E.,  {Saar} S.~H.,   {Cuntz} M.,  2004, \mn@doi [\aj]
  {10.1086/423438}, \href
  {https://ui.adsabs.harvard.edu/abs/2004AJ....128.1802O} {128, 1802}

\bibitem[\protect\citeauthoryear{{Oranje}}{{Oranje}}{1983}]{Oranje(1983b)}
{Oranje} B.~J.,  1983, aap, \href
  {https://ui.adsabs.harvard.edu/abs/1983A&A...124...43O} {124, 43}

\bibitem[\protect\citeauthoryear{{Ortiz} \& {Rast}}{{Ortiz} \&
  {Rast}}{2005}]{Ortiz(2005)}
{Ortiz} A.,  {Rast} M.,  2005, memsai, \href
  {http://adsabs.harvard.edu/abs/2005MmSAI..76.1018O} {76, 1018}

\bibitem[\protect\citeauthoryear{{Pasquini}}{{Pasquini}}{1992}]{Pasquini(1992)}
{Pasquini} L.,  1992, \aap, \href
  {https://ui.adsabs.harvard.edu/abs/1992A&A...266..347P} {266, 347}

\bibitem[\protect\citeauthoryear{{Pastor Yabar}, {Martinez Gonzalez}  \&
  {Collados}}{{Pastor Yabar} et~al.}{2015}]{Yabar2015}
{Pastor Yabar} A.,  {Martinez Gonzalez} M.~J.,   {Collados} M.,  2015, \mn@doi
  [\mnras] {10.1093/mnrasl/slv108}, \href
  {https://ui.adsabs.harvard.edu/abs/2015MNRAS.453L..69P} {453, L69}

\bibitem[\protect\citeauthoryear{{Penza}, {Caccin}  \& {Del Moro}}{{Penza}
  et~al.}{2004}]{Penza(2004)}
{Penza} V.,  {Caccin} B.,   {Del Moro} D.,  2004, \mn@doi [aap]
  {10.1051/0004-6361:20041151}, \href
  {https://ui.adsabs.harvard.edu/abs/2004A&A...427..345P} {427, 345}

\bibitem[\protect\citeauthoryear{{Pepe} et~al.,}{{Pepe}
  et~al.}{2021}]{Pepe(2021)}
{Pepe} F.,  et~al., 2021, \mn@doi [aap] {10.1051/0004-6361/202038306}, \href
  {https://ui.adsabs.harvard.edu/abs/2021A&A...645A..96P} {645, A96}

\bibitem[\protect\citeauthoryear{{Pesnell}, {Thompson}  \&
  {Chamberlin}}{{Pesnell} et~al.}{2012}]{Pesnell2012}
{Pesnell} W.~D.,  {Thompson} B.~J.,   {Chamberlin} P.~C.,  2012, \mn@doi
  [\solphys] {10.1007/s11207-011-9841-3}, \href
  {https://ui.adsabs.harvard.edu/abs/2012SoPh..275....3P} {275, 3}

\bibitem[\protect\citeauthoryear{{Pietrow}}{{Pietrow}}{2022}]{Pietrow2022}
{Pietrow} A.~G.~M.,  2022, PhD thesis, Stockholm University

\bibitem[\protect\citeauthoryear{{Pietrow}, {Kiselman}, {de la Cruz
  Rodr{\'\i}guez}, {D{\'\i}az Baso}, {Pastor Yabar}  \& {Yadav}}{{Pietrow}
  et~al.}{2020}]{Pietrow2020}
{Pietrow} A.~G.~M.,  {Kiselman} D.,  {de la Cruz Rodr{\'\i}guez} J.,
  {D{\'\i}az Baso} C.~J.,  {Pastor Yabar} A.,   {Yadav} R.,  2020, \mn@doi
  [\aap] {10.1051/0004-6361/202038750}, \href
  {https://ui.adsabs.harvard.edu/abs/2020A&A...644A..43P} {644, A43}

\bibitem[\protect\citeauthoryear{{Pietrow}, {Druett}, {de la Cruz Rodriguez},
  {Calvo}  \& {Kiselman}}{{Pietrow} et~al.}{2022}]{Pietrow2022a}
{Pietrow} A.~G.~M.,  {Druett} M.~K.,  {de la Cruz Rodriguez} J.,  {Calvo} F.,
  {Kiselman} D.,  2022, \mn@doi [\aap] {10.1051/0004-6361/202142346}, \href
  {https://ui.adsabs.harvard.edu/abs/2022A&A...659A..58P} {659, A58}

\bibitem[\protect\citeauthoryear{{Pietrow} et~al.,}{{Pietrow}
  et~al.}{2023a}]{Pietrow23flare}
{Pietrow} A.~G.~M.,  et~al., 2023a, arXiv e-prints

\bibitem[\protect\citeauthoryear{{Pietrow}, {Hoppe}, {Bergemann}  \&
  {Calvo}}{{Pietrow} et~al.}{2023b}]{Pietrow2023b}
{Pietrow} A.~G.~M.,  {Hoppe} R.,  {Bergemann} M.,   {Calvo} F.,  2023b, \mn@doi
  [arXiv e-prints] {10.48550/arXiv.2304.01048}, \href
  {https://ui.adsabs.harvard.edu/abs/2023arXiv230401048P} {p. arXiv:2304.01048}

\bibitem[\protect\citeauthoryear{{Pietrow}, {Kiselman}, {Andriienko}, {Petit
  dit de la Roche}, {D{\'\i}az Baso}  \& {Calvo}}{{Pietrow}
  et~al.}{2023c}]{Pietrow2023}
{Pietrow} A.~G.~M.,  {Kiselman} D.,  {Andriienko} O.,  {Petit dit de la Roche}
  D.~J.~M.,  {D{\'\i}az Baso} C.~J.,   {Calvo} F.,  2023c, \mn@doi [\aap]
  {10.1051/0004-6361/202244811}, \href
  {https://ui.adsabs.harvard.edu/abs/2023A&A...671A.130P} {671, A130}

\bibitem[\protect\citeauthoryear{{Rajhans}, {Tripathi}, {Kashyap}, {Klimchuk}
  \& {Ghosh}}{{Rajhans} et~al.}{2023}]{Rajhans(2023)}
{Rajhans} A.,  {Tripathi} D.,  {Kashyap} V.~L.,  {Klimchuk} J.~A.,   {Ghosh}
  A.,  2023, \mn@doi [\apj] {10.3847/1538-4357/acb4ed}, \href
  {https://ui.adsabs.harvard.edu/abs/2023ApJ...944..158R} {944, 158}

\bibitem[\protect\citeauthoryear{{Ramelli}, {Setzer}, {Engelhard}, {Bianda},
  {Paglia}, {Stenflo}, {K{\"u}veler}  \& {Plewe}}{{Ramelli}
  et~al.}{2017}]{ramelli17}
{Ramelli} R.,  {Setzer} M.,  {Engelhard} M.,  {Bianda} M.,  {Paglia} F.,
  {Stenflo} J.~O.,  {K{\"u}veler} G.,   {Plewe} R.,  2017, arXiv e-prints,
  \href {https://ui.adsabs.harvard.edu/abs/2017arXiv170803284R} {p.
  arXiv:1708.03284}

\bibitem[\protect\citeauthoryear{{Ramelli}, {Bianda}, {Setzer}, {Enegelhard},
  {Paglia}, {Stenflo}, {K{\"u}veler}  \& {Plewe}}{{Ramelli}
  et~al.}{2019}]{ramelli19}
{Ramelli} R.,  {Bianda} M.,  {Setzer} M.,  {Enegelhard} M.,  {Paglia} F.,
  {Stenflo} J.~O.,  {K{\"u}veler} G.,   {Plewe} R.,  2019, in {Belluzzi} L.,
  {Casini} R.,  {Romoli} M.,   {Trujillo Bueno} J.,  eds,  Astronomical Society
  of the Pacific Conference Series Vol. 526, Solar Polariation Workshop 8.
  p.~287

\bibitem[\protect\citeauthoryear{{Reiners}, {Shulyak}, {Anglada-Escud{\'e}},
  {Jeffers}, {Morin}, {Zechmeister}, {Kochukhov}  \& {Piskunov}}{{Reiners}
  et~al.}{2013}]{Reiners(2013)}
{Reiners} A.,  {Shulyak} D.,  {Anglada-Escud{\'e}} G.,  {Jeffers} S.~V.,
  {Morin} J.,  {Zechmeister} M.,  {Kochukhov} O.,   {Piskunov} N.,  2013,
  \mn@doi [aap] {10.1051/0004-6361/201220437}, \href
  {http://adsabs.harvard.edu/abs/2013A%26A...552A.103R} {552, A103}

\bibitem[\protect\citeauthoryear{{Robertson}, {Bender}, {Mahadevan}, {Roy}  \&
  {Ramsey}}{{Robertson} et~al.}{2016}]{Robertson(2016)}
{Robertson} P.,  {Bender} C.,  {Mahadevan} S.,  {Roy} A.,   {Ramsey} L.~W.,
  2016, \mn@doi [\apj] {10.3847/0004-637X/832/2/112}, \href
  {https://ui.adsabs.harvard.edu/abs/2016ApJ...832..112R} {832, 112}

\bibitem[\protect\citeauthoryear{{Robinson}}{{Robinson}}{1980}]{Robinson(1980)}
{Robinson} R.~D. J.,  1980, \mn@doi [apj] {10.1086/158184}, \href
  {https://ui.adsabs.harvard.edu/abs/1980ApJ...239..961R} {239, 961}

\bibitem[\protect\citeauthoryear{{Roettenbacher}, {Monnier}, {Harmon},
  {Barclay}  \& {Still}}{{Roettenbacher} et~al.}{2013}]{Roettenbacher(2013)}
{Roettenbacher} R.~M.,  {Monnier} J.~D.,  {Harmon} R.~O.,  {Barclay} T.,
  {Still} M.,  2013, \mn@doi [\apj] {10.1088/0004-637X/767/1/60}, \href
  {https://ui.adsabs.harvard.edu/abs/2013ApJ...767...60R} {767, 60}

\bibitem[\protect\citeauthoryear{Romano, Elmhamdi, Falco, Costa, Kordi,
  Al-Trabulsy  \& Al-Shammari}{Romano et~al.}{2018}]{Romano2018}
Romano P.,  Elmhamdi A.,  Falco M.,  Costa P.,  Kordi A.~S.,  Al-Trabulsy
  H.~A.,   Al-Shammari R.~M.,  2018, \mn@doi [The Astrophysical Journal]
  {10.3847/2041-8213/aaa1df}, 852, L10

\bibitem[\protect\citeauthoryear{{Rutten}, {de Wijn}  \&
  {S{\"u}tterlin}}{{Rutten} et~al.}{2004}]{Rutten04}
{Rutten} R.~J.,  {de Wijn} A.~G.,   {S{\"u}tterlin} P.,  2004, \mn@doi [\aap]
  {10.1051/0004-6361:20035636}, \href
  {https://ui.adsabs.harvard.edu/abs/2004A&A...416..333R} {416, 333}

\bibitem[\protect\citeauthoryear{{Sasso}, {Andretta}, {Terranegra}  \&
  {Gomez}}{{Sasso} et~al.}{2017}]{Sasso(2017)}
{Sasso} C.,  {Andretta} V.,  {Terranegra} L.,   {Gomez} M.~T.,  2017, \mn@doi
  [aap] {10.1051/0004-6361/201730676}, \href
  {https://ui.adsabs.harvard.edu/abs/2017A&A...604A..50S} {604, A50}

\bibitem[\protect\citeauthoryear{{Sch{\"o}fer} et~al.,}{{Sch{\"o}fer}
  et~al.}{2019}]{Schofer(2019)}
{Sch{\"o}fer} P.,  et~al., 2019, \mn@doi [aap] {10.1051/0004-6361/201834114},
  \href {https://ui.adsabs.harvard.edu/abs/2019A&A...623A..44S} {623, A44}

\bibitem[\protect\citeauthoryear{{Schrijver} \& {Zwaan}}{{Schrijver} \&
  {Zwaan}}{2000}]{Schrijver(2000)}
{Schrijver} C.~J.,  {Zwaan} C.,  2000, {Solar and Stellar Magnetic Activity}

\bibitem[\protect\citeauthoryear{{Schroeder}}{{Schroeder}}{2000}]{Schroeder(2000)}
{Schroeder} D.,  2000, San Diego : Academic Press, \href
  {https://ui.adsabs.harvard.edu/abs/2000asop.conf.....S} {}

\bibitem[\protect\citeauthoryear{{Shapiro}, {Solanki}, {Krivova}, {Schmutz},
  {Ball}, {Knaack}, {Rozanov}  \& {Unruh}}{{Shapiro}
  et~al.}{2014}]{Shapiro(2014)}
{Shapiro} A.~I.,  {Solanki} S.~K.,  {Krivova} N.~A.,  {Schmutz} W.~K.,  {Ball}
  W.~T.,  {Knaack} R.,  {Rozanov} E.~V.,   {Unruh} Y.~C.,  2014, \mn@doi [aap]
  {10.1051/0004-6361/201323086}, \href
  {http://adsabs.harvard.edu/abs/2014A%26A...569A..38S} {569, A38}

\bibitem[\protect\citeauthoryear{{Sheminova}, {Rutten}  \& {Rouppe van der
  Voort}}{{Sheminova} et~al.}{2005}]{Sheminova(2005)}
{Sheminova} V.~A.,  {Rutten} R.~J.,   {Rouppe van der Voort} L.~H.~M.,  2005,
  \mn@doi [\aap] {10.1051/0004-6361:20042593}, \href
  {https://ui.adsabs.harvard.edu/abs/2005A&A...437.1069S} {437, 1069}

\bibitem[\protect\citeauthoryear{{Shen}, {Diercke}  \& {Denker}}{{Shen}
  et~al.}{2018}]{Shen(2018)}
{Shen} Z.,  {Diercke} A.,   {Denker} C.,  2018, \mn@doi [Astronomische
  Nachrichten] {10.1002/asna.201813536}, \href
  {https://ui.adsabs.harvard.edu/abs/2018AN....339..661S} {339, 661}

\bibitem[\protect\citeauthoryear{{Shine}, {Milkey}  \& {Mihalas}}{{Shine}
  et~al.}{1975}]{Shine1975}
{Shine} R.~A.,  {Milkey} R.~W.,   {Mihalas} D.,  1975, \mn@doi [\apj]
  {10.1086/153744}, \href
  {https://ui.adsabs.harvard.edu/abs/1975ApJ...199..724S} {199, 724}

\bibitem[\protect\citeauthoryear{{Sim{\~o}es}, {Reid}, {Milligan}  \&
  {Fletcher}}{{Sim{\~o}es} et~al.}{2019}]{Simoes2019}
{Sim{\~o}es} P. J.~A.,  {Reid} H. A.~S.,  {Milligan} R.~O.,   {Fletcher} L.,
  2019, \mn@doi [\apj] {10.3847/1538-4357/aaf28d}, \href
  {https://ui.adsabs.harvard.edu/abs/2019ApJ...870..114S} {870, 114}

\bibitem[\protect\citeauthoryear{{Skumanich}, {Smythe}  \&
  {Frazier}}{{Skumanich} et~al.}{1975}]{Skumanich(1975)}
{Skumanich} A.,  {Smythe} C.,   {Frazier} E.~N.,  1975, \mn@doi [apj]
  {10.1086/153846}, \href {http://adsabs.harvard.edu/abs/1975ApJ...200..747S}
  {200, 747}

\bibitem[\protect\citeauthoryear{{Skumanich}, {Lites}  \& {Mart{\'\i}nez
  Pillet}}{{Skumanich} et~al.}{1994}]{Skumanich(1994)}
{Skumanich} A.,  {Lites} B.~W.,   {Mart{\'\i}nez Pillet} V.,  1994, in {Rutten}
  R.~J.,  {Schrijver} C.~J.,  eds,  NATO Advanced Study Institute (ASI) Series
  C Vol. 433, Solar Surface Magnetism. p.~99

\bibitem[\protect\citeauthoryear{{Smith}}{{Smith}}{1960}]{Smith(1960)}
{Smith} E. V.~P.,  1960, \mn@doi [apj] {10.1086/146913}, \href
  {https://ui.adsabs.harvard.edu/abs/1960ApJ...132..202S} {132, 202}

\bibitem[\protect\citeauthoryear{{Socas-Navarro} \&
  {Uitenbroek}}{{Socas-Navarro} \& {Uitenbroek}}{2004}]{SocasNavarro(2004)}
{Socas-Navarro} H.,  {Uitenbroek} H.,  2004, \mn@doi [apjl] {10.1086/383147},
  \href {https://ui.adsabs.harvard.edu/abs/2004ApJ...603L.129S} {603, L129}

\bibitem[\protect\citeauthoryear{{Solanki}}{{Solanki}}{1986}]{Solanki1986}
{Solanki} S.~K.,  1986, \aap, \href
  {https://ui.adsabs.harvard.edu/abs/1986A&A...168..311S} {168, 311}

\bibitem[\protect\citeauthoryear{{Solanki}}{{Solanki}}{2003}]{Solanki2003}
{Solanki} S.~K.,  2003, \mn@doi [\aapr] {10.1007/s00159-003-0018-4}, \href
  {https://ui.adsabs.harvard.edu/abs/2003A&ARv..11..153S} {11, 153}

\bibitem[\protect\citeauthoryear{{Solanki} \& {Brigljevic}}{{Solanki} \&
  {Brigljevic}}{1992}]{Solanki1992}
{Solanki} S.~K.,  {Brigljevic} V.,  1992, \aap, \href
  {https://ui.adsabs.harvard.edu/abs/1992A&A...262L..29S} {262, L29}

\bibitem[\protect\citeauthoryear{{Sowmya}, {Shapiro}, {Rouppe van der Voort},
  {Krivova}  \& {Solanki}}{{Sowmya} et~al.}{2023}]{Sowmya(2023)}
{Sowmya} K.,  {Shapiro} A.~I.,  {Rouppe van der Voort} L.~H.~M.,  {Krivova}
  N.~A.,   {Solanki} S.~K.,  2023, \mn@doi [arXiv e-prints]
  {10.48550/arXiv.2309.03690}, \href
  {https://ui.adsabs.harvard.edu/abs/2023arXiv230903690S} {p. arXiv:2309.03690}

\bibitem[\protect\citeauthoryear{{Stenflo}}{{Stenflo}}{2015}]{Stenflo2015}
{Stenflo} J.~O.,  2015, \mn@doi [\aap] {10.1051/0004-6361/201424685}, \href
  {https://ui.adsabs.harvard.edu/abs/2015A&A...573A..74S} {573, A74}

\bibitem[\protect\citeauthoryear{{Stenflo} \& {Lindegren}}{{Stenflo} \&
  {Lindegren}}{1977}]{Stenflo(1977)}
{Stenflo} J.~O.,  {Lindegren} L.,  1977, aap, \href
  {https://ui.adsabs.harvard.edu/abs/1977A&A....59..367S} {59, 367}

\bibitem[\protect\citeauthoryear{{Stix}}{{Stix}}{2002}]{Stix2002}
{Stix} M.,  2002, {The sun: an introduction}

\bibitem[\protect\citeauthoryear{{Teixeira}, {Sousa}, {Tsantaki}, {Monteiro},
  {Santos}  \& {Israelian}}{{Teixeira} et~al.}{2016}]{Teixeira(2016)}
{Teixeira} G.~D.~C.,  {Sousa} S.~G.,  {Tsantaki} M.,  {Monteiro}
  M.~J.~P.~F.~G.,  {Santos} N.~C.,   {Israelian} G.,  2016, \mn@doi [aap]
  {10.1051/0004-6361/201525783}, \href
  {https://ui.adsabs.harvard.edu/abs/2016A&A...595A..15T} {595, A15}

\bibitem[\protect\citeauthoryear{{Thompson}}{{Thompson}}{2006}]{Thompson(2006)}
{Thompson} W.~T.,  2006, \mn@doi [\aap] {10.1051/0004-6361:20054262}, \href
  {https://ui.adsabs.harvard.edu/abs/2006A&A...449..791T} {449, 791}

\bibitem[\protect\citeauthoryear{{Thompson}, {Watson}, {de Mooij}  \&
  {Jess}}{{Thompson} et~al.}{2017}]{Thompson(2017)}
{Thompson} A.~P.~G.,  {Watson} C.~A.,  {de Mooij} E.~J.~W.,   {Jess} D.~B.,
  2017, \mn@doi [mnras] {10.1093/mnrasl/slx018}, \href
  {http://adsabs.harvard.edu/abs/2017MNRAS.468L..16T} {468, L16}

\bibitem[\protect\citeauthoryear{{Title} \& {AIA Team}}{{Title} \& {AIA
  Team}}{2006}]{Title(2006)}
{Title} A.~M.,  {AIA Team} 2006, in AAS/Solar Physics Division Meeting \#37. p.
  36.05

\bibitem[\protect\citeauthoryear{{Tlatov} \& {Pevtsov}}{{Tlatov} \&
  {Pevtsov}}{2014}]{Tlatov(2014)}
{Tlatov} A.~G.,  {Pevtsov} A.~A.,  2014, \mn@doi [\solphys]
  {10.1007/s11207-013-0382-9}, \href
  {https://ui.adsabs.harvard.edu/abs/2014SoPh..289.1143T} {289, 1143}

\bibitem[\protect\citeauthoryear{{Toriumi}, {Airapetian}, {Hudson},
  {Schrijver}, {Cheung}  \& {DeRosa}}{{Toriumi} et~al.}{2020}]{Toriumi(2020)}
{Toriumi} S.,  {Airapetian} V.~S.,  {Hudson} H.~S.,  {Schrijver} C.~J.,
  {Cheung} M. C.~M.,   {DeRosa} M.~L.,  2020, \mn@doi [apj]
  {10.3847/1538-4357/abadf9}, \href
  {https://ui.adsabs.harvard.edu/abs/2020ApJ...902...36T} {902, 36}

\bibitem[\protect\citeauthoryear{{Trelles Arjona}, {Mart{\'\i}nez Gonz{\'a}lez}
   \& {Ruiz Cobo}}{{Trelles Arjona} et~al.}{2021}]{Arjona2021}
{Trelles Arjona} J.~C.,  {Mart{\'\i}nez Gonz{\'a}lez} M.~J.,   {Ruiz Cobo} B.,
  2021, \mn@doi [\apjl] {10.3847/2041-8213/ac0af2}, \href
  {https://ui.adsabs.harvard.edu/abs/2021ApJ...915L..20T} {915, L20}

\bibitem[\protect\citeauthoryear{{Unruh}, {Solanki}  \& {Fligge}}{{Unruh}
  et~al.}{1999}]{Unruh(1999)}
{Unruh} Y.~C.,  {Solanki} S.~K.,   {Fligge} M.,  1999, aap, \href
  {https://ui.adsabs.harvard.edu/abs/1999A&A...345..635U} {345, 635}

\bibitem[\protect\citeauthoryear{Varun et~al.,}{Varun
  et~al.}{2018}]{NagaVarun2018}
Varun Y.~N.,  et~al., 2018, in Proceedings of the 22nd All-Russia Conference on
  Solar and Solar-Terrestrial Physics. The Central Astronomical Observatory of
  the Russian Academy of Sciences at Pulkovo,
  \mn@doi{10.31725/0552-5829-2018-303-306}, \url
  {https://doi.org/10.31725/0552-5829-2018-303-306}

\bibitem[\protect\citeauthoryear{{Verma}}{{Verma}}{2018}]{Verma2018}
{Verma} M.,  2018, \mn@doi [\aap] {10.1051/0004-6361/201732214}, \href
  {https://ui.adsabs.harvard.edu/abs/2018A&A...612A.101V} {612, A101}

\bibitem[\protect\citeauthoryear{{Verma} \& {Denker}}{{Verma} \&
  {Denker}}{2014}]{Verma2014}
{Verma} M.,  {Denker} C.,  2014, \mn@doi [\aap] {10.1051/0004-6361/201322476},
  \href {https://ui.adsabs.harvard.edu/abs/2014A&A...563A.112V} {563, A112}

\bibitem[\protect\citeauthoryear{{Vernazza}, {Avrett}  \& {Loeser}}{{Vernazza}
  et~al.}{1981}]{Vernazza(1981)}
{Vernazza} J.~E.,  {Avrett} E.~H.,   {Loeser} R.,  1981, \mn@doi [apjs]
  {10.1086/190731}, \href
  {https://ui.adsabs.harvard.edu/abs/1981ApJS...45..635V} {45, 635}

\bibitem[\protect\citeauthoryear{{Vissers} et~al.,}{{Vissers}
  et~al.}{2021}]{2021Vissers}
{Vissers} G.~J.~M.,  et~al., 2021, \mn@doi [\aap]
  {10.1051/0004-6361/202038900}, \href
  {https://ui.adsabs.harvard.edu/abs/2021A&A...645A...1V} {645, A1}

\bibitem[\protect\citeauthoryear{{Wallace}, {Hinkle}  \&
  {Livingston}}{{Wallace} et~al.}{2005}]{Wallace(2005)}
{Wallace} L.,  {Hinkle} K.,   {Livingston} W.~C.,  2005, {An atlas of sunspot
  umbral spectra in the visible from 15,000 to 25,500 cm-{\textonesuperior}
  (3920 to 6664 {\r{A}})}

\bibitem[\protect\citeauthoryear{{Wang}, {Yurchyshyn}, {Liu}, {Ahn}, {Toriumi}
  \& {Cao}}{{Wang} et~al.}{2018}]{Wang2018}
{Wang} H.,  {Yurchyshyn} V.,  {Liu} C.,  {Ahn} K.,  {Toriumi} S.,   {Cao} W.,
  2018, \mn@doi [Research Notes of the American Astronomical Society]
  {10.3847/2515-5172/aaa670}, \href
  {https://ui.adsabs.harvard.edu/abs/2018RNAAS...2....8W} {2, 8}

\bibitem[\protect\citeauthoryear{{White} \& {Livingston}}{{White} \&
  {Livingston}}{1978}]{White(1978)}
{White} O.~R.,  {Livingston} W.,  1978, \mn@doi [\apj] {10.1086/156650}, \href
  {https://ui.adsabs.harvard.edu/abs/1978ApJ...226..679W} {226, 679}

\bibitem[\protect\citeauthoryear{{Wilson} \& {Vainu Bappu}}{{Wilson} \& {Vainu
  Bappu}}{1957}]{Wilson(1957)}
{Wilson} O.~C.,  {Vainu Bappu} M.~K.,  1957, \mn@doi [apj] {10.1086/146339},
  \href {https://ui.adsabs.harvard.edu/abs/1957ApJ...125..661W} {125, 661}

\bibitem[\protect\citeauthoryear{{Worden}, {White}  \& {Woods}}{{Worden}
  et~al.}{1998}]{Worden1998}
{Worden} J.~R.,  {White} O.~R.,   {Woods} T.~N.,  1998, \mn@doi [\solphys]
  {10.1023/A:1004921707249}, \href
  {https://ui.adsabs.harvard.edu/abs/1998SoPh..177..255W} {177, 255}

\bibitem[\protect\citeauthoryear{{Worden}, {Woods}, {Neupert}  \&
  {Delaboudini{\`e}re}}{{Worden} et~al.}{1999}]{Worden1999}
{Worden} J.,  {Woods} T.~N.,  {Neupert} W.~M.,   {Delaboudini{\`e}re} J.-P.,
  1999, \mn@doi [\apj] {10.1086/306693}, \href
  {https://ui.adsabs.harvard.edu/abs/1999ApJ...511..965W} {511, 965}

\bibitem[\protect\citeauthoryear{{Yadav}, {de la Cruz Rodr{\'\i}guez}, {Kerr},
  {D{\'\i}az Baso}  \& {Leenaarts}}{{Yadav} et~al.}{2022}]{Yadav2022}
{Yadav} R.,  {de la Cruz Rodr{\'\i}guez} J.,  {Kerr} G.~S.,  {D{\'\i}az Baso}
  C.~J.,   {Leenaarts} J.,  2022, \mn@doi [\aap] {10.1051/0004-6361/202243440},
  \href {https://ui.adsabs.harvard.edu/abs/2022A&A...665A..50Y} {665, A50}

\bibitem[\protect\citeauthoryear{Yang, Zhang, Zhu  \& Song}{Yang
  et~al.}{2017}]{Yang2017}
Yang S.,  Zhang J.,  Zhu X.,   Song Q.,  2017, \mn@doi [The Astrophysical
  Journal] {10.3847/2041-8213/aa9476}, 849, L21

\bibitem[\protect\citeauthoryear{{Yoon}, {Yun}  \& {Kim}}{{Yoon}
  et~al.}{1995}]{Yoon(1995)}
{Yoon} T.-S.,  {Yun} H.~S.,   {Kim} J.-H.,  1995, Journal of Korean
  Astronomical Society, \href
  {https://ui.adsabs.harvard.edu/abs/1995JKAS...28..245Y} {28, 245}

\bibitem[\protect\citeauthoryear{{Zechmeister} et~al.,}{{Zechmeister}
  et~al.}{2020}]{Zechmeister(2020)}
{Zechmeister} M.,  et~al., 2020, {SERVAL: SpEctrum Radial Velocity AnaLyser}
  (\mn@eprint {ascl} {2006.011})

\bibitem[\protect\citeauthoryear{{Zhao} \& {Dumusque}}{{Zhao} \&
  {Dumusque}}{2023}]{Zhao(2023)}
{Zhao} Y.,  {Dumusque} X.,  2023, \mn@doi [\aap] {10.1051/0004-6361/202244568},
  \href {https://ui.adsabs.harvard.edu/abs/2023A&A...671A..11Z} {671, A11}

\bibitem[\protect\citeauthoryear{{Zhu}, {Lin}, {Wang}, {Liu}  \& {Yang}}{{Zhu}
  et~al.}{2019}]{Zhu(2019)}
{Zhu} G.,  {Lin} G.,  {Wang} D.,  {Liu} S.,   {Yang} X.,  2019, \mn@doi
  [\solphys] {10.1007/s11207-019-1517-4}, \href
  {https://ui.adsabs.harvard.edu/abs/2019SoPh..294..117Z} {294, 117}

\bibitem[\protect\citeauthoryear{Zou, Jiang, Feng, Zuo, Wang  \& Wei}{Zou
  et~al.}{2019}]{Zou2019}
Zou P.,  Jiang C.,  Feng X.,  Zuo P.,  Wang Y.,   Wei F.,  2019, \mn@doi [The
  Astrophysical Journal] {10.3847/1538-4357/aaf3b7}, 870, 97

\bibitem[\protect\citeauthoryear{Zou, Jiang, Wei, Feng, Zuo  \& Wang}{Zou
  et~al.}{2020}]{Zou_2020}
Zou P.,  Jiang C.,  Wei F.,  Feng X.,  Zuo P.,   Wang Y.,  2020, \mn@doi [The
  Astrophysical Journal] {10.3847/1538-4357/ab6aa8}, 890, 10

\bibitem[\protect\citeauthoryear{{da Silva Santos}, {de la Cruz Rodr{\'\i}guez}
   \& {Leenaarts}}{{da Silva Santos} et~al.}{2018}]{Joao2018}
{da Silva Santos} J.~M.,  {de la Cruz Rodr{\'\i}guez} J.,   {Leenaarts} J.,
  2018, \mn@doi [\aap] {10.1051/0004-6361/201833664}, \href
  {https://ui.adsabs.harvard.edu/abs/2018A&A...620A.124D} {620, A124}

\bibitem[\protect\citeauthoryear{{da Silva Santos}, {Danilovic}, {Leenaarts},
  {de la Cruz Rodr{\'\i}guez}, {Zhu}, {White}, {Vissers}  \& {Rempel}}{{da
  Silva Santos} et~al.}{2022}]{Joao2022}
{da Silva Santos} J.~M.,  {Danilovic} S.,  {Leenaarts} J.,  {de la Cruz
  Rodr{\'\i}guez} J.,  {Zhu} X.,  {White} S.~M.,  {Vissers} G.~J.~M.,
  {Rempel} M.,  2022, \mn@doi [\aap] {10.1051/0004-6361/202243191}, \href
  {https://ui.adsabs.harvard.edu/abs/2022A&A...661A..59D} {661, A59}

\bibitem[\protect\citeauthoryear{{de Beurs} et~al.,}{{de Beurs}
  et~al.}{2022}]{Beurs(2022)}
{de Beurs} Z.~L.,  et~al., 2022, \mn@doi [\aj] {10.3847/1538-3881/ac738e},
  \href {https://ui.adsabs.harvard.edu/abs/2022AJ....164...49D} {164, 49}

\bibitem[\protect\citeauthoryear{{de la Cruz Rodr{\'\i}guez}, {De Pontieu},
  {Carlsson}  \& {Rouppe van der Voort}}{{de la Cruz Rodr{\'\i}guez}
  et~al.}{2013}]{Jaime13}
{de la Cruz Rodr{\'\i}guez} J.,  {De Pontieu} B.,  {Carlsson} M.,   {Rouppe van
  der Voort} L.~H.~M.,  2013, \mn@doi [\apjl] {10.1088/2041-8205/764/1/L11},
  \href {https://ui.adsabs.harvard.edu/abs/2013ApJ...764L..11D} {764, L11}

\bibitem[\protect\citeauthoryear{{de la Cruz Rodr{\'\i}guez}, {Leenaarts},
  {Danilovic}  \& {Uitenbroek}}{{de la Cruz Rodr{\'\i}guez}
  et~al.}{2019}]{Jaime2019}
{de la Cruz Rodr{\'\i}guez} J.,  {Leenaarts} J.,  {Danilovic} S.,
  {Uitenbroek} H.,  2019, \mn@doi [\aap] {10.1051/0004-6361/201834464}, \href
  {https://ui.adsabs.harvard.edu/abs/2019A&A...623A..74D} {623, A74}

\makeatother
\end{thebibliography}




\appendix

\section{SDO segmentation maps example}
\label{appendix:a}

\begin{figure}

	\centering
	\includegraphics[width=8.5cm]{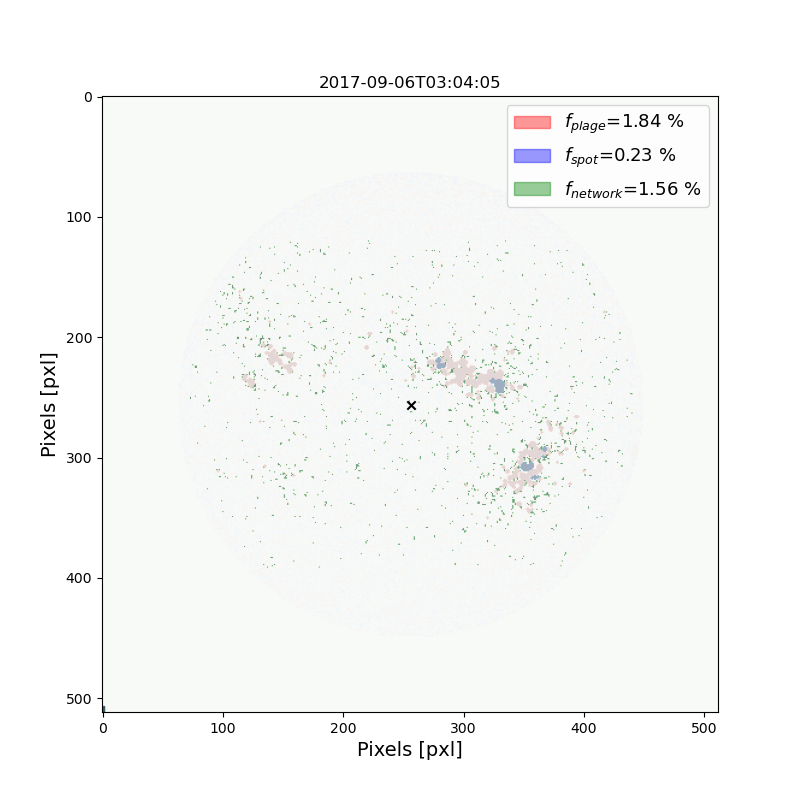}
	\caption{Example of the segmentation of SDO AIA1700 filtergrams for the same event as displayed with the Meudon spectroheliograph (see Fig.\ref{FigMeudon1}): the 4 September 2017 event (exact time in the title). The CLV of the quiet Sun has already been removed. Plages (red regions) are considered as regions brighter than 25\% the quiet chromosphere, spots (blue regions) as 25\% darker than the quiet chromosphere. The network is defined as regions brighter than $I>I_{50}+3\cdot(I_{50}-I_{16})$, with $I_{p}$ the intensity value corresponding to the $p$ percentile of the intensity distribution. Filling factor of the three components are indicated in the legend. A large plage ($x=300,y=220$) connecting two sunspots and another at ($x=350,y=320$) are the main AR at that time. 
        }
	\label{FigSDO2}
 
\end{figure}

We displayed in Fig.\ref{FigSDO2} an example of the SDO AIA 1700 filtergrams segmentation. The image is a 515 $\times$ 512 resolution image for which the Sun radius covers in average 220 pixels. The zones flagged as plages, spots and network based on the threshold values (see Sect.~\ref{sec:sdo}) are displayed in dark red, dark blue color and dark green colors.

\section{Meudon segmentation maps example}
\label{appendix:a}

\begin{figure}

	\centering
	\includegraphics[width=8.5cm]{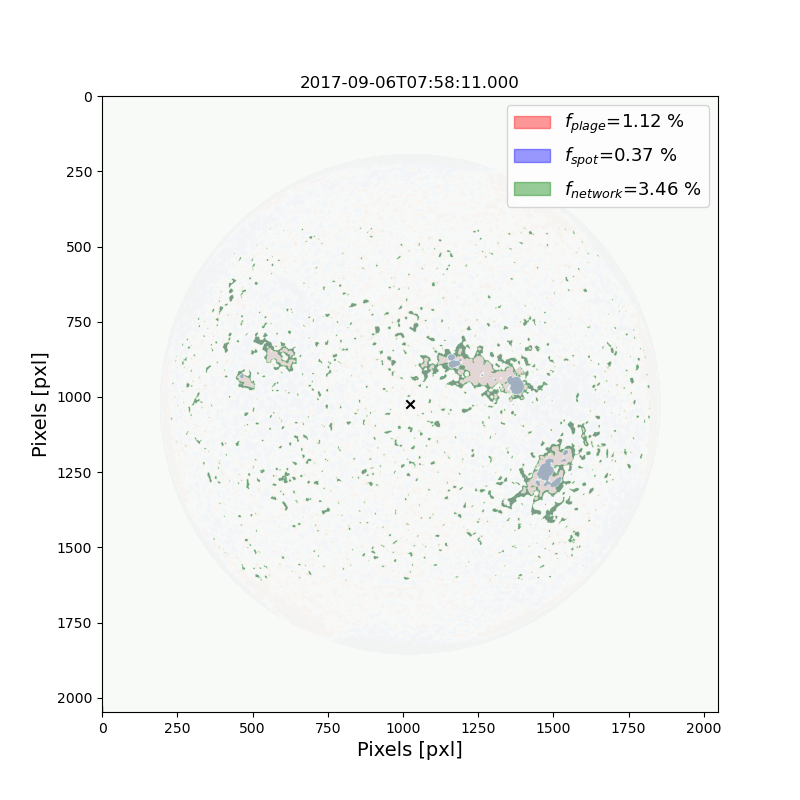}
	\caption{Same as Fig.\ref{FigSDO2} but for the Meudon segmentation. The closest date matching the SDO observation was selected to compare the segmentation map. Both segmentation maps are closely in agreement for spots and plages components, but tend to differ for the diffuse network component.  
        }
	\label{FigSDO2_bis}
 
\end{figure}

We displayed in Fig.\ref{FigSDO2_bis} an example of the Meudon Ca II K segmentation map for a similar epoch as in Fig.\ref{FigSDO2}. The image is a 2048 $\times$ 2048 resolution image for which the Sun radius covers in average 836 pixels. The zones flagged as plages, spots and network based on the threshold values (see Sect.~\ref{sec:meudon}) are displayed in dark red, dark blue color and dark green colors. By comparing the SDO and Meudon segmentation maps, a close agreement can be found. However, the network filling factor of Meudon appears as slightly overestimated compared to SDO that can be due to a threshold plage value too high. This is notably visible by a larger amount of network surrounding the plages compared to SDO. Also, it is unclear if part of the difference is due to the better S/N of SDO or to its lower spatial resolution. Note that a perfect match between both frames are not necessarily expected since the segmentation criterion are performed on filters probing different solar atmospheric layers.

\section{Correction of the ISS spectra}
\label{appendix:b}

ISS high resolution spectra are contaminated by background scattering light and imperfect correction. Such effects are visible by clear offset flux values with weird curvature of the photospheric wings and anomalous core emission. In order to correct for the effect, a reference spectrum made of the 10 percent of the quietest observations was constructed (mainly made of 2006 observations as visible in Fig.\ref{FigISS_sindex}). The background correction was obtained by fitting a polynomial relation between the spectra and the reference spectra: 

\begin{equation}
    I(\lambda,t)-I_{\text{ref}}(\lambda) = A(t)+B(t)\cdot I_{\text{ref}}(\lambda)+C(t)\cdot(I_{\text{ref}}(\lambda))^2
\end{equation}

\begin{figure}

	\centering
	\includegraphics[width=8.5cm]{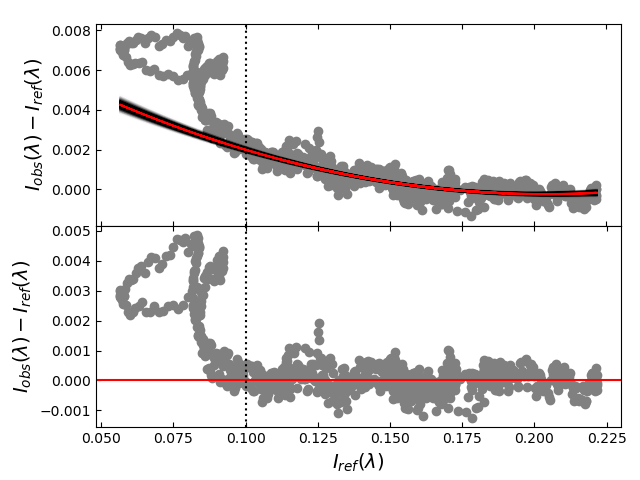}
	\caption{Example of light scattering correction on an ISS spectrum of the Ca II K line taken the 2015-12-02. The correction is obtained by adjusting the photospheric flux ($I_{\text{ref}}(\lambda)>0.1$) between the observed spectrum $I_{\text{obs}}(\lambda)$ and a reference spectrum at solar minima $I_{\text{ref}}(\lambda)$. \textbf{Top panel}: a quadratic relation (red curve) is assumed and fit on the photospheric component (right side of the vertical dotted line). The relation is then extrapolated on the chromospheric part corresponding to the line core ($I_{\text{ref}}(\lambda)<0.1$). Uncertainties from the quadratic fit are visualised by 100 hundreds random realisations in the $1\sigma$ of the parameters distributions. \textbf{Bottom panel}: Corrected spectra. The core emission is reduced from 0.0070, to half its value at 0.0035. }
	\label{FigISS_correction}
 
\end{figure}

Such correction is equivalent to scale the photospheric flux level of the observations to a reference value. The quadratic fit was performed for $I_{\text{ref}}(\lambda)>0.10$ in order to avoid the core emission of the line.  The coefficients of the fit were then used to correct the spectra. An example for the correction is showed in Fig.\ref{FigISS_correction}. In that example the emission is corrected to half its raw value. 

\section{Extension of the analyses for the Ca II H line}
\label{appendix:c}

The paper mainly focused the analyses on the Ca II K line for simplicity. However, we noted that qualitatively a similar behaviour was observed for the Ca II H. Previous authors have already showed the almost perfect linear relation between both lines \citep{Kneer(1981),Lemaire(1984)}. We show the result of the Gaussian fit on the binned $\mu$ profiles of the Meudon spectroheliograph in Fig.\ref{FigMeudon4}. Similar behaviour are observed for spots and plages compared to the Ca II K lines in Fig.\ref{FigMeudon4}. A notable difference with the Ca II K line concerns the CLV of the network which is not visible in any Gaussian parameter.

\begin{figure*}

	\centering
	\includegraphics[width=18cm]{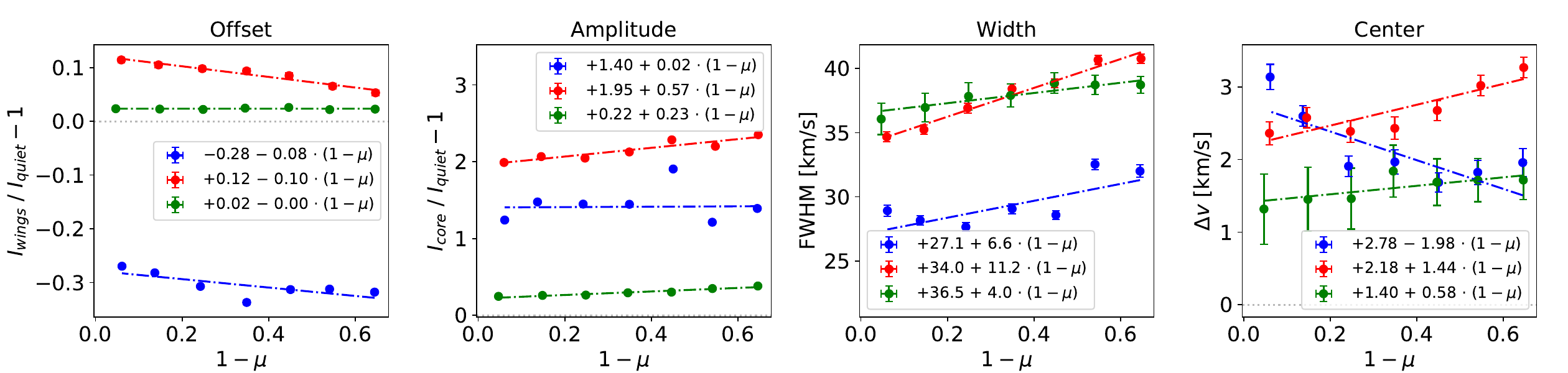}
	\caption{Same as Fig.\ref{FigMeudon4} for the Ca II H line. Even if parameters are slightly different, similar trend are observed. }
	\label{FigMeudon5}
 
\end{figure*}
 
\begin{figure*}

	\centering
	\includegraphics[width=18cm]{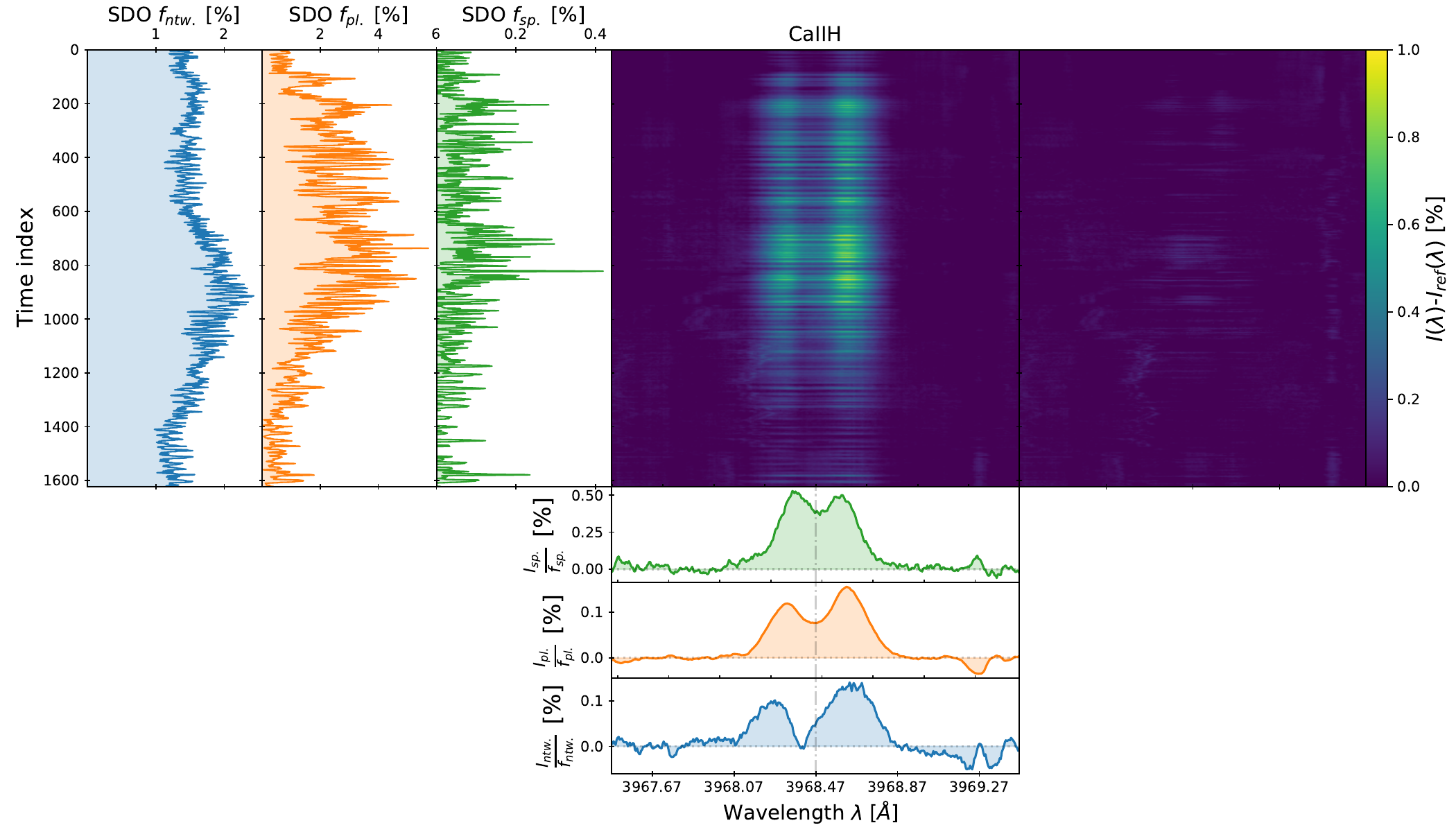}
	\caption{Same as Fig.\ref{FigComp1} for the Ca II H line. The emission profiles are almost identical to the Ca II K line. A thin profile for the spot, a broad asymmetric for the plage and a broader profile with core center depletion for the network.}
	\label{FigComp2}
 
\end{figure*}

In a similar way, we performed the ISS decomposition of the Ca II H line with SDO filling factors (see Fig.\ref{FigComp2}). The emission profiles are almost identical to the Ca II K line (see Fig.\ref{FigComp1}).


\bsp	
\label{lastpage}
\end{document}